\documentclass[letterpaper,english,aps, nofootinbib, pr, twocolumn, lengthcheck, superscriptaddress]{revtex4-2}

\usepackage{amsmath}

\usepackage[T1]{fontenc}
\usepackage[colorlinks=true,linktocpage=true,linkcolor=blue,citecolor=blue]{hyperref}
\usepackage{color,txfonts}
\usepackage{verbatim}
\usepackage{bm}

\usepackage{hyperref}
\usepackage{amssymb}
\usepackage{graphicx}
\usepackage{slashed}
\usepackage{wrapfig}
\usepackage{bbm}
\usepackage[caption=false]{subfig}
\usepackage{verbatim}

\begin{document}
\title{Resolving the Stellar-Collapse and Hierarchical-Merger origins of the Coalescing Black Holes}

\author{Yin-Jie Li}
\email{Contributed equally.}
\affiliation{Key Laboratory of Dark Matter and Space Astronomy, Purple Mountain Observatory, Chinese Academy of Sciences, Nanjing 210023, People's Republic of China}

\author{Yuan-Zhu Wang}
\email{Contributed equally.}
\affiliation{Institute for Theoretical Physics and Cosmology, Zhejiang University of Technology, Hangzhou, 310032, People's Republic of China}
\affiliation{Key Laboratory of Dark Matter and Space Astronomy, Purple Mountain Observatory, Chinese Academy of Sciences, Nanjing 210023, People's Republic of China}

\author{Shao-Peng Tang}
\affiliation{Key Laboratory of Dark Matter and Space Astronomy, Purple Mountain Observatory, Chinese Academy of Sciences, Nanjing 210023, People's Republic of China}

\author{Yi-Zhong Fan}
\email{The corresponding author: yzfan@pmo.ac.cn}
\affiliation{Key Laboratory of Dark Matter and Space Astronomy, Purple Mountain Observatory, Chinese Academy of Sciences, Nanjing 210023, People's Republic of China}
\affiliation{School of Astronomy and Space Science, University of Science and Technology of China, Hefei, Anhui 230026, People's Republic of China}

\def\aj{AJ}                   
\def\actaa{Acta Astron.}      
\def\araa{ARA\&A}             
\def\apj{Astrophys.~J.}                 
\def\apjl{Astrophys.~J.~Lett.}                
\def\apjs{ApJS}               
\def\ao{Appl.~Opt.}           
\def\apss{Ap\&SS}             
\def\aap{A\&A}                
\def\aapr{A\&A~Rev.}          
\def\aaps{A\&AS}              
\def\azh{AZh}                 
\def\baas{BAAS}               
\def\bac{Bull. astr. Inst. Czechosl.}
\def\caa{Chinese Astron. Astrophys.}
\def\cjaa{Chinese J. Astron. Astrophys.}
\def\icarus{Icarus}           
\def\jcap{J. Cosmology Astropart. Phys.}
\def\jrasc{JRASC}             
\def\memras{MmRAS}            
\def\mnras{MNRAS}             
\def\na{New A}                
\def\nar{New A Rev.}          
\def\pra{Phys.~Rev.~A}        
\def\prb{Phys.~Rev.~B}        
\def\prc{Phys.~Rev.~C}        
\def\prd{Phys.~Rev.~D}        
\def\pre{Phys.~Rev.~E}        
\def\prl{Phys.~Rev.~Lett.}    
\def\pasa{PASA}               
\def\pasp{PASP}               
\def\pasj{PASJ}               
\def\rmxaa{Rev. Mexicana Astron. Astrofis.}%
\def\qjras{QJRAS}             
\def\skytel{S\&T}             
\def\solphys{Sol.~Phys.}      
\def\sovast{Soviet~Ast.}      
\def\ssr{Space~Sci.~Rev.}     
\def\zap{ZAp}                 
\def\nat{Nature}              
\def\iaucirc{IAU~Circ.}       
\def\aplett{Astrophys.~Lett.} 
\def\apspr{Astrophys.~Space~Phys.~Res.}
\def\bain{Bull.~Astron.~Inst.~Netherlands} 
\def\fcp{Fund.~Cosmic~Phys.}  
\def\gca{Geochim.~Cosmochim.~Acta}   
\def\grl{Geophys.~Res.~Lett.} 
\def\jcp{J.~Chem.~Phys.}      
\def\jgr{J.~Geophys.~Res.}    
\def\jqsrt{J.~Quant.~Spec.~Radiat.~Transf.}
\def\memsai{Mem.~Soc.~Astron.~Italiana}
\def\nphysa{Nucl.~Phys.~A}   
\def\physrep{Phys.~Rep.}   
\def\physscr{Phys.~Scr}   
\def\planss{Planet.~Space~Sci.}   
\def\procspie{Proc.~SPIE}   

\let\astap=\aap
\let\apjlett=\apjl
\let\apjsupp=\apjs
\let\applopt=\ao

\begin{abstract}
Spin and mass properties provide essential clues in distinguishing the origins of coalescing black holes (BHs). 
With a dedicated semiparametric population model for the coalescing binary black holes (BBHs), 
we identify two distinct categories of BHs among the GWTC-3 events, which is {favored over the one population scenario by} a logarithmic Bayes factor ($\ln\mathcal{B}$) of 7.5. One category, with a mass ranging from $\sim 25M_\odot$ to $\sim 80M_\odot$, is distinguished by the high spin magnitudes ($\sim0.75$) and consistent with the hierarchical merger origin. The other category, characterized by low spins, has a sharp mass cutoff at $\sim 40M_\odot$, which is natural for the stellar-collapse origin and in particular the pair-instability explosion of massive stars. 
We infer the local hierarchical merger rate density as $0.46^{+0.61}_{-0.24}~{\rm Gpc^{-3}yr^{-1}}$. 
Additionally, we find that a fraction of the BBHs has a cosine-spin-tilt-angle distribution concentrated preferentially around $1$, and the fully isotropic distribution for spin orientation is disfavored by a $\ln\mathcal{B}$ of -6.3, suggesting that the isolated field evolution channels are contributing to the total population.
\end{abstract}


\maketitle

{\it Introduction.}---Thanks to the excellent performance of the LIGO/Virgo network, about $90$ gravitational wave (GW) events have been officially reported so far, and most of them are coalescing binary black holes (BBHs) \cite{2019PhRvX...9c1040A,2021PhRvX..11b1053A,2021arXiv210801045T,2021arXiv211103606T}. The origins and evolution paths of these binaries, however, are still under debate
\citep{2022PhR...955....1M,2022LRR....25....1M,2023PhRvX..13a1048A}.

As a result of the (pulsational) pair-instability supernova [(P)PISN] explosions \cite{2017ApJ...836..244W,2021ApJ...912L..31W}, the BHs formed by stellar evolutions are expected to be absent in the so-called upper-mass gap (UMG), which is widely anticipated to start at $\sim 40-55~M_{\odot}$, though the threshold may be shifted under some special circumstances \cite{2019ApJ...887...53F,2020ApJ...902L..36F,2020MNRAS.493.4333R,2021ApJ...912L..31W,2020ApJ...890..113B,2021MNRAS.504..146V,2021MNRAS.501.4514C}.
However, the hierarchical mergers in dynamical environments \citep{2021NatAs...5..749G,2019PhRvL.123r1101Y,2022PhR...955....1M,2022LRR....25....1M}, the stellar mergers \citep{2020MNRAS.497.1043D,2020ApJ...904L..13R} and primordial BHs \cite{1974MNRAS.168..399C,2018PhRvL.121h1306C} may populate the UMG, and make it invisible.
In general, the hierarchical merger-formed BHs are distinguishable from those born in stellar explosions for their high spin magnitudes (with a typical value of $\sim 0.7$) \cite{2017ApJ...840L..24F,2017PhRvD..95l4046G,2021NatAs...5..749G}.
Therefore, it is possible to distinguish the category of higher-generation (HG) BHs via analyzing the mass versus spin-magnitude distribution for BHs from GW observations, and simultaneously constrain the lower-edge of the UMG (if it was not contaminated significantly by other exotic BHs). 
A UMG-like high-mass cutoff at $\sim 45 M_{\odot}$ favored by the GWTC-1 data \citep{2019ApJ...882L..24A,2019ApJ...887...53F} had been challenged by further observations, in particular, GW190521 \citep{2020PhRvL.125j1102A}, which suggested the absence of mass cutoff till $\gtrsim 80 M_{\odot}$ \citep{2021ApJ...913L...7A, 2022ApJ...924..101E, 2021ApJ...917...33L}. 
Some parametric investigations anyhow found that the UMG may still exist \citep{2021ApJ...907L...9N,2021ApJ...916L..16B,2021ApJ...913L..23E,2020ApJ...904L..26F,2021ApJ...913...42W} and there may be hierarchical mergers \citep{2020RNAAS...4....2K,2021ApJ...915L..35K,2022ApJ...941L..39W}, under some special astrophysical assumptions (e.g., single channel for all the BBHs). 
Such conclusions are however not confirmed by the non- or semiparametric approaches (with minimal assumptions) \citep[][]{2023PhRvX..13a1048A,2023PhRvD.108j3009G,2022ApJ...928..155T,2023ApJ...946...16E,2023arXiv230207289C}.
To reliably clarify the situation, {\it for the first time,} we propose a mixture flexible population model 
incorporating the correlation in the component-mass versus spin-magnitude distribution, to explore the subpopulations or groups of the coalescing BHs, and hence determine the origins and evolution channels of the BBHs.

{\it Population model.}---Since coalescing BHs with diverse origins may have different component-mass ($m$) and spin-magnitude ($\chi$) distributions, we use a mixture model for $m$-$\chi$ distribution, which reads

\begin{equation}
\begin{aligned}
\pi(m, \chi | {\bf \Lambda})=\sum_{i=1}^{N}{\pi_i(m, \chi | {\bf \Lambda}_i)\times r_i},\\
\end{aligned}
\end{equation}
 where $r_i$ is the mixing fraction of the $i$th component. 
 Different from previous approaches modeling on the chirp-mass, mass-ratio, and aligned-spin distributions \citep{2021ApJ...910..152Z,2021CQGra..38o5007T,2022ApJ...928..155T}, our model is more appropriate to investigate the subpopulations of {\it individual} BHs (in coalescing systems), and hence to identify the subpopulations in the BBHs. The $m$-$\chi$ distribution of $i$-th component is

\begin{equation}
\begin{aligned}
\pi_i(m, \chi | {\bf \Lambda}_i)&=\mathcal{PS}[m|\alpha_i,m_{{\rm min},i},m_{{\rm max},i},\delta_i,f_i(m;\{f_i^j\}_{j=0}^{N_{\rm knot}})]\\
&\times  \mathcal{G}(\chi|\chi_{{\rm min},i}, \chi_{{\rm max},i}, \mu_{\chi,i}, \sigma_{\chi,i}),
\end{aligned}
\end{equation}
where $\mathcal{PS}$ is a one-dimensional \textsc{PowerLawSpline} model \citep{2022ApJ...924..101E}, which provides a flexible and continuous function, thus is appropriate to determine the underlying mass distributions of the BHs in different categories if distinguishable; $f_i(m)$ is the cubic-spline perturbation function interpolated between ${N_{\rm knot}}$ knots placed in the mass range. 
We use 12 knots (The number of the knots does not affect our conclusions, see Fig.~S27 of the Supplemental Material \citep{supplemental}) to interpolate the perturbation function $f_i$, located linearly in the logarithm space within [6, 80] $M_{\odot}$, and restrict the perturbation to zero at the minimum and maximum knots. 
$\mathcal{G}$ is a truncated Gaussian within the range of [$\chi_{{\rm min},i}$, $\chi_{{\rm max},i}$] with a central value ($\mu_{\chi,i}$) and a standard deviation ($\sigma_{\chi,i}$).  

{As found previously, the two objects in a BBH are not randomly paired \cite{2020ApJ...891L..27F,2022ApJ...933L..14L}. The pairing functions for different channels (e.g., the dynamical and field channels) may not be the same \citep{2021ApJ...910..152Z,2022ApJ...941L..39W}. However,  currently we can not robustly differentiate the formation channel of most BBHs. Therefore, we take a simple pairing function of $(m_2/m_1)^\beta$. Anyhow, even for the unpaired (or randomly paired, i.e., $\beta=0$) scenario, our results are unchanged, implying that our findings are insensitive to the pairing function (see Sec.~\uppercase\expandafter{\romannumeral7} in the Supplemental Material \citep{supplemental}).}

{The cosine-tilt-angle distribution reads \citep{2023PhRvX..13a1048A}
\begin{equation}
\begin{aligned}
&\mathcal{GU}(\cos\theta_1, \cos\theta_2|  \zeta, \sigma_{{t}})=\\
&\mathcal{G}(\cos\theta_1|-1, 1,1 , \sigma_{t})\times \mathcal{G}(\cos\theta_2|-1, 1,1 , \sigma_{t})\times \zeta+\\
&(1-\zeta)\times \mathcal{U}(\cos\theta_1|-1,1)\times \mathcal{U}(\cos\theta_2|-1,1),
\end{aligned}
\end{equation} 
where $\mathcal{U}$ is the uniform distribution between $-1$ and 1, and $\mathcal{G}$ is the (normalized) Gaussian distribution truncated between $-1$ and 1.}
{Following \citet{2023PhRvX..13a1048A}, the merger rate density evolution function is taken as  $\mathcal{R}\propto(1 + z)^{2.7}$. 
The final population model takes the form of 
\begin{equation}\label{eq:main_pop}
\begin{aligned}
&\pi({\bf \lambda} | {\bf \Lambda}; \zeta, \sigma_{{t}}, \beta) =\\
& C({\bf \Lambda};\beta) \times \pi(m_1, \chi_1| {\bf \Lambda})\times\pi(m_2, \chi_2| {\bf \Lambda})\times (m_2/m_1)^\beta\\
&\times \Theta(m_1-m_2) \times \mathcal{GU}(\cos\theta_1, \cos\theta_2|  \zeta, \sigma_{{t}}) \times p(z|\gamma=2.7),
\end{aligned}
\end{equation} 
where $C({\bf \Lambda};\beta)$ is the normalized function, and $\Theta(m_1-m_2)$ is Heaviside function. All hyperparameters are described in Table~\uppercase\expandafter{\romannumeral1} in the Supplemental Material \citep{supplemental}.}

\begin{figure*}
	\centering  
\includegraphics[width=0.96\linewidth]{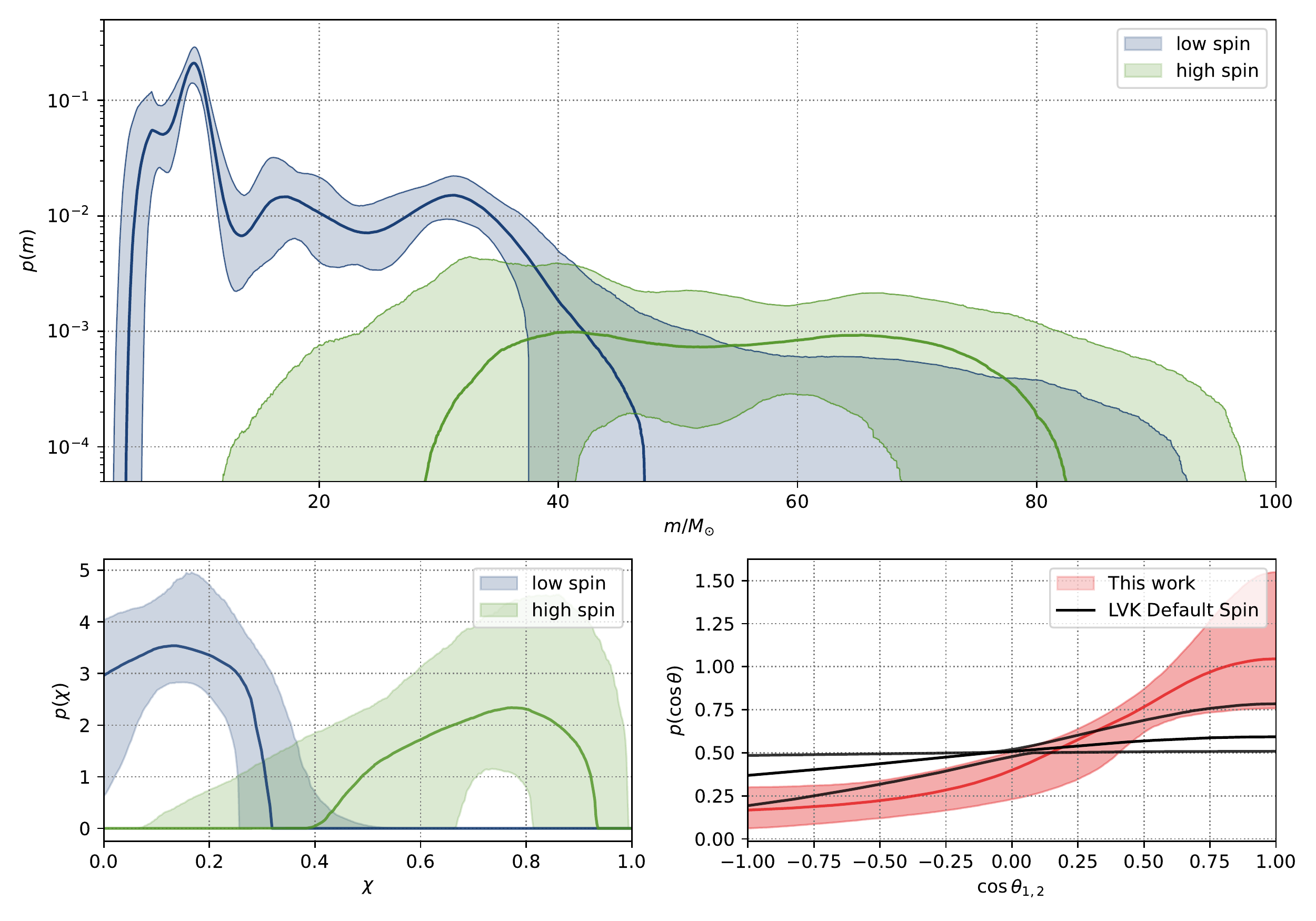}
\caption{Reconstructed mass (top) and spin (bottom) distribution of BHs; the solid curves are the medians and the colored bands are the 90\% credible intervals; note the spin-magnitude distribution of each subpopulation is normalized. 
}
\label{fig:dist}
\end{figure*}

\begin{figure*}
	\centering  
\includegraphics[width=0.96\linewidth]{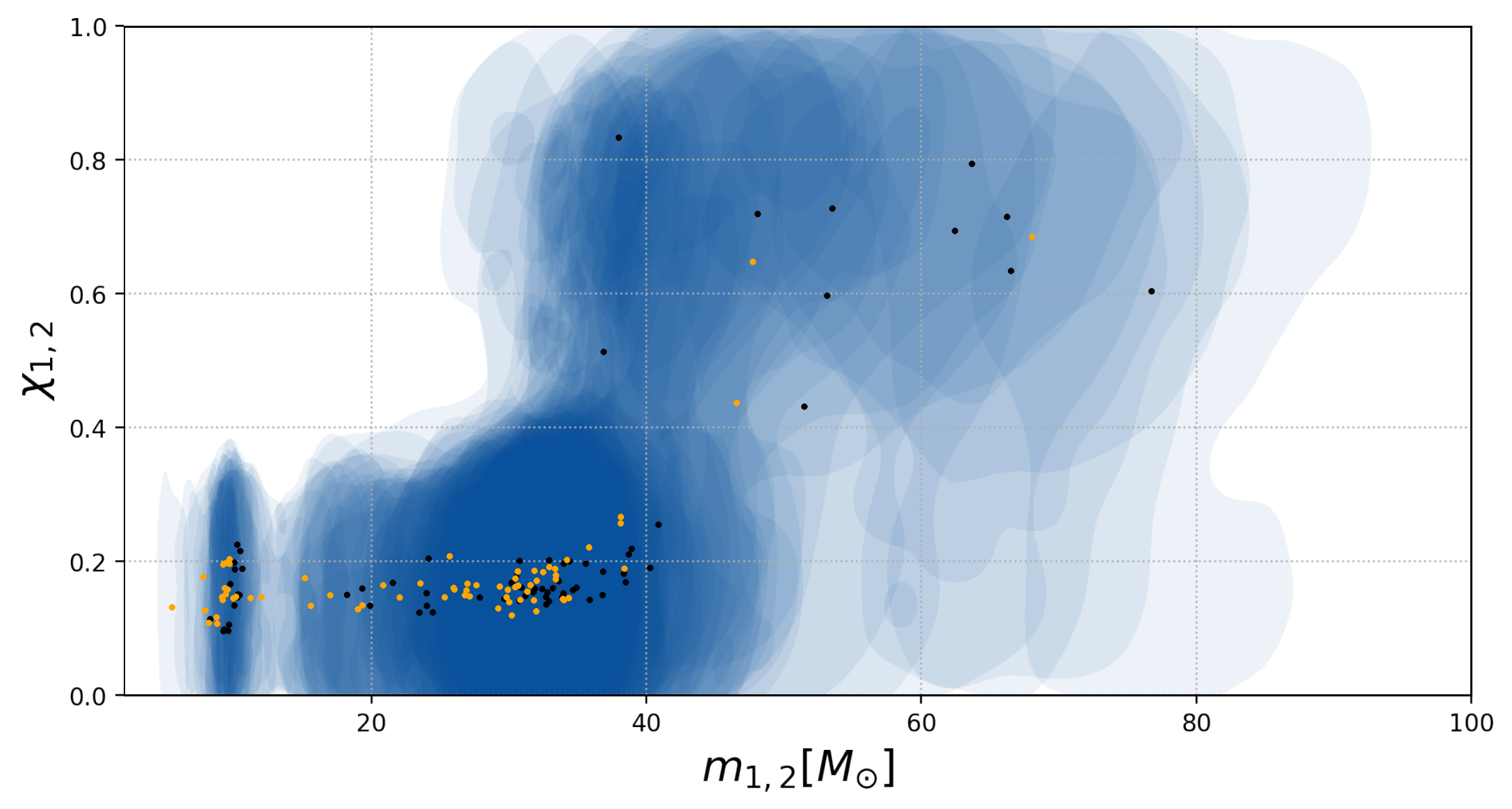}
\caption{Posteriors of individual component masses and spin magnitudes of BBHs in GWTC-3 reweighted to a population-informed prior inferred by the fiducial model. The shaded areas mark the $90\%$ credible regions and the black (orange) points stand for the mean values for the primary (secondary) BHs. The difference between the two groups is evident: the first group is lighter than $\sim50M_{\odot}$ with spin magnitudes $\lesssim0.3$, while the second group extends to the higher mass range and has a typical spin magnitude $\sim0.7$.}
\label{fig:reweighted}
\end{figure*}

{\it Results.}---We have performed hierarchical inferences and Bayesian model comparisons 
(see Secs.~\uppercase\expandafter{\romannumeral1},~\uppercase\expandafter{\romannumeral2},~\uppercase\expandafter{\romannumeral3}, and ~\uppercase\expandafter{\romannumeral4} of the Supplemental Material \citep{supplemental}). It reveals two groups of BHs with significantly different spin-magnitude distributions and mass distributions (see Fig.\ref{fig:dist}), which are evidently illustrated by the posterior distributions of the observed events weighted by the population-informed priors obtained by our two-component model (see Fig.\ref{fig:reweighted}). The two categories or groups of BHs are clear and identifiable.
For the spin-magnitude distributions, the first group (hereafter the low-spin group; LSG) peaks at $0.14^{+0.13}_{-0.12}$ and terminates at {$0.32_{-0.06}^{+0.43}$} (hereafter the values are for the median value and the 90\% symmetric interval, unless otherwise noted), while in the second group (hereafter the high-spin group; HSG), the spin-magnitude distribution starts at {$0.36_{-0.31}^{+0.30}$}, peaks at {$0.75^{+0.17}_{-0.27}$} and ends at $0.93^{+0.06}_{-0.12}$. The significantly different spin-magnitude distributions between the two groups indicates the different physical origins \citep{2021NatAs...5..749G}.
For the mass distributions, the LSG (HSG) starts at {$3.26^{+1.53}_{-0.108}M_{\odot}$ ($27.16^{+12.98}_{-19.49}M_{\odot}$)} and terminates at {$47.26^{+45.90}_{-9.81}M_{\odot}$ ($82.74^{+14.86}_{-14.29}M_{\odot}$)} with an overall power-law index of {$2.10^{+1.05}_{-1.17}$ ($0.82^{+4.22}_{-3.40}$)}. The posterior of other parameters for the mass and spin distributions is displayed in the Fig.~S20 and Fig.~S21 in the Supplemental Material \citep{supplemental}.

{\it (i)---Evidence for hierarchical mergers.}---As shown in the Fig.~\ref{fig:dist}, the spin-magnitude distribution of HSG (blue dashed region) is peaked at $\sim 0.6-0.8$, which could be naturally associated with the spin distribution of the remnants of BBH mergers \cite{2017ApJ...840L..24F,2017PhRvD..95l4046G}.
With the identification of the HSG as the HG category, we estimate that hierarchical mergers (containing at least one HG BH) take a fraction of {$2.6^{+3.7}_{-1.5}\%$}, and {$0.39^{+0.74}_{-0.27}\%$} of the sources have two HG BHs (see Fig.~S26 in Supplemental Material \citep{supplemental}). Such fractions may be too high to be realistic for the globular clusters alone \citep{2023MNRAS.522..466A}, thus the contributions from the nuclear star clusters \citep{2016ApJ...831..187A} and/or the accretion disks of active galactic nuclei (AGN) \citep{2019PhRvL.123r1101Y,2021MNRAS.507.3362T,2021ApJ...920L..42G} may be needed. 
Assuming the merger rate of BBHs evolves with the redshift as $\propto (z+1)^{2.7}$, as obtained by \citet{2023PhRvX..13a1048A}, we infer the local hierarchical merger rate density as {$0.46^{+0.61}_{-0.24}~{\rm Gpc^{-3}yr^{-1}}$}. 
Additionally, we have identified events with probabilities $>50\%$ to be hierarchical mergers (summarized in Table~\uppercase\expandafter{\romannumeral4} of Supplemental Material \citep{supplemental}). In particular, GW190521 and GW191109\_010717 have probabilities $>50\%$ to host double HG BHs. 

\begin{figure}
	\centering  
\includegraphics[width=0.96\linewidth]{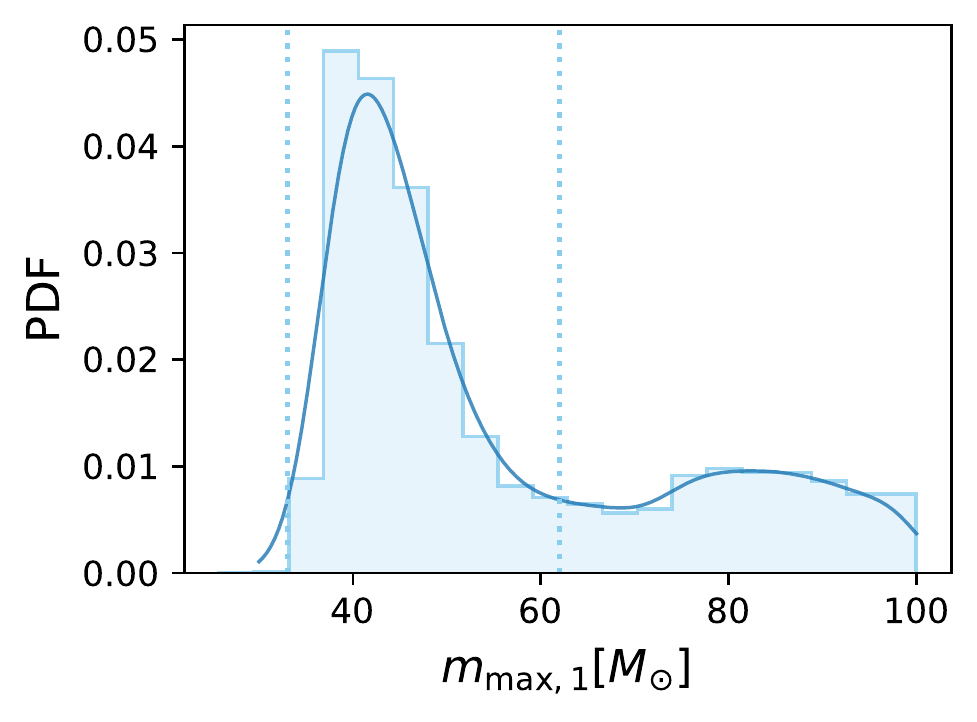}
\caption{Posterior of the maximum-mass cutoff of the LSG, the 68.3\% ($1\sigma$) minimal credible intervals are indicated by dashed line; $m_{\rm max,1}$ has a tail extending to the higher mass range, which may be caused by the flexibility of the \textsc{PowerLawSpline} mass model, see detailed discussion in Sec.~\uppercase\expandafter{\romannumeral8}~B. of the Supplemental Material \citep{supplemental}.}
\label{fig:mmax1}
\end{figure}

{\it (ii)---Evidence for pair-instability supernovae explosions.}---Apart from the HG category, the remaining BHs (i.e., the LSG) belongs to the stellar-collapse category. There is a rapid decline after $\sim 40M_{\odot}$ in the mass function of the LSG, as shown in Fig.~\ref{fig:dist}, which supports the existence of the UMG caused by the (P)PISN \cite{2017ApJ...836..244W,2021ApJ...912L..31W}. Indeed, we have {$m_{\rm max,1}=41.56_{-8.55}^{+20.46}M_{\odot}$} (the $1\sigma$ minimal credible interval, as shown in Fig.\ref{fig:mmax1}), which is in good agreement with the maximum mass of the stellar-formed BHs predicted by stellar evolution theories \cite{2019ApJ...887...53F,2020ApJ...902L..36F,2020MNRAS.493.4333R} {, as well as the results inferred with astrophysical-motivated parametric model and the deep-learning model \cite{2022ApJ...941L..39W,2022PhRvD.106j3013M}.}
{The mass of the 99\% (99.5\%) percentile for the first-generation is $37.42^{+4.29}_{-2.96}M_{\odot}$ ($39.61^{+8.70}_{-3.79}M_{\odot}$), see Fig.~S24 in the Supplemental Material \citep{supplemental}. The $m_{\rm max,1}$ has a tail caused by the flexibility of the \textsc{PowerLaw Spline} model, as illustrated by the Fig.~S23 in the Supplemental Material \citep{supplemental}.} \citet{2023PhRvX..13a1048A} suggest that either the UMG presents above $75M_{\odot}$ or there is a non-negligible fraction of the high-mass binaries formed in a way that avoids pair instability. The  latter is in agreement with our interpretation.

\begin{figure}
	\centering  
\includegraphics[width=0.96\linewidth]{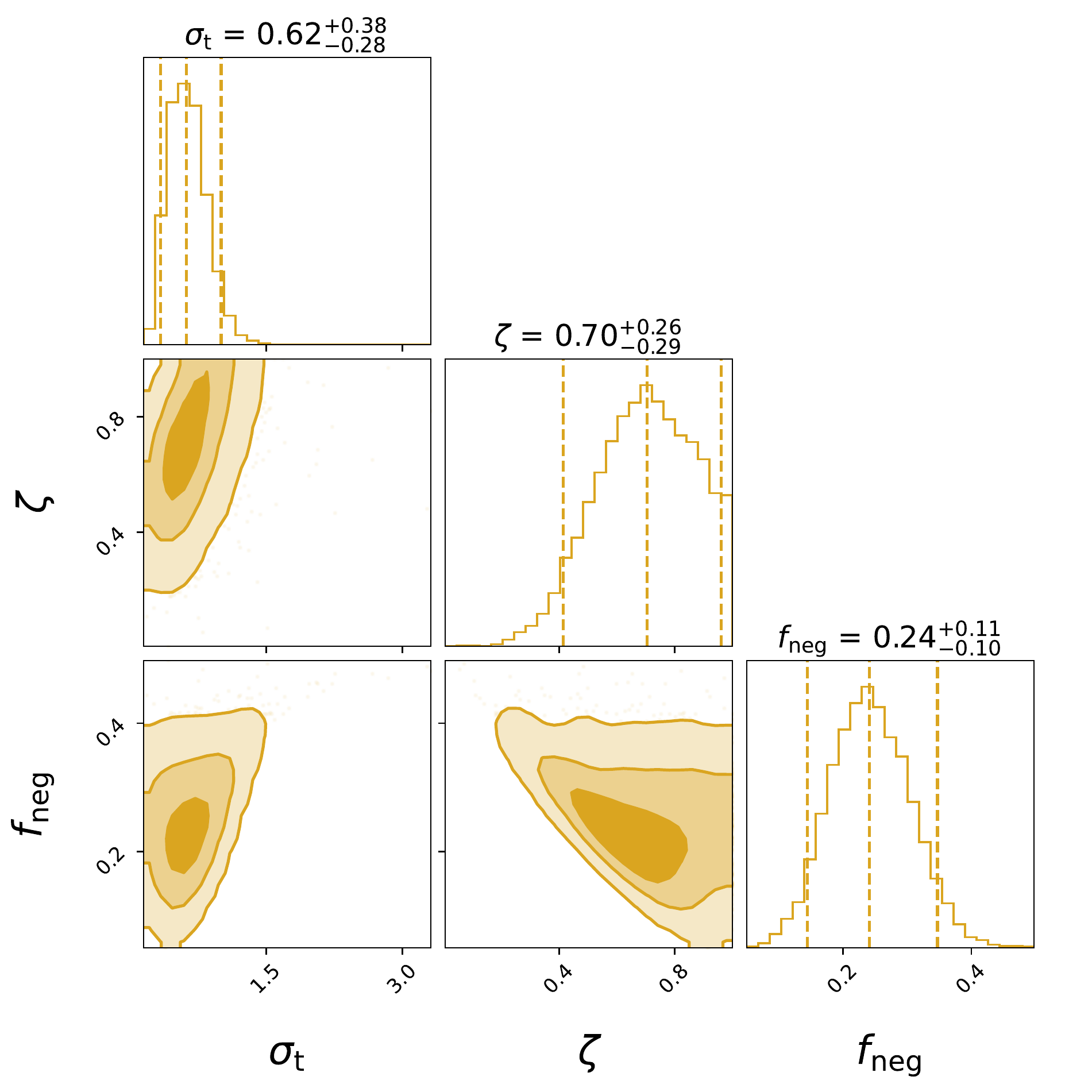}
\caption{Posterior distribution of $\sigma_{\rm t}$, $\zeta$, and the antialigned fraction $f_{\rm neg}$; the dashed and solid contours mark the central 50\% and 90\% posterior credible regions, respectively.}
\label{fig:sigmat}
\end{figure}

{\it (iii)---The spin-orientation distribution of BBHs.}---We find a fraction ($\zeta=0.70^{+0.26}_{-0.29}$) of the BBHs has a cosine-spin-tilt-angle distribution peaking at 1 with width of $\sigma_{\rm t} =0.62^{+0.38}_{-0.28}$ (see Fig.~\ref{fig:sigmat}), consistent with the isolated field BBHs \cite{2016ApJ...832L...2R}. 
We estimate that $24^{+11}_{-10}\%$ of BHs have spin-tilt angles $>90^{\circ}$.
Flatter distributions of $\cos\theta$ have been reported in \citep{2023PhRvX..13a1048A}, whereas we find the fully isotropic spin-orientation distribution is disfavored by $\ln\mathcal{B} =-6.3$.
The difference between our results and the previous \citep{2023PhRvX..13a1048A,2022A&A...668L...2V} is mainly attributed to the different configurations for the Monte Carlo integral in likelihood estimation and the different modeling of spin-magnitude versus mass distribution (see Sec.~\uppercase\expandafter{\romannumeral6} of the Supplemental Material \citep{supplemental} for more details).
 \citet{2022A&A...668L...2V} found that the $\cos\theta$ distribution may not peak at $\sim 1$. For comparison, we perform an inference with a variable $\mu_{\rm t}$ for the Two-component model, and obtain $0.79^{+0.19}_{-0.29}$ (see Fig.~S14 in the Supplemental Material \citep{supplemental}).
It is worth noticing that, the modeling of the tilt-angle distribution does not affect the identification of the two subpopulations of BHs in this work (see Sec.~\uppercase\expandafter{\romannumeral7} of the Supplemental Material \citep{supplemental}).

We have further investigate the $\cos\theta$ distributions in the two subpopulations, it shows that, beside the first-generation subpopulation, the HG subpopulation may also have a fraction of nearly aligned assembly.  
Nevertheless, the fully isotropic distribution for HG subpopulation cannot be ruled out yet (see Sec.~\uppercase\expandafter{\romannumeral9} of the Supplemental Material \citep{supplemental} for details).

{\it Conclusion and discussion.}---We investigate the population properties of BBHs in the GWTC-3, with a flexible semi-parametric population model. {For the first time,} we identify two subpopulations of coalescing BHs (with a $\ln\mathcal{B}=7.5$ comparing to the one-component model), which are nicely consistent with the widely discussed hierarchical mergers \cite{2021NatAs...5..749G} and the pair-instability mass gap (PIMG) \cite{2016A&A...594A..97B,2019ApJ...882..121S}. These two issues are currently of great concern in gravitational-wave astronomy \citep{2021NatAs...5..749G,2021ApJ...912L..31W,2022PhR...955....1M,2023PhRvX..13a1048A}.

It was found that the unequal-mass BBHs have larger effective spins \citep{2021ApJ...922L...5C,2023PhRvX..13a1048A}. 
The hierarchical mergers identified in this work may explain such an anticorrelation between $q$-$\chi_{\rm eff}$ (see Fig.~S30 of the Supplemental Material \citep{supplemental}). 
\citet{2022ApJ...929L..26F} found that the observed spins in x-ray binaries (XRBs) \citep{2006ARA&A..44...49R} are in tension with the BBH spin distribution at the $>99.9\%$ level. When we only account for the first-generation BBHs (i.e., the LSG), such tension would be more significant, 
indicating the lack of shared evolutionary paths between the two types of systems \citep{2022ApJ...938L..19G}.
The spin-magnitude distribution of HSG can partially overlap with that of the XRB BHs, however, there is significant tension between their mass distributions \citep{2023ApJ...946....4L}.

\citet{2021ApJ...915L..35K} reported evidence for hierarchical mergers in GWTC-2. However their population model consists of a specific mass function and a fixed spin-magnitude distribution for higher-generation BHs, assuming that all BBHs were formed dynamically in gravitationally bound clusters, which seems unnatural \citep{2021ApJ...910..152Z}.
Many other flexible semi- or nonparametric population models \cite{2022ApJ...928..155T,2023PhRvX..13a1048A,2023ApJ...946...16E,2023arXiv230207289C,2023ApJ...957...37R} have been used to fit the GWTC-3 data, but none of these approaches has taken into account the correlation in the component-mass versus spin-magnitude distribution and hence drawn different conclusions from us. 
 
After posting our initial version online, 
other semi- or nonparametric population models have also been developed to explore the subpopulations within the GWTC-3, see \citet{2023arXiv230401288G} and \citet{2024arXiv240403166R}. 
Although some findings of these two studies align with ours, neither provides evidence for a subpopulation with $\chi\sim0.7$ or a mass cutoff at $\sim 40M_{\odot}$. 
Different from us, \citet{2023arXiv230401288G} assumes that both spin magnitudes of a BBH follow the same distribution. While discrepancies between our findings and those of \citet{2024arXiv240403166R} likely arise from different approaches in modeling population parameters: we focus on spin magnitudes and orientations, while they concentrate on effective spins.

\citet{2021NatAs...5..749G} proposed that investigating the occurrence of hierarchical mergers constitutes an orthogonal and complementary direction to the usual ``field versus dynamics'' formation-channel debate.
The route to the former issue usually depends on the spin-magnitude distributions \citep{2017ApJ...840L..24F,2017PhRvD..95l4046G}, while the latter mainly relies on the spin-orientation distributions \citep{2016ApJ...832L...2R,2018PhRvD..98h4036G}. 
Our subsequent research \citep{2024arXiv240409668L}, building on the identification of hierarchical mergers in this study, 
shows that BBH components in the $10M_{\odot}$ and $35M_{\odot}$ peaks of the primary-mass distribution exhibit nearly aligned and isotropic spin orientations, respectively. These orientations correspond to field and dynamic formation channels, aligning with the properties reported in \citet{2023arXiv230401288G,2024arXiv240403166R} for the respective peaks. 

Encouragingly, hundreds of observations are promising after the next observing run \citep{2020LRR....23....3A}, so that the origins and formation channels may be clearly identified.
Additionally, the location of the PIMG will be measured more precisely, which can help constrain the $^{12}{\rm C}(\alpha,\gamma)^{16}{\rm O}$ in the stellar evolution theories \cite{2020ApJ...902L..36F}. Subsequently, one can even measure the lower edge of the PIMG evolving with the redshift, and hence the metallicity evolving  with the age of the Universe \cite{2021MNRAS.504..146V}.
Moreover, the lower edge of PIMG provides essential ingredients for measuring the expansion rate ($H(z)$) of the Universe, via so-called ``Spectral sirens'' \citep{2019ApJ...883L..42F,2022PhRvL.129f1102E,2023ApJ...949...76A,2024arXiv240611607L}.
With the significantly enriched GW data, more subpopulations of BHs/BBHs and features in each subpopulation may be revealed, which are not identified with current data using more complex population models as introduced in the Sec. ~\uppercase\expandafter{\romannumeral2} of the Supplemental Material \citep{supplemental}.

{\it Data availability}---The codes used for this work are publicly available in \href{https://github.com/JackLee0214/Resolving-the-stellar-collapse-and-hierarchical-merger-origins-of-the-coalescing-black-holes}{GitHub}.

\acknowledgments

This work was supported in part by the NSFC under grants of No. 12233011, No. 11921003 and No. 12203101, the Strategic Priority Research Program of the Chinese Academy of Sciences (No. XDB0550400), and the General Fund of the China Postdoctoral Science Foundation (No. 2023M733736). We thank Yong-Jia Huang for discussion. This research has made use of data and software obtained from the Gravitational Wave Open Science Center (https://www.gw-openscience.org), a service of LIGO Laboratory, the LIGO Scientific Collaboration and the Virgo Collaboration. LIGO is funded by the U.S. National Science Foundation. Virgo is funded by the French Centre National de Recherche Scientifique (CNRS), the Italian Istituto Nazionale della Fisica Nucleare (INFN) and the Dutch Nikhef, with contributions by Polish and Hungarian institutes. {The  publicly available code \textsc{GWPopulation} (https://github.com/ColmTalbot/gwpopulation) is referenced in calculating the variance of log-likelihood in the Monte Carlo integrals.}

\bibliographystyle{apsrev4-2}
\bibliography{aeireferences,loca}
\clearpage

\end{document}


\title{Resolving the stellar-collapse and hierarchical-merger origins of the coalescing black holes (Supplemental Material)}
\author{Yin-Jie Li$^{1}$, Yuan-Zhu Wang$^{1,2}$, Shao-Peng Tang$^{1}$, Yi-Zhong~Fan$^{1,3}$.}
\noaffiliation

\affiliation{Key Laboratory of Dark Matter and Space Astronomy, Purple Mountain Observatory, Chinese Academy of Sciences, Nanjing 210008, China}
\affiliation{Institute for Theoretical Physics and Cosmology, Zhejiang University of Technology, Hangzhou, 310032, China}
\affiliation{School of Astronomy and Space Science, University of Science and Technology of China, Hefei 230026, China}

\maketitle

\def\aj{AJ}                   
\def\actaa{Acta Astron.}      
\def\araa{ARA\&A}             
\def\apj{Astrophys.~J.}                 
\def\apjl{Astrophys.~J.~Lett.}                
\def\apjs{ApJS}               
\def\ao{Appl.~Opt.}           
\def\apss{Ap\&SS}             
\def\aap{A\&A}                
\def\aapr{A\&A~Rev.}          
\def\aaps{A\&AS}              
\def\azh{AZh}                 
\def\baas{BAAS}               
\def\bac{Bull. astr. Inst. Czechosl.}
\def\caa{Chinese Astron. Astrophys.}
\def\cjaa{Chinese J. Astron. Astrophys.}
\def\icarus{Icarus}           
\def\jcap{J. Cosmology Astropart. Phys.}
\def\jrasc{JRASC}             
\def\memras{MmRAS}            
\def\mnras{MNRAS}             
\def\na{New A}                
\def\nar{New A Rev.}          
\def\pra{Phys.~Rev.~A}        
\def\prb{Phys.~Rev.~B}        
\def\prc{Phys.~Rev.~C}        
\def\prd{Phys.~Rev.~D}        
\def\pre{Phys.~Rev.~E}        
\def\prl{Phys.~Rev.~Lett.}    
\def\pasa{PASA}               
\def\pasp{PASP}               
\def\pasj{PASJ}               
\def\rmxaa{Rev. Mexicana Astron. Astrofis.}%
\def\qjras{QJRAS}             
\def\skytel{S\&T}             
\def\solphys{Sol.~Phys.}      
\def\sovast{Soviet~Ast.}      
\def\ssr{Space~Sci.~Rev.}     
\def\zap{ZAp}                 
\def\nat{Nature}              
\def\iaucirc{IAU~Circ.}       
\def\aplett{Astrophys.~Lett.} 
\def\apspr{Astrophys.~Space~Phys.~Res.}
\def\bain{Bull.~Astron.~Inst.~Netherlands} 
\def\fcp{Fund.~Cosmic~Phys.}  
\def\gca{Geochim.~Cosmochim.~Acta}   
\def\grl{Geophys.~Res.~Lett.} 
\def\jcp{J.~Chem.~Phys.}      
\def\jgr{J.~Geophys.~Res.}    
\def\jqsrt{J.~Quant.~Spec.~Radiat.~Transf.}
\def\memsai{Mem.~Soc.~Astron.~Italiana}
\def\nphysa{Nucl.~Phys.~A}   
\def\physrep{Phys.~Rep.}   
\def\physscr{Phys.~Scr}   
\def\planss{Planet.~Space~Sci.}   
\def\procspie{Proc.~SPIE}   

\let\astap=\aap
\let\apjlett=\apjl
\let\apjsupp=\apjs
\let\applopt=\ao

\section{Hierarchical Bayesian inference}

We perform hierarchical Bayesian inferences to fit the data of the observed events $\{d\}$ with the population models. Following the framework described in \citet{2019MNRAS.486.1086M,2021ApJ...913L...7A,2023PhRvX..13a1048A}, the likelihood of the hyper-parameters $\boldsymbol{\Lambda}$, given data $\{d\}$ from $N_{\rm det}$ GW detections, can be expressed as 
\begin{equation}\label{eq_llh}
\mathcal{L}(\{d\} |\boldsymbol{\Lambda})\propto N^{N_{\rm det}}e^{-N{\xi(\boldsymbol{\Lambda})}}\prod_{i=1}^{N_{\rm det}}\int{\mathcal{L}(d_i|\theta_i)\pi(\theta_i|\boldsymbol{\Lambda})d\theta_i},
\end{equation} 
where $N$ is the number of mergers in the Universe over the observation period, which is related to the merger rate, and $\xi(\boldsymbol{\Lambda})$ means the detection fraction.
The single-event likelihood $\mathcal{L}(d_i|\theta_i)$ can be estimated using the posterior samples (see \citet{2021ApJ...913L...7A} for the detail), and $\xi(\boldsymbol{\Lambda})$ is estimated using a Monte Carlo integral over detected injections as introduced in the Appendix of ref.\cite{2021ApJ...913L...7A}. { We only take into account the mass-dependent Malmquist bias without spin-dependent selection effects due to the finite number of injections. This practice has negligible impact on the inferred spin distribution, and will not affect the conclusion of this work, see Section~\ref{app:sel} for the detailed illustration.}
The injection campaigns can be adopted from (https://zenodo.org/doi/10.5281/zenodo.5636815), where they combine O1, O2, and O3 injection sets, ensuring a constant rate of injections across the total observing time.

Following \citet{2023PhRvX..13a1048A}, we use the data of BBHs from GWTC-3 \cite{2019PhRvX...9c1040A,2021PhRvX..11b1053A,2021arXiv210801045T,2021arXiv211103606T}, and adopt a false-alarm rate (FAR) of $1yr^{-1}$ as the threshold to select the events. What's more, GW190814 is excluded from the analysis, because it is a significant outlier \cite{2023PhRvX..13a1048A,2022ApJ...926...34E} in the BBH populations, and may be associated with neutron star$-$BH binary systems \cite{2021ApJ...922....3T}. Therefore in total, there are 69 events selected for our analysis. The posterior samples for each BBH event are released on (https://zenodo.org/doi/10.5281/zenodo.5546662), the `C01:Mixed' samples are used for analysis in this work, {and the `\textsc{IMRPhenomXPHM}' samples are used for cross checking.} {In the `C01:Mixed' posterior samples, a few events  (i.e., 3 of 69) have posterior points less than 5000. For practice, we extend the sample sizes of these events to 5000 by random choice (i.e., some posterior points are reused), otherwise only 1993 (i.e., the minimum sample size among all the events, for GW200129\_065458) points per event can be used for the calculation of (log-)likelihood. Though the effective numbers of samples for these events do not increase as the samples extended, the effective numbers for other events indeed increase (comparing to 1993-size samples), hence the total variance of (log-)likelihood for observed events will be reduced given hyper-parameters $\boldsymbol{\Lambda}$, see Section~\ref{sec:sample_size} for more details.} 

The priors of all the parameters are summarized in Tab.~\ref{prior}. To check the reliability of our model and the validation of the findings, we also introduce some other population models for comparison, like PS\&DoubleSpin, and PS\&LinearCorrelation (as defined below). 
In this work, we apply the sampler \textit{Pymultinest} \cite{2016ascl.soft06005B} to obtain the posterior of the hierarchical hyper-parameters.

\begin{table*}[htpb]
\caption{Hyper-parameters and Priors for the fiducial model and other models}\label{prior}
\begin{tabular}{lccc}
\hline
\hline
\multirow{2}{*}{descriptions}   & \multirow{2}{*}{parameters}  & \multicolumn{2}{c}{priors}  \\
\cline{3-4}
&&1st component & 2nd component \\
\cline{1-4}
slope index of the mass function  & $\alpha_i$ & U(-4,8) & U(-4,8) \\
smooth scale of the $i$-th mass lower edge& $\delta_i$ &U(1,10)&U(1,10)\\
 minimum mass of the $i$-th mass function&$m_{{\rm min},i}[M_{\odot}]$ & U(2,60)&U(2,60)\\
 maximum mass of the $i$-th mass function&$m_{{\rm max},i}[M_{\odot}]$ & U(20,100) &U(20,100)\\
interpolation values of perturbation for $i$-th mass function &$\{f_i^j\}_{j=2}^{11}$ &$\mathcal{N}(0,1)$ &$\mathcal{N}(0,1)$ \\
Mass Constraints & & \multicolumn{2}{c}{$m_{{\rm max},i}-m_{{\rm min},i}>20M_{\odot}$} \\
\cline{1-4}
mean of $\chi$ distribution in $i$-th component& $\mu_{{\chi},i}$ &U(0,1) & U(0,1)  \\
standard deviation of $\chi$ distribution in $i$-th component& $\sigma_{{\chi},i}$ &U(0.05, 0.5) & U(0.05, 0.5)  \\
minimum value of $\chi$ distribution in $i$-th component& $\chi_{{\rm min},i}$ &0 & U(0,0.8)  \\
maximum value of $\chi$ distribution in $i$-th component & $\chi_{{\rm max},i}$ &U(0.1,1) & U(0.1,1)  \\
Spin Constraints & & \multicolumn{2}{c}{$\chi_{{\rm max},i}>\mu_{{\chi},i}>\chi_{{\rm min},i}$} \\
\cline{1-4}
mixing fraction of the $i$-th component& $r_i$ &\multicolumn{2}{c}{{$\mathcal{D}$}}\\
\cline{1-4}
standard deviation of nearly aligned $\cos\theta$ 
& $\sigma_{{\rm t}}$ &\multicolumn{2}{c}{U(0.1, 4)}  \\
mixing fraction of nearly aligned assembly & $\zeta$ &\multicolumn{2}{c}{U(0,1)} \\
\cline{1-4}
pairing function & $\beta$ &\multicolumn{2}{c}{U(0,8)}\\
\cline{1-4}
local merger rate density & ${\rm log}_{10}R_0[{\rm Gpc^{-3}yr^{-1}}]$ &\multicolumn{2}{c}{U(0,3)}\\
\hline
\\
\multicolumn{4}{c}{Parameters For PS\&LinearCorrelation}\\
\hline
 $\sigma_{\chi}$ at the lower-mass  edge & $\sigma_{\chi}^{\rm left}$ & \multicolumn{2}{c}{U(0.05,0.5)} \\
$\sigma_{\chi}$ at the higher-mass  edge &$\sigma_{\chi}^{\rm right}$ &  \multicolumn{2}{c}{U(0.05,0.5)} \\
$\mu_{\chi}$ at the lower-mass  edge & $\mu_{\chi}^{\rm left}$ &  \multicolumn{2}{c}{U(0,1)} \\
$\mu_{\chi}$ at the higher-mass  edge & $\mu_{\chi}^{\rm right}$ & \multicolumn{2}{c}{U(0,1)} \\
\hline
\\
\multicolumn{4}{c}{Parameters For Two-component\&LinearCorrelation}\\
\hline
$\mu_{\chi}$ at the lower-mass edge in $i$-th component& $\mu_{{\chi},i}^{\rm left}$   &U(0,1) & U(0,1)  \\
$\mu_{\chi}$ at the higher-mass edge in $i$-th component& $\mu_{{\chi},i}^{\rm right}$ &U(0,1) & U(0,1)  \\
Spin Constraints & & \multicolumn{2}{c}{$\mu_{{\chi},2}^{\rm left}>\mu_{{\chi},1}^{\rm right}$} \\
\hline
\hline
\end{tabular}
\begin{tablenotes}
        \footnotesize
        \item[1] {\bf Note.} Here, U ($\mathcal{N}$) represents the uniform (normal) distribution, and {$\mathcal{D}$ is for the Dirichlet distribution.}
      \end{tablenotes}
\end{table*}

\section{Model comparison}

\begin{table}[htpb]
\centering
\caption{Model comparison results}\label{tab:bf}
\begin{tabular}{lcc}
\hline
\hline
Models     &  $\ln{\mathcal{B}}$    \\
\hline
Two-component  & 0  \\
One-component &{-7.5}  \\
PS\&LinearCorrelation & {-4.7} \\
PS\&DoubleSpin & {-9.2} \\
PS\&DefaultSpin & {-10.8} \\
PP\&DefaultSpin & {-18.3} \\
{Two-component\&LinearCorrelation}  & {-0.6}  \\
{Two-component\&DoubleSpin}  & {-0.9}  \\
{One-component\&Step} & {-1.3}  \\
\hline
{Two-component (variable $\mu_{\rm t}$)} & {0.1}  \\
{Two-component (isotropic spin or without $\cos\theta_{1,2}$ information)} & {-6.3}  \\
{Two-component ($\beta=0$)} & {-14.9}  \\
\hline
\hline
\end{tabular}
\\
\begin{tabular}{l}
Note: the log Bayes factors are relative to the Two-component model.
\end{tabular}
\end{table}

For context, $\ln\mathcal{B}>(2.3,~3.5,~4.6)$  can be interpreted as a (strong, very strong, decisive) preference for one model over another \cite{HaroldJeffreysWrited1961Theory}.
We perform inferences with our model of single component, and two components, respectively; the Two-component model is significantly more favored by the current GW data, as summarized in Tab.~\ref{tab:bf}. 
We have also inferred with a Three-component model, and it has lower evidence than the Two-component model, where the distributions of the second and third components are highly degenerate.
We further compare the Two-component model to other models in order to find out whether the two components adopted in the modeling are indeed necessary. Finally, all of the other alternative models {(with single population) are disfavored, and the Two-component model with more complex features are not necessary with current data (see below for details).}   

\subsection{Only one peak in the spin-magnitude distribution?}
We first compare the Two-component model to the other models with only one peak in the spin-magnitude distribution, i.e., PS\&DefaultSpin and PP\&DefaultSpin adopted in \citet{2023PhRvX..13a1048A}, where the spin-magnitude distribution and the mass functions are those of the Default model and the PS / PP model in \citet{2023PhRvX..13a1048A}. Note that in their PS model, the secondary mass distribution follows $m_2^{\beta_{\rm q}}$ conditioned by $m_1$, where ${\beta_{\rm q}}$ is the power law index of the mass ratio distribution.   
As shown in Tab.~\ref{tab:bf}, the PS\&DefaultSpin, and the PP\&DefaultSpin are definitively disfavored by the log Bayes factors of $\ln\mathcal{B}=-10.8$, and $\ln\mathcal{B}=-18.3$, respectively.

\subsection{Is double-peak spin-magnitude distribution uncorrelated with the mass distribution?}
We have found that double peaks in the spin-magnitude distribution are required, but are they still consistent in the whole mass range? To answer this question, we carry out another inference with the PS\&DoubleSpin model, where the mass distribution (i.e., PS model adopted from \citet{2023PhRvX..13a1048A}) and the spin distribution (i.e., DoubleSpin) are constructed independently. For the spin-magnitude distribution, we apply two components to catch the underlying peaks; therefore the DoubleSpin 
model reads 
\begin{equation}
\pi(\chi | {\bf \Lambda})=\pi_1(\chi | {\bf \Lambda}_1)\times(1-r_2)+\pi_2(\chi | {\bf \Lambda}_2)\times r_2,
\end{equation}
where $r_2$ is the mixing fraction of the secondary component, and the spin-magnitude distribution of $i$-th component is 
\begin{equation}
\pi_i(\chi | {\bf \Lambda}_i)=\mathcal{G}(\chi |\chi_{{\rm min},i}, \chi_{{\rm max},i}, \mu_{\chi,i}, \sigma_{\chi,i}),
\end{equation}
Then the spin distribution of the BBHs should be
\begin{equation}
\begin{aligned}
&\pi(\chi_1,\chi_2, \cos\theta_1,\cos\theta_2 | {\bf \Lambda}; \zeta, \sigma_{{\rm t}})=\\
& \pi( \chi_1 | {\bf \Lambda})\times\pi( \chi_2 | {\bf \Lambda})\times \mathcal{GU}(\cos\theta_1, \cos\theta_2|, \zeta, \sigma_{{\rm t}}). 
 \end{aligned}
 \end{equation}
The priors of the spin model are the same as those summarized in Tab.\ref{prior}. We find that the PS\&DoubleSpin is disfavored by a log Bayes factor of $\ln\mathcal{B}=-9.2$ compared to the Two-component model, but is slightly better than the PS\&DefaultSpin model (see Tab.\ref{tab:bf}). We therefore conclude that, for coalescing BBHs, the spin-magnitude distribution is not independent of but correlated with the mass distribution. 

\subsection{Is spin-magnitude distribution linearly changed with the mass distribution?}
\begin{figure}
\centering  
\includegraphics[width=0.96\linewidth]{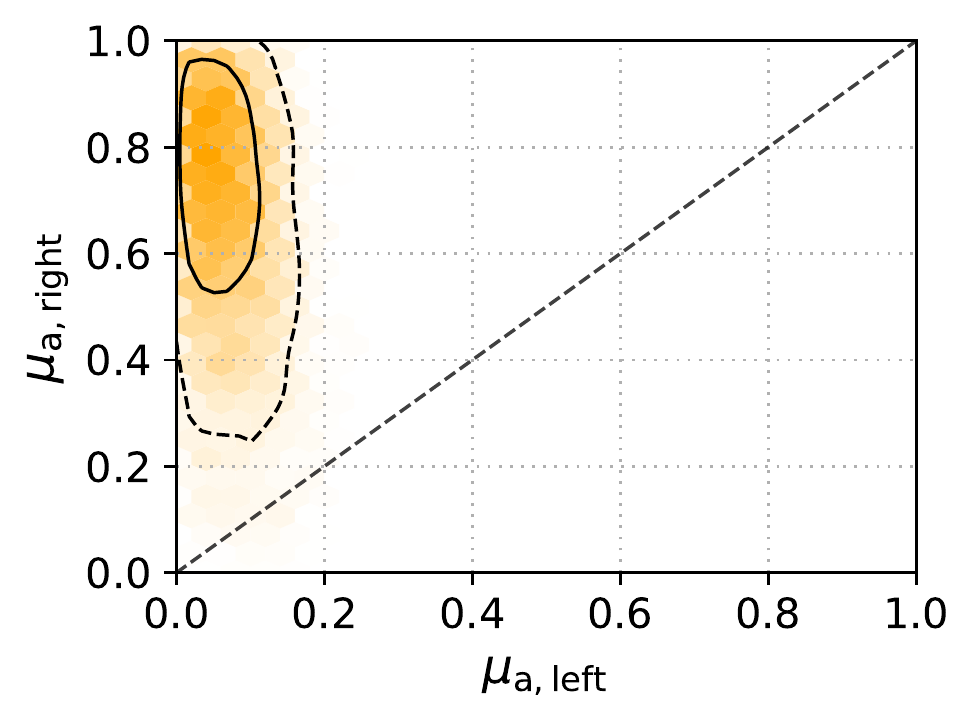}
\includegraphics[width=0.96\linewidth]{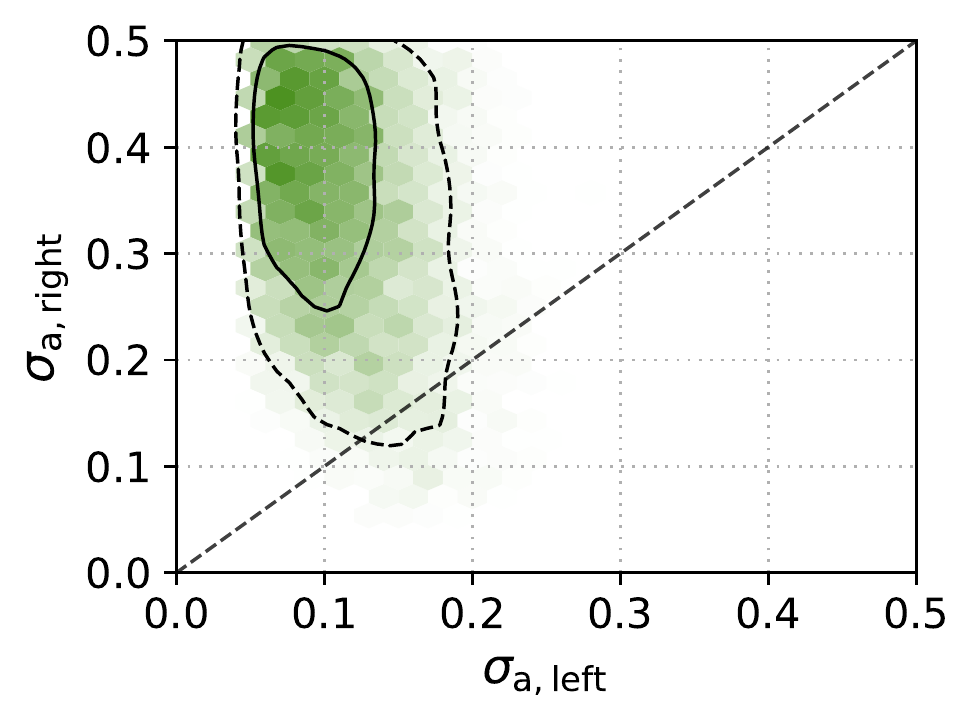}
\caption{Posterior distributions of $\mu_{\chi}^{\rm right}$ vs. $\mu_{\chi}^{\rm left}$ (Top) and $\sigma_{\chi}^{\rm right}$ vs. $\sigma_{\chi}^{\rm left}$ (Bottom). The dashed and solid contours mark the central 50\% and 90\% posterior credible regions, respectively. It turns out that $\mu_{\chi}^{\rm right}>\mu_{\chi}^{\rm left}$ at 99\% credible level and $\sigma_{\chi}^{\rm right}>\sigma_{\chi}^{\rm left}$ at 93\% credible level.}
\label{app:spin}
\end{figure}

As we have found that the spin-magnitude distribution is not consistent in the whole mass range, then does the spin-magnitude distribution gradually change as the mass increases? So we introduce a mass-spin distribution where the spin-magnitude distribution (i.e., LinearCorrelation) linearly correlates with the mass distribution (i.e., PS adopted from \citet{2023PhRvX..13a1048A}). Therefore, the PS\&LinearCorrelation model reads
\begin{equation}
\begin{aligned}
&\pi(\chi_1,\chi_2, \cos\theta_1,\cos\theta_2 | {\bf \Lambda}; \zeta, \sigma_{{\rm t}})=\\
&\mathcal{G}(\chi_1|0,1, \mu_{\chi}(m), \sigma_{\chi}(m))\times \mathcal{G}(\chi_2|0,1, \mu_{\chi}(m), \sigma_{\chi}(m))\times \\
& \mathcal{GU}(\cos\theta_1, \cos\theta_2| \zeta, \sigma_{{\rm t}}),
 \end{aligned}
\end{equation}
where $\mu_{\chi}(m)$ and $\sigma_{\chi}(m)$ are the linear function of the component mass $m$,
\begin{equation}
\begin{aligned}
 \mu_{\chi}(m)=\frac{(m-m_{\rm min})\mu_{\chi}^{\rm right}+(m_{\rm max}-m)\mu_{\chi}^{\rm left}}{m_{\rm max}-m_{\rm min}},\\
 \sigma_{\chi}(m)=\frac{(m-m_{\rm min})\sigma_{\chi}^{\rm right}+(m_{\rm max}-m)\sigma_{\chi}^{\rm left}}{m_{\rm max}-m_{\rm min}}.
 \end{aligned}
\end{equation}
where the priors and the descriptions of the hyper-parameters for the model are summarized in Table~\ref{prior}. We find that the PS\&LinearCorrelation model is disfavored by a log Bayes factor of $\ln\mathcal{B}={-4.7}$ compared to the Two-component model. Therefore the spin-magnitude distribution does not linearly change with the masses of the BHs, though we find that $\mu_{\chi}^{\rm right}>\mu_{\chi}^{\rm left}$ at 99.9\% credible level and $\sigma_{\chi}^{\rm right}>\sigma_{\chi}^{\rm left}$ at 93\% credible level (see Figure~\ref{app:spin}), which means that the heavier BHs have larger spin magnitude, and the spin-magnitude distribution broadens with the increasing mass.

\subsection{One-component\&Step}
\begin{figure*}
	\centering  
\includegraphics[width=0.8\linewidth]{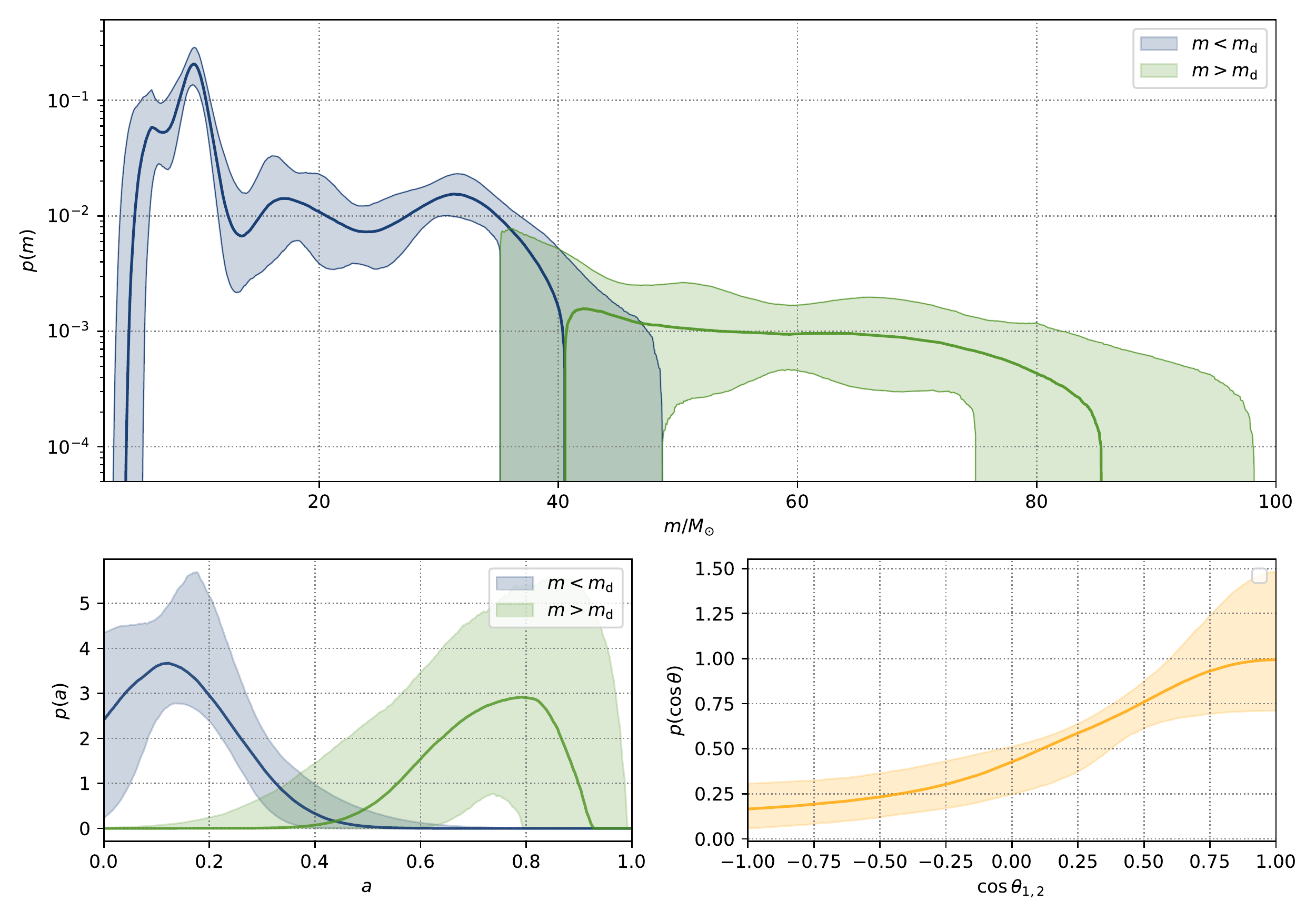}
\caption{Reconstructed mass (top) and spin (bottom) distributions inferred with One-component\&Step mode; The solid curves are the medians and the colored bands are the 90\% credible intervals. 
}
\label{fig:step_dist}
\end{figure*}

\begin{figure}[!h]
	\centering  
\includegraphics[width=0.9\linewidth]{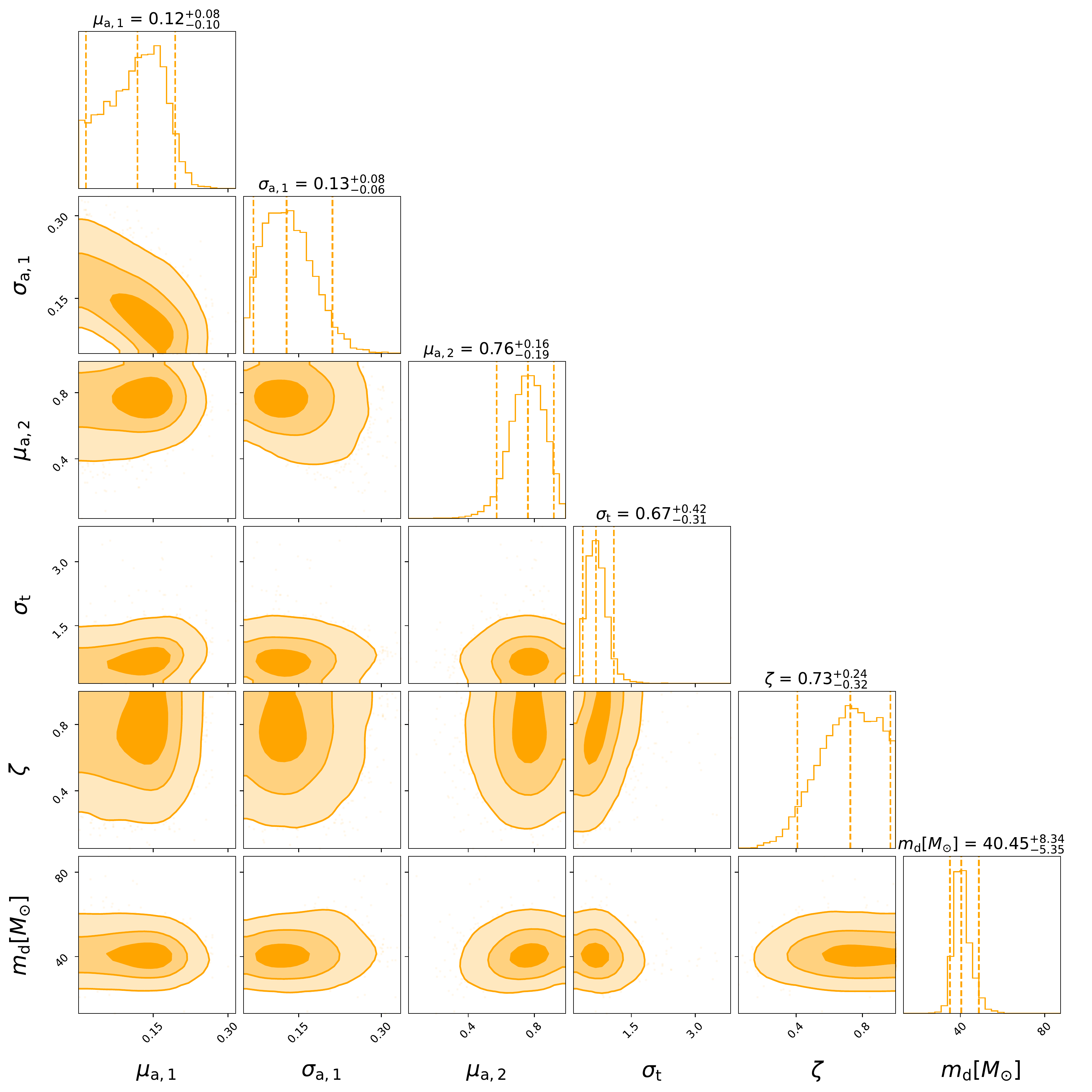}
\caption{Posterior distribution of hyper-parameters inferred with One-component\&Step mode; The dashed and solid contours mark the central 50\% and 90\% posterior credible regions, respectively.}
\label{fig:step_corner}
\end{figure}

{As the Two-component model is  favored over the one-component model and the PS\&LinearCorrelation model. Here we modify the one-component model, and use a step function to describe the evolution of spin-magnitude distribution with the masses of BHs,
\begin{equation}
\pi(\chi | m;m_{\rm d}, {\bf \Lambda})=\mathcal{G}(\chi |0, 1,  \mu_{\chi}(m|m_{\rm d}), \sigma_{\chi}),
\end{equation}
where
\begin{equation}
 \mu_{\chi}(m|m_{\rm d})=
\left\{
	\begin{aligned}
	\mu_{\chi,1} \quad m<m_{\rm d},\\
        \mu_{\chi,2} \quad m>m_{\rm d},\\
 \end{aligned}
	\right	.
\end{equation}

We find the One-component\&Step model is only slightly less favored compared to the Two-component model by $\ln\mathcal{B}=-1.3$. However the inferred mass and spin distributions are similar to those inferred by the Two-component model, as shown in Figure~\ref{fig:step_dist}. Additionally, the divided point  is well constrained to be $m_{\rm d}=40.45^{+8.34}_{-5.35}M_{\odot}$, as shown in Figure~\ref{fig:step_corner}, which is consistent with the 
$m_{\rm max,1}$ inferred with the Two-component model.}
Now the spin-magnitude distributions of BHs are sharply changed at $40M_{\odot}$, and the BHs with masses $>40M_{\odot}$ have spin magnitudes $\sim0.7$, which are more consistent with the higher-generation BHs. Evidently, even with the One-component model we have identified two distinct sub-populations. 

\subsection{{Is there a third sub-population?}}\label{sec:app_post}
\begin{figure*}
	\centering  
\includegraphics[width=0.8\linewidth]{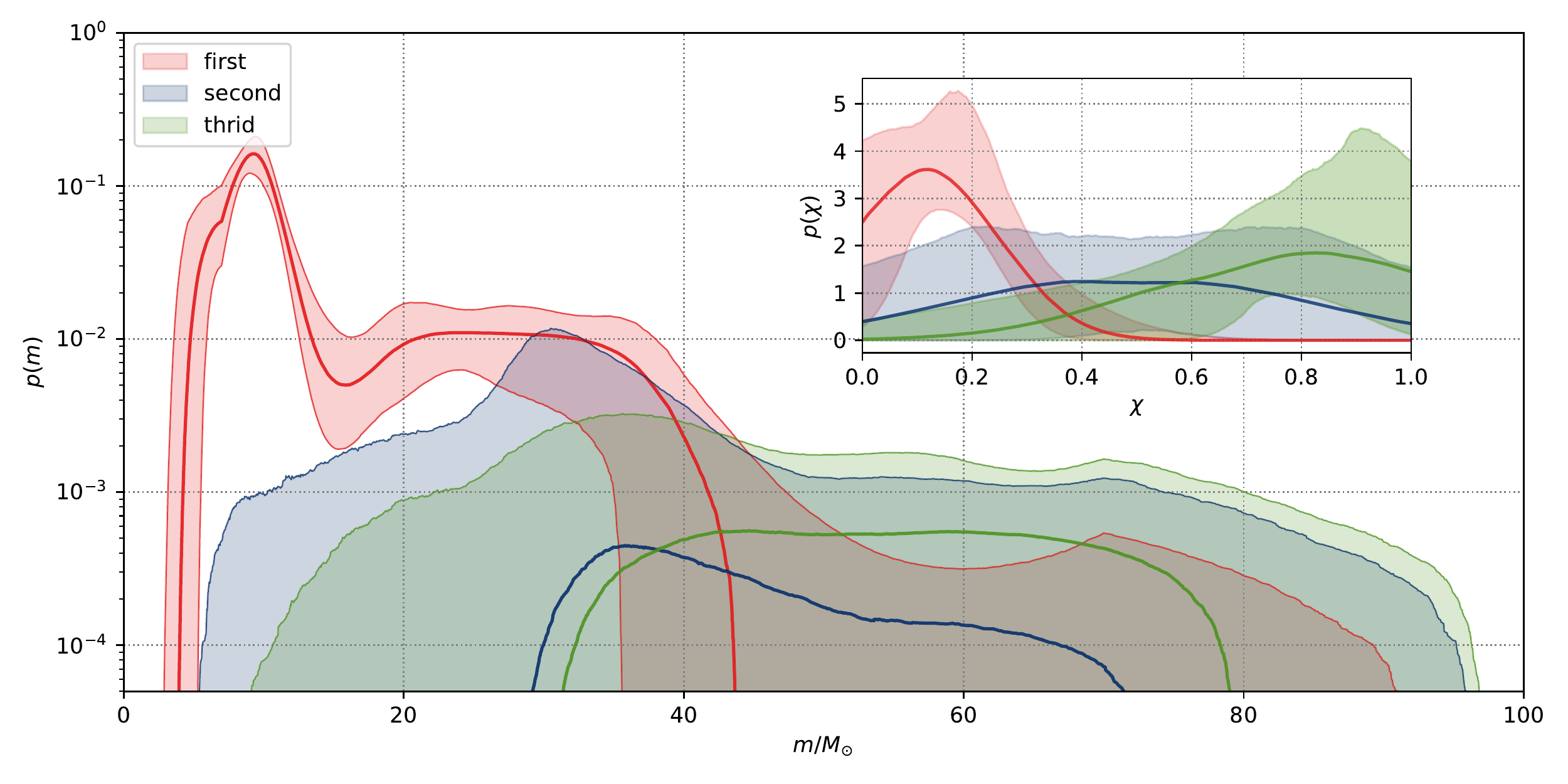 }
\caption{ Reconstructed mass distribution of BHs for each component, and the inset is for the spin-magnitude distribution of the each component. The solid curves are the medians and the colored bands are the 90\% credible intervals. }
\label{fig:third}
\end{figure*}

{Many evolution/formation channels are proposed, which are expected to generate BBHs with different (maybe not significantly) mass and spin distributions \citep{2021ApJ...910..152Z}. Therefore, more than two sub-populations may be recognized with different mass versus spin-magnitude distributions (if significant). We have performed inference with a Three-component model in order to find out another sub-population. }
{In practice, the inference of the Three-component model is extremely computationally intensive, since many parameters are degenerate. Therefore, for the comparison between the Three-component model and the Two-component model, we set only 8 knots linearly in the logarithm space within [7, 70] $M_{\odot}$ for the cubic-spline perturbation function, and fixed $\chi_{{\rm min},i}=0$, $\chi_{{\rm max},i}=1$ (Note that these configurations are implemented in both Two-component model and Three-component model, when they are compared). It turns out that the Three-component model is slightly less favored compared to the Two-component model by $\ln\mathcal{B}_{\rm Two}^{\rm Three}=-1.7$, and the distributions of the second and third components are highly degenerate with each other, as shown in Figure~\ref{fig:third}. Therefore, more than two components in our model is not necessary to fit the current GW data, and more detailed sub-populations may be identified after the fourth and fifth Observing Runs of LIGO/Virgo/KAGRA \cite{2020LRR....23....3A}.}

\subsection{{Does the spin-magnitude distribution evolve with the mass in each sub-population ?}}
{ As the Two-component model is favored over both the one-component model and the PS\&LinearCorrelation model, one may ask whether the spin-magnitude distributions are evolving with mass in each component (sub-population). 
For such a purpose, we describe the spin-magnitude distribution of each component as 
\begin{equation}
\pi_i(\chi | m; {\bf \Lambda}_i)=\mathcal{G}(\chi |\chi_{{\rm min},i}, \chi_{{\rm max},i}, \mu_{\chi,i}(m), \sigma_{\chi,i}),
\end{equation}
with
\begin{equation}
 \mu_{\chi,i}(m)=\frac{(m-m_{\rm min,i})\mu_{\chi,i}^{\rm right}+(m_{\rm max,i}-m)\mu_{\chi,i}^{\rm left}}{m_{\rm max,i}-m_{\rm min,i}}.
\end{equation}
}

\begin{figure*}
\centering  
\includegraphics[width=0.4\linewidth]{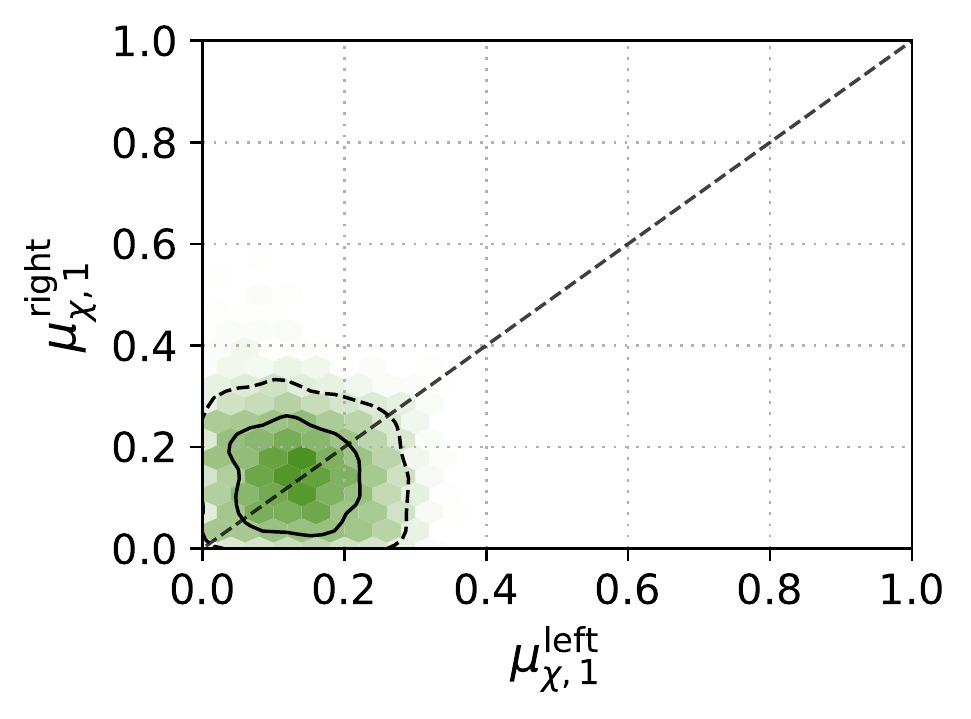}
\includegraphics[width=0.4\linewidth]{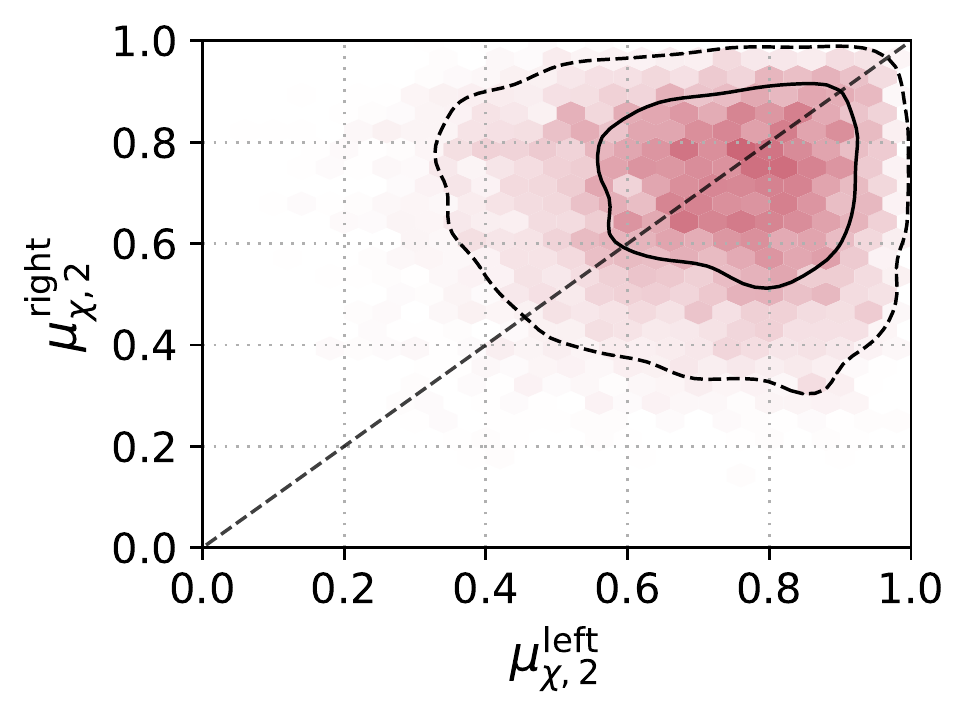}
\caption{Posterior distributions of $\mu_{\chi,1}^{\rm right}$ vs. $\mu_{\chi,1}^{\rm left}$ (left) and $\mu_{\chi,2}^{\rm right}$ vs. $\mu_{\chi,2}^{\rm left}$ (right). The dashed and solid contours mark the central 50\% and 90\% posterior credible regions, respectively. It turns out that there are no evidence for $\mu_{\chi}^{\rm right}>\mu_{\chi}^{\rm left}$ or $\mu_{\chi}^{\rm right}<\mu_{\chi}^{\rm left}$ in either component.}
\label{fig:Doublelinear}
\end{figure*}

To distinguish between the two components, we set $\mu_{\chi,2}^{\rm left}>\mu_{\chi,1}^{\rm right}$. 
The Figure~\ref{fig:Doublelinear} shows that there is no evidence for the dependence of the spin magnitudes on mass in each sub-population.
Additionally, we find that the two sub-populations identified here are consistent with those in the Two-component model (see Figure~\ref{fig:compare_m_ct}).

\begin{figure}
	\centering  
\includegraphics[width=0.8\linewidth]{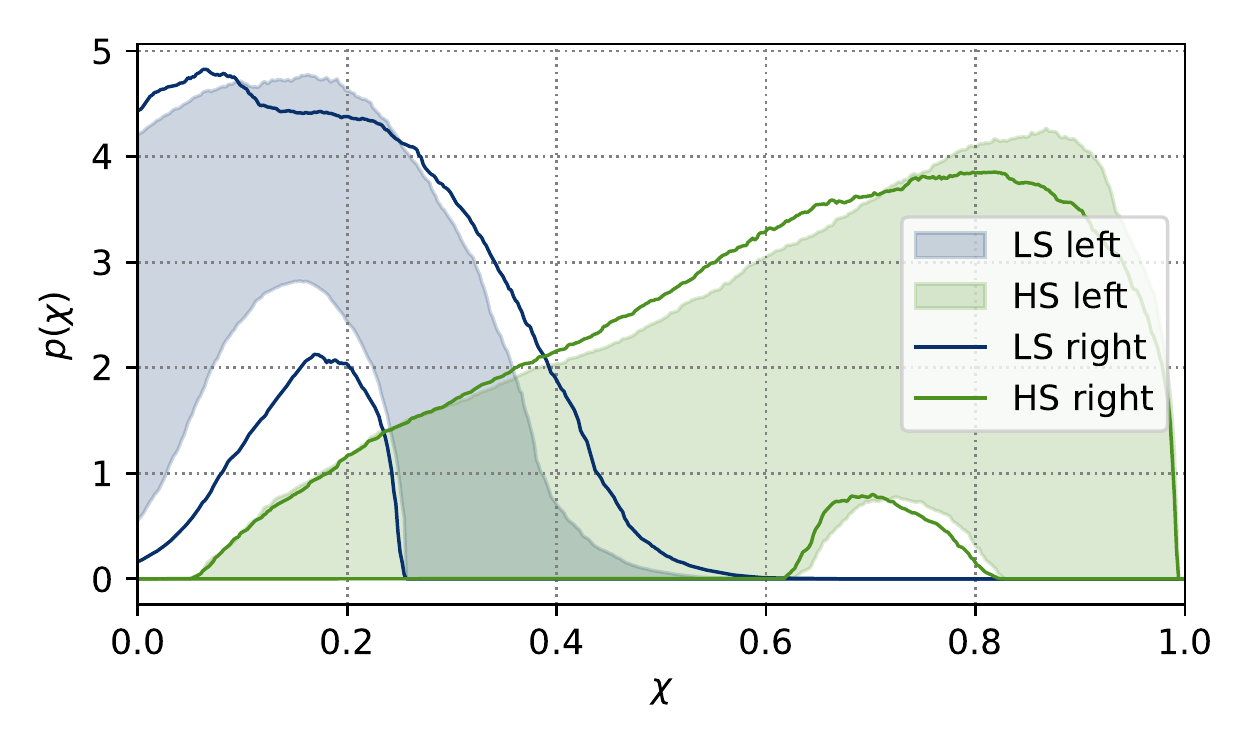}
\caption{Reconstructed spin distributions of BHs inference with Two-component\&LinearCorrelation, the solid curves and the dash bands are the spin-magnitude distributions of BHs at right and left sides in the mass range for the two sub-populations.
}
\label{fig:dist_Doublelinear}
\end{figure}

\subsection{Are there multiple peaks in the spin-magnitude distribution for each sub-population?}
{As the Two-component model is favored over both the one-component model and the PS\&DoubleSpin model, one may ask: are there multiple peaks in the spin-magnitude distribution for each component (sub-population)?
Since the BHs for HSG is too few, and most BHs belong to the LSG. For simplicity, we only use a DoubleSpin model for the spin-magnitude distribution in the first component (sub-population), which reads
\begin{equation}
\begin{aligned}
&\pi_1(\chi | {\bf \Lambda}_1)=\\
&\mathcal{G}(\chi |\chi_{{\rm min},1}, \chi_{{\rm max},1}, \mu_{\chi,1,\rm I}, \sigma_{\chi,1,\rm I})\times(1-r_{\rm II})+\\
&\mathcal{G}(\chi |\chi_{{\rm min},1}, \chi_{{\rm max},1}, \mu_{\chi,1,\rm II}, \sigma_{\chi,1,\rm II})\times r_{\rm II}.
\end{aligned}
\end{equation}}

{The Bayes factor of Two-component\&DoubleSpin sceanrio is comparable to the Two-component model. In other words, there is no evidence for multiple peaks in the spin-magnitude distribution of the first-generation BHs (see 
Figure~\ref{fig:twopeak_dist}). Again, the two sub-populations identified here are similar to those found in the pure Two-component model see Figure~\ref{fig:compare_m_ct}.

\begin{figure}
	\centering  
\includegraphics[width=0.8\linewidth]{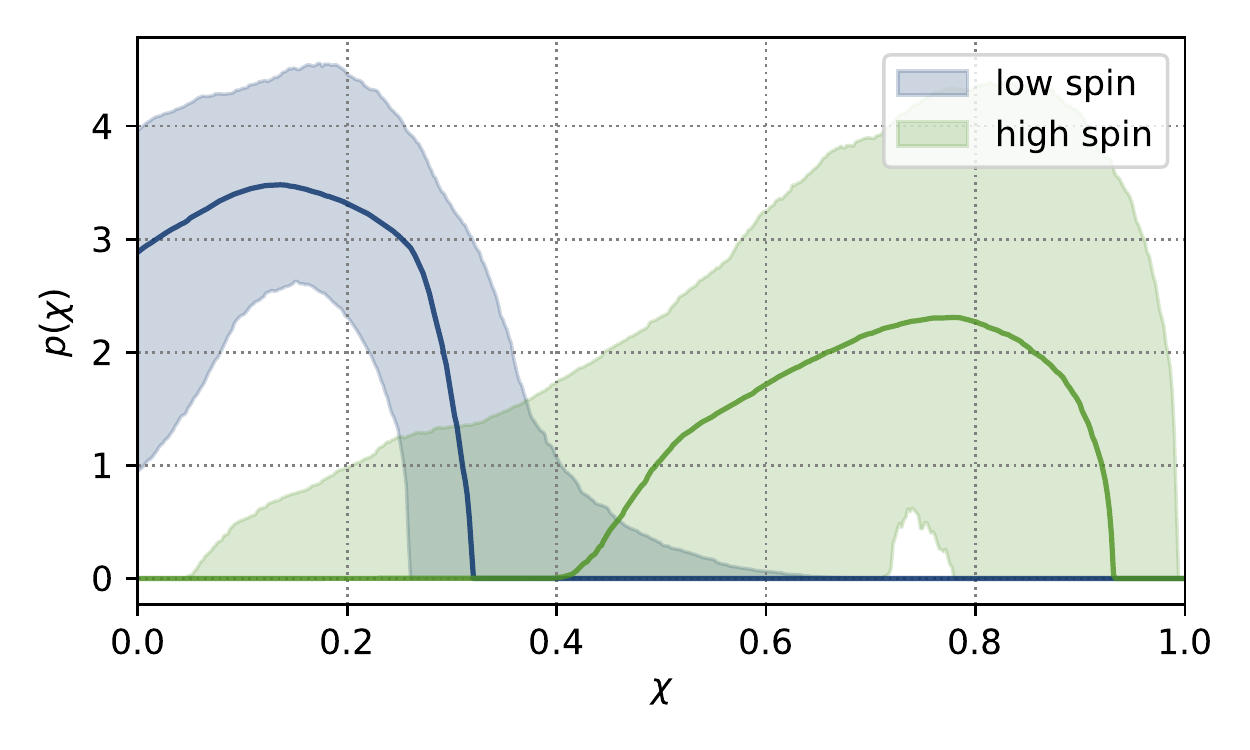}
\caption{The same as Figure~\ref{fig:dist_Doublelinear}, but for the inference with Two-component\&DoubleSpin.
}
\label{fig:twopeak_dist}
\end{figure}

\begin{figure*}
	\centering  
\includegraphics[width=0.8\linewidth]{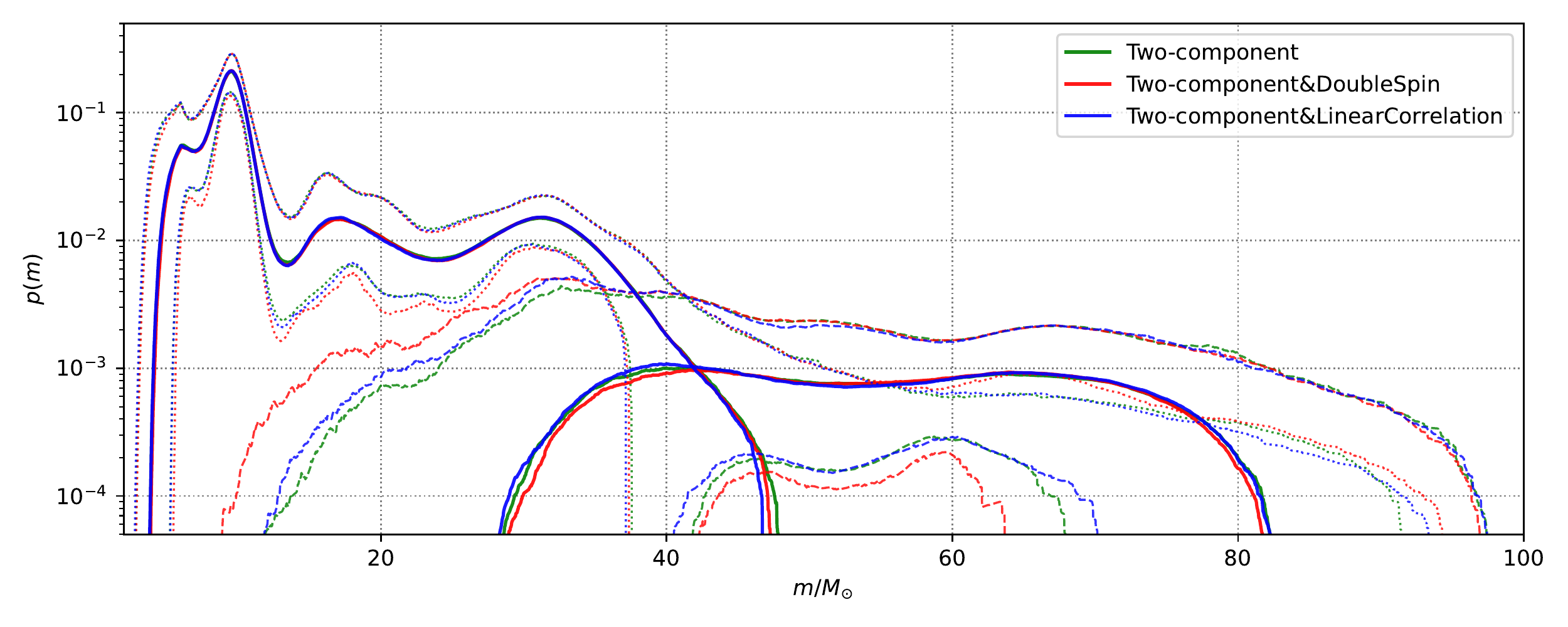}
\includegraphics[width=0.4\linewidth]{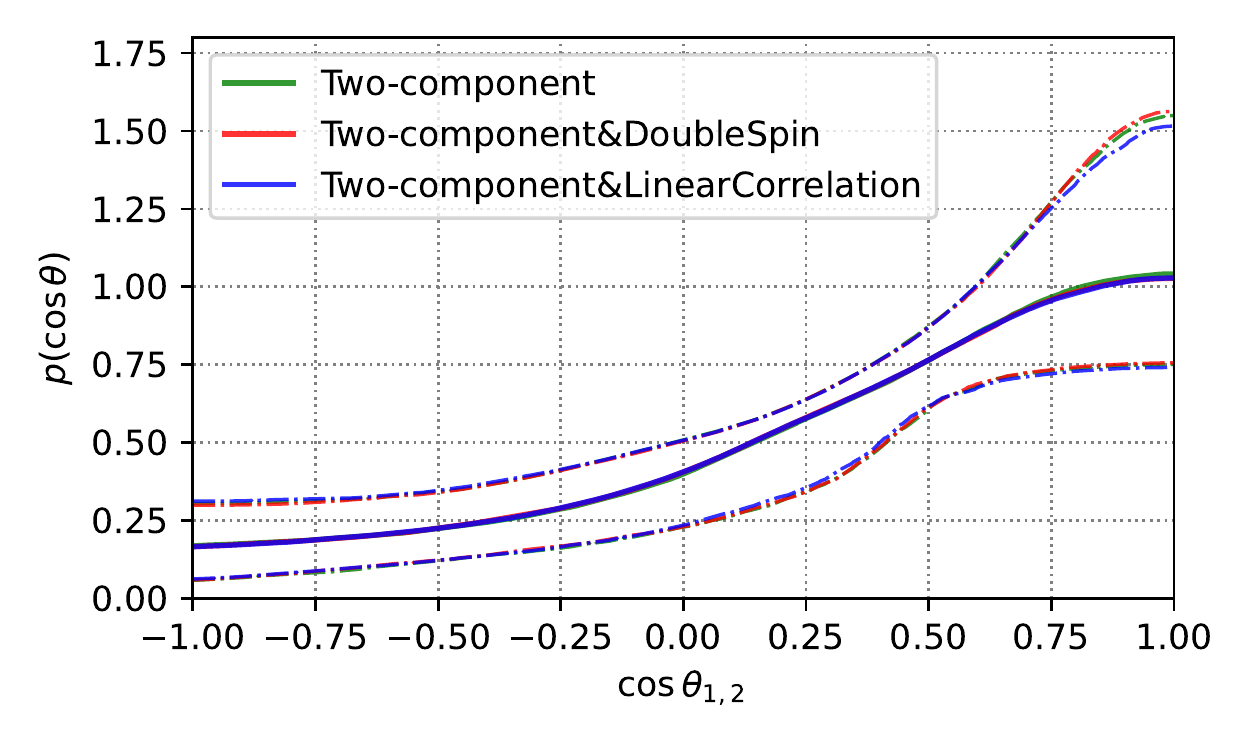}
\caption{Mass and spin-orientation distributions inferred with Two-component\&DoubleSpin and Two-component\&LinearCorrelation models comparing to those inferred with Two-component model.
}
\label{fig:compare_m_ct}
\end{figure*}

In view of the above facts, we conclude that there are two distinct sub-population of BHs via the mass versus spin-magnitude distributions. More sub-populations/features may exist, but significantly enriched data are needed to clarify.

\section{Uncertainties in the likelihood estimation.}

{As introduced in the main text, we use the Monte Carlo summations over samples to approximate the integrals in the likelihood given in Eq. \ref{eq_llh}, which will bring statistical error in the likelihood estimations \cite{2019RNAAS...3...66F,2020arXiv201201317T,2022arXiv220400461E,2022arXiv221012287G,2022ApJ...926...79G}. Following Ref. \cite{2023PhRvX..13a1048A,2023MNRAS.526.3495T}, we define effective number of samples for the $i$-th event in the Monte Carlo integral as $N_{{\rm eff},i}=\frac{[\sum_{j}{w_{i,j}]^2}}{\sum_j{w_{i,j}^2}}$, where $w_{i,j}$ is the weight of $j$-th sample in $i$-th event. Essick \& Farr\cite{2022arXiv220400461E} argue that $N_{{\rm eff},i}$ $\sim 10$ is sufficiently high to ensure accurate marginalization over $i$-th event, therefore we constrain the prior of hyper-parameters to ensure $N_{{\rm eff},i} > 10$ for all the events. Following Refs.\cite{2019RNAAS...3...66F,2021ApJ...913L...7A}, we further constrain the effective number of found injections remaining after population reweighting as $N_{\rm eff,sel}> 4 N_{\rm det}$, to ensure an accurate estimation of $\xi(\boldsymbol{\Lambda})$.}

{Following Refs.\cite{2021ApJ...913L...7A,2021ApJ...921L..15G}, we only account for mass-dependent Malmquist bias without spin-dependent selection effects due to the finite number of injections, so we fixed the spin distribution to be the same as that of the injection campaigns. 
 This approximation is expected to have a negligible impact on the inferred spin distribution compared to the statistical uncertainties \cite{2021ApJ...913L...7A}, though it is slightly easier to detect BBH signals with larger $\chi_{\rm eff}$, as was illustrated in \citet{2018PhRvD..98h3007N}.}

{The variance in the estimate of the total (log-)likelihood can be expressed as $\sigma^2_{\ln\hat{\mathcal{L}}({\bf \Lambda})} = \sigma^2_{\rm obs} + \sigma^2_{\rm sel}$, with $\sigma^2_{\rm obs}\equiv \sum_{i}^{N_{\rm det}}{\sigma^2_{{\ln\hat{\mathcal{L}_i}({\bf \Lambda})}}}$ and  $\sigma^2_{\rm sel}\equiv N_{\rm det}^2 \sigma^2_{\ln\hat{\xi}({\bf \Lambda})}$ \cite{2023MNRAS.526.3495T}, where $\sigma^2_{{\ln\hat{\mathcal{L}_i}({\bf \Lambda})}}$ is the per-event variance, and $ \sigma^2_{\ln\hat{\xi}({\bf \Lambda})}$ is the variance of the selection function. }
{As introduced in the main text, when calculating the (log-)likelihood, we use a size of 5000 for per event samples instead of the minimum sample size across all events (i.e., 1993 of GW200129\_065458).  Such manipulation will not increase the effective numbers for the events which initially have sample sizes smaller than 5000 (i.e., GW150914, GW200112\_155838, and GW200129\_065458 have sample sizes of 3337, 4323, and 1993, respectively), but it will increase the effective numbers for the events which initially have more than 5000 samples given hyper-parameters $\boldsymbol{\Lambda}$. Therefore, this 5000-size sample will allow a larger parameter space for the two-component model (which may be closer to the true distribution), under the constraint of $N_{{\rm eff},i}>10$.} 

{The variance of log-likelihood from events' samples that averaged over the posteriors is $\left \langle  \sigma^2_{\rm obs} \right \rangle_{\rm post} =0.64$, and the uncertainty from injection samples $\left \langle \sigma^2_{\rm sel}\right \rangle_{\rm post} =0.53$, so the total variance in the estimate of the log-likelihood is $\left \langle \sigma^2_{\ln\hat{\mathcal{L}}({\bf \Lambda})}  \right \rangle_{\rm post} = 1.21$, which indicates that our result is almost unimpacted by Monte Carlo convergence \cite{2023MNRAS.526.3495T}. 
For cross checking, we also perform inference with the `\textsc{IMRPhenomXPHM}' samples, and find the results are nearly identical; the average variance of log-likelihood from events' sample is $\left \langle  \sigma^2_{\rm obs} \right \rangle_{\rm post} =0.37$, since a larger sample-size for per-event is adopted (6913; i.e., the minimum sample size of the events in the `\textsc{IMRPhenomXPHM}' samples), and the total average variance reduced to $\left \langle \sigma^2_{\ln\hat{\mathcal{L}}({\bf \Lambda})}  \right \rangle_{\rm post} = 0.99$.
}

\section{The selection effects}\label{app:sel}

We have performed inference with spin-induced selection bias, however the variance of selection function $\sigma_{\rm sel}$ is too large to obtain a reliable result. As shown in Figure~\ref{fig:spin_vars}, the $N_{\rm eff,sel}$ trends to $4 N_{\rm det}$, implying that the injection samples are not enough for an accurate analysis \cite{2019RNAAS...3...66F}.
The spin-magnitude distribution of the first component peaks sharply at 0.21, as shown in Figure~\ref{fig:spin_vars} and Figure~\ref{fig:compare_selection_dist} , a similar circumference is also reported in \citet{2023PhRvD.108j3009G}. Anyhow, the resulting mass distribution of the two sub-populations are similar to those found in the main text.

\begin{figure*}
	\centering  
\includegraphics[width=0.6\linewidth]{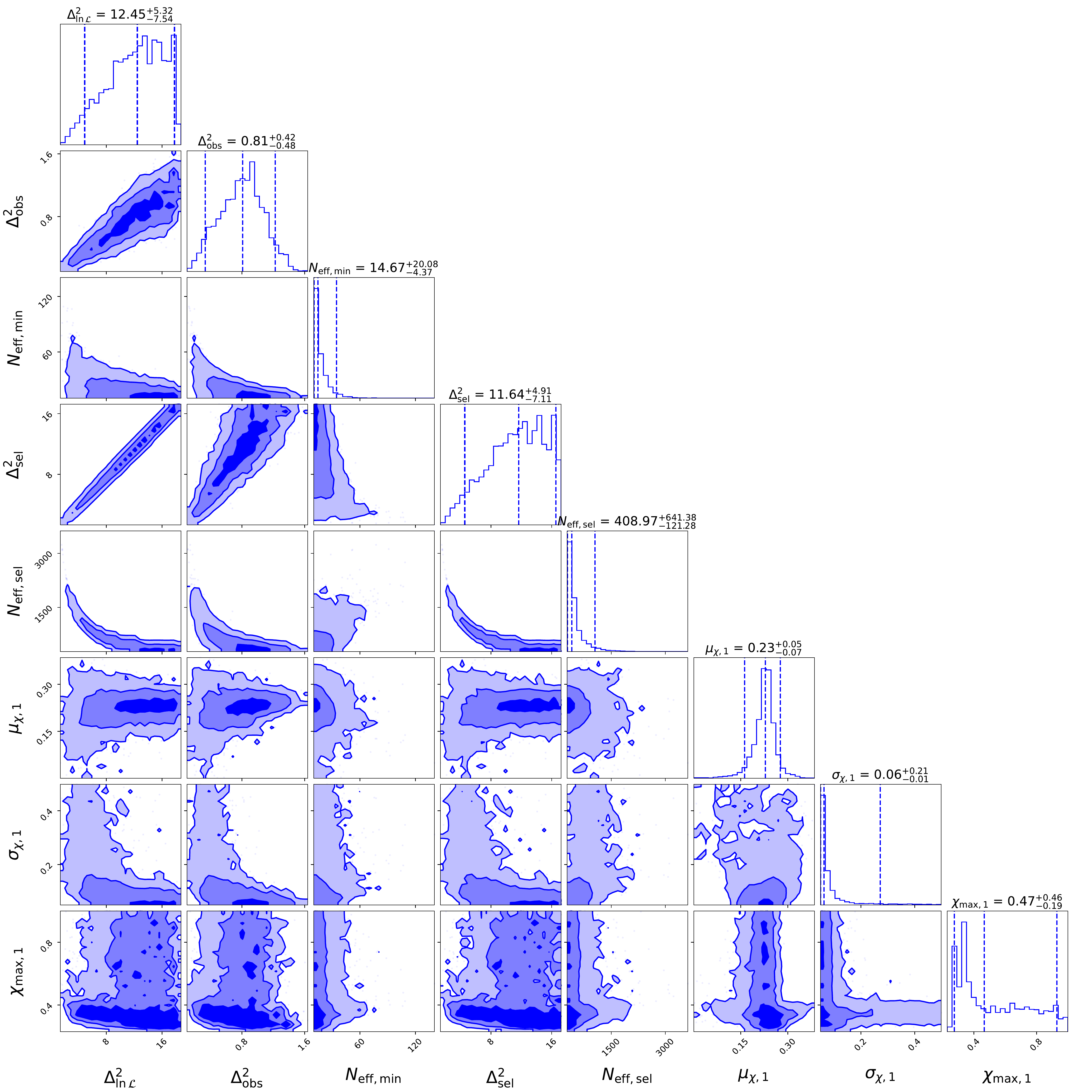}
\caption{ Distributions of the variance in the estimate of log-likelihood with spin-induced selection function. The dashed lines in the marginal distribution represent the $90\%$ credible intervals. }
\label{fig:spin_vars}
\end{figure*}

We have also performed inference with spin-induced selection bias and set $N_{\rm eff,sel}> 4000$ to ensure small variance of selection function $\sigma_{\rm sel}$.
Since the current size of the injection samples is too small, a narrow spin distribution will not pass the threshold of $N_{\rm eff,sel}> 4000$, and only the flat distribution can be accepted,
as shown in Figure~\ref{fig:compare_selection_dist}. 
Even so, we still find two sub-populations, with mass and spin distributions similar to those found in the main text (see Figure~\ref{fig:compare_selection_dist}), which supports our main finding.
However, these distributions may be biased by the preference for broader distribution, due to
the limited injection samples and $N_{\rm eff,sel}> 4000$.
Therefore, for the main analysis in this work, we do not include the spin-induced selection effects.

{To display the impacts of neglecting the spin-induced selection effects on the inferred spin distribution more intuitively,
in Figure~\ref{fig:spin_inj}, we plot the spin-magnitude distributions of the observable injections 
with BH masses in the ranges of $(5,~45)M_{\odot}$ and $(25,~80)M_{\odot}$.} {The differences between the two sub-populations are minor, which should not dominate the cluster of the two sub-populations. Additionally, the detecting preference for the larger spin magnitudes is really weak, so that neglecting the spin-induced selection effects only has minor impact on the spin-magnitude distributions.}

\begin{figure*}
	\centering  
\includegraphics[width=0.8\linewidth]{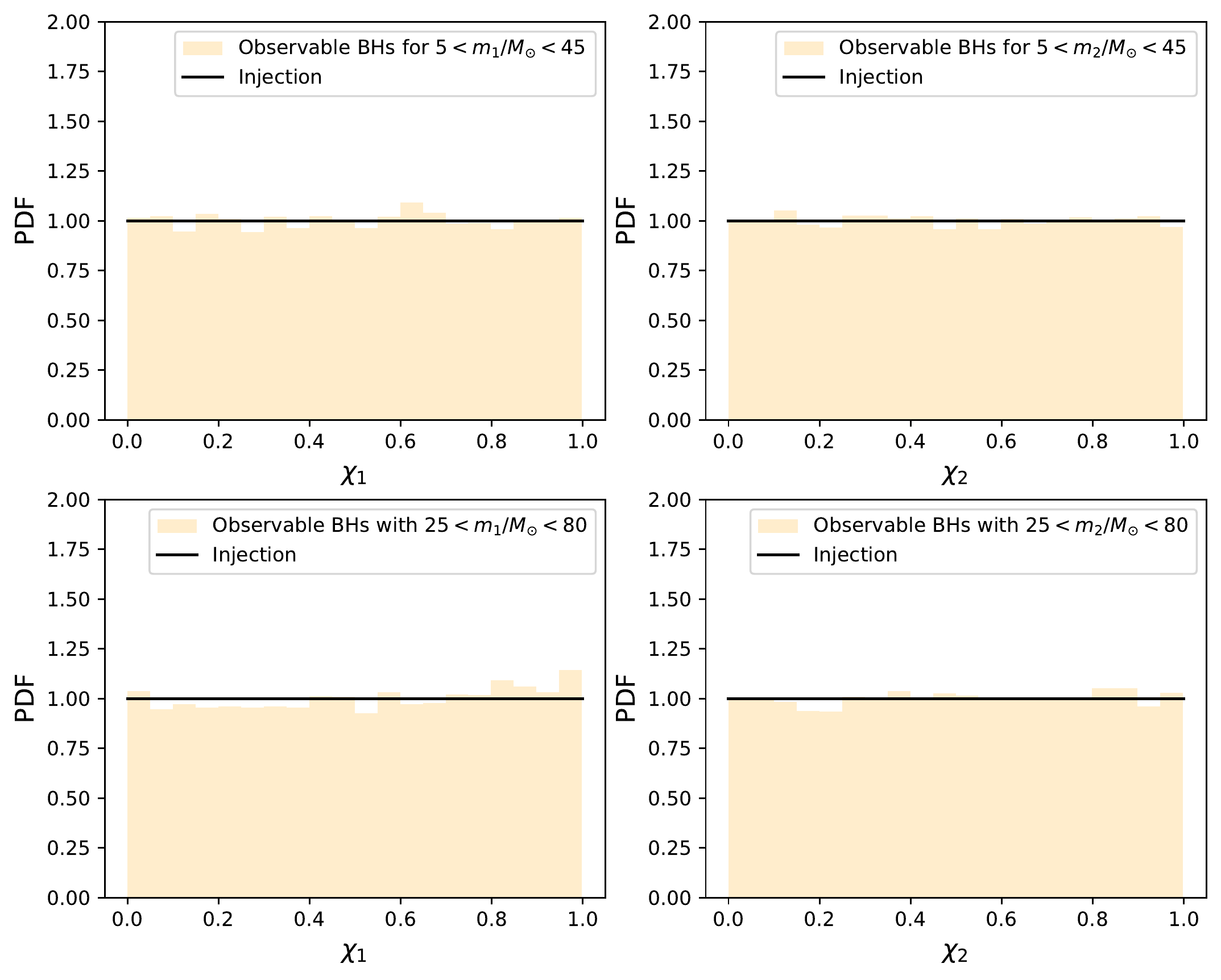}
\caption{{The distributions of the spin magnitudes of the observable BHs in the injection samples of LVK \cite{2023PhRvX..13a1048A}, adopted from (https://zenodo.org/doi/10.5281/zenodo.5636815).
}}
\label{fig:spin_inj}
\end{figure*}

{As a test (to check the validation of identifying the two sub-populations of BHs), we have performed inferences without whole selection effects. 
Since neglecting the selection effects 
will change the quantitative results, the presence of clustering within the mass versus spin magnitude distribution is expected to persist.
In other words, if there are two sub-populations of BHs in the underlying BBHs, then they should be also identifiable in the observed BBHs. 
As displayed in Figure~\ref{fig:compare_selection_dist}, we found the same population properties as those in the main text, except for the power-law slopes of the mass distribution.} 

\begin{figure*}
	\centering  
\includegraphics[width=0.8\linewidth]{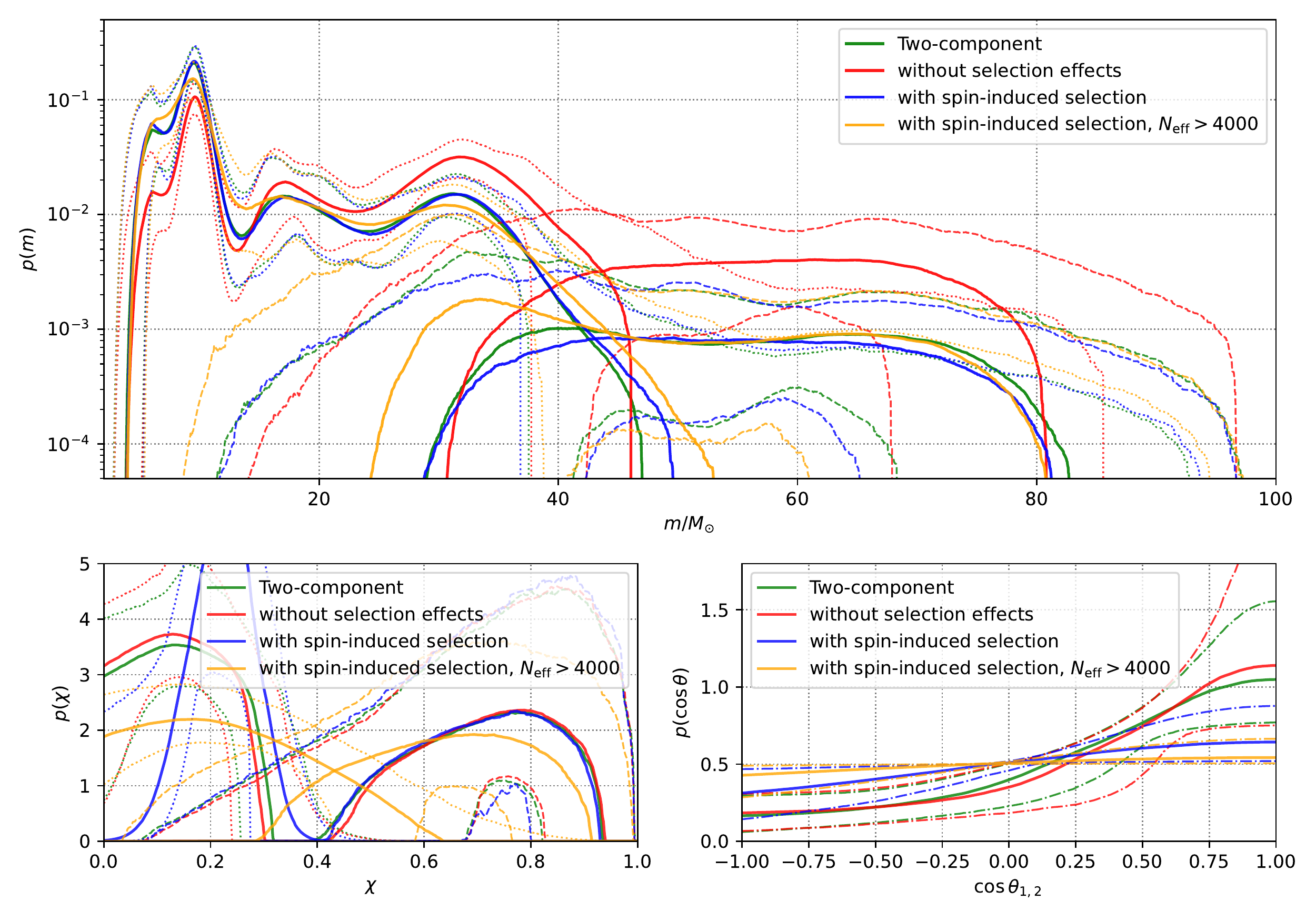}
\caption{Comparison of the results inferred with different selection effects and the convergence conditionson $N_{\rm eff}$. 
}
\label{fig:compare_selection_dist}
\end{figure*}

{In view of the above facts, we conclude that with the limited injection samples \cite{2023PhRvX..13a1048A} we have to leave out the spin-induced selection effects in our analysis, which has negligible impacts on the identification of the two sub-populations presented in the main text}.

\section{The impacts of sample sizes and convergence conditions}\label{sec:sample_size}
{We find that the sample sizes of observed events and injections, together with the convergence conditions ($N_{{\rm eff},i} > N_{\rm obs}^{\rm thr}$), potentially influence the spin populations of BBHs, especially the spin-orientation distribution. 
We have performed analysis with several configurations, as summarized in Table~\ref{tab:config}.}

\begin{table}[htpb]
\centering
\caption{Log Bayes factors for different configurations.}\label{tab:config}
\begin{tabular}{lccc}
\hline
\hline
\multicolumn{2}{c}{configurations} & $\ln{\mathcal{B}}$  \\
per-event sample size & $N_{\rm obs}^{\rm thr}$ &\\
\hline
500  & 10 &  5.1 \\
1000  & 10 &   6.0 \\
10000  & 10 &   7.0 \\
5000  & 69 &  5.4 \\
\hline
\hline
\end{tabular}
\\
\begin{tabular}{l}
Note: the log Bayes factors are between the Two-component model\\
 and the One-component model.
\end{tabular}
\end{table}

For each configuration, we compare the log Bayes factors (lnBFs) between the Two-component model and the One-component model as summarized in Table~\ref{tab:config}. All the distributions (obtained by the Two-component model) are shown in Figure~\ref{fig:compare_size_dist}.
We find that when the per-event sample size is not too small (like 500 and 1000) or the $N_{\rm obs}^{\rm thr}$ is too large for the limited sample size, the lnBF between the Two-component model and the One-component model will be smaller, and the posterior distributions for the spin magnitudes (orientations) will be flat or broader. The same happens, when the injection samples are not enough for the spin-induced selection effects, as shown in Figure~\ref{fig:compare_selection_dist}.
This is because the model with a narrow distribution (which may be close to the true distribution and has a high likelihood) is less likely to pass the threshold of ($N_{{\rm eff},i} > N_{\rm obs}^{\rm thr}$) than the model with a flat distribution. Therefore, with insufficient event / injection samples, the inferred spin-magnitude and spin-orientation distributions may be biased, which have preference for flatter/ broader/featureless distributions.

\begin{figure*}
	\centering  
\includegraphics[width=0.8\linewidth]{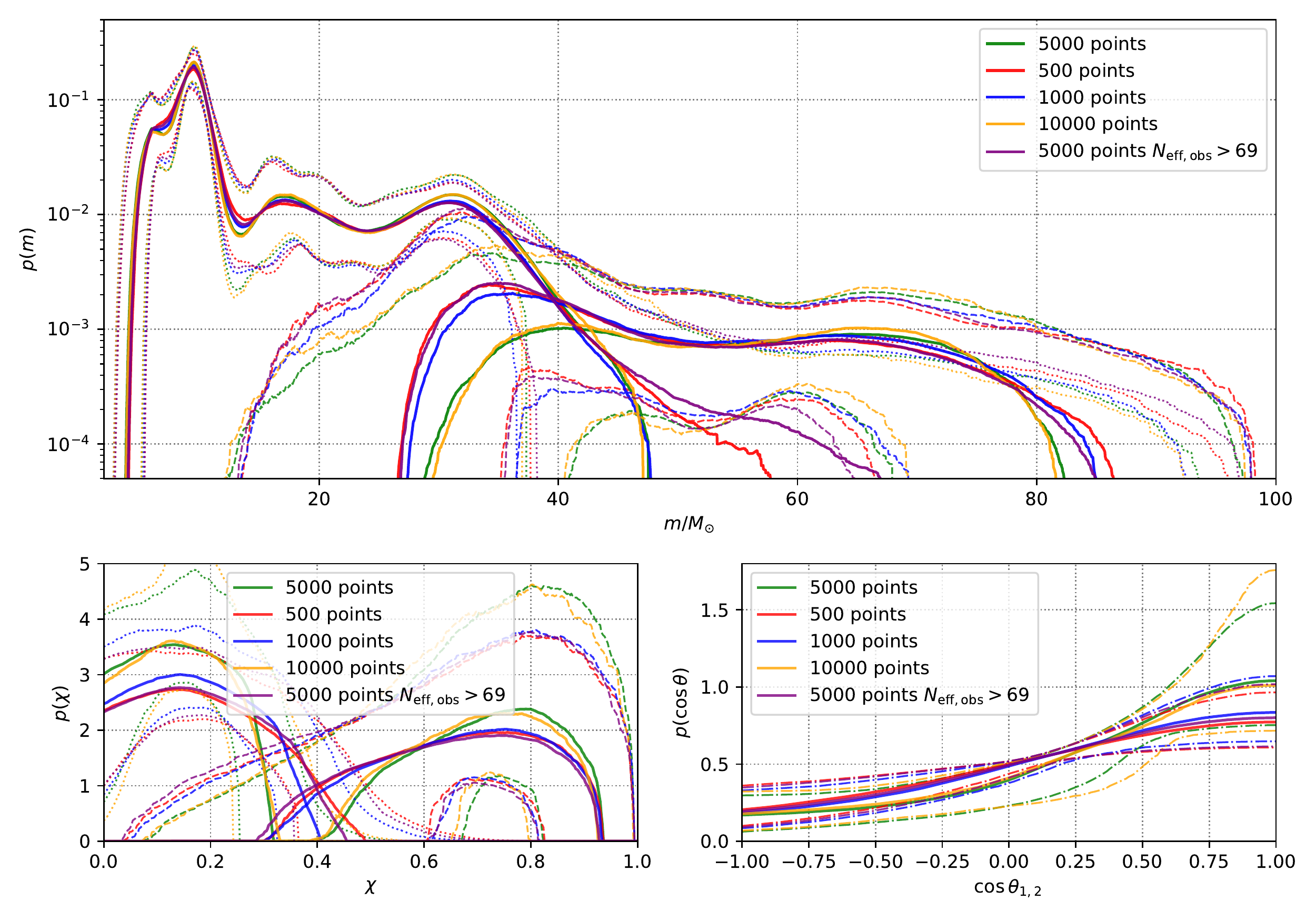}
\caption{Comparison of the results inferred with different sizes of per-event samples and thresholds for effective numbers. It shows that the smaller sample size and higher threshold result in flatter distributions.}
\label{fig:compare_size_dist}
\end{figure*}

In view of above facts, it is important to choose an appropriate configuration for analysis in this work. In practice, we adopt the sample size of 5000 per event and $N_{\rm obs}^{\rm thr}=10$, since empirical tests have demonstrated that increasing the sample size beyond 5000 does not significantly alter the results. For instance, the outcomes with a sample size of 10000 are remarkably consistent with those obtained from a 5,000 sample (refer to Figure~\ref{fig:compare_size_dist}).
Additionally, the $N_{\rm obs}^{\rm thr}=10$ is safe for inference with 5000-size per-event samples, given that the total likelihood uncertainty remains acceptably low (see \citet{2023MNRAS.526.3495T} and Figure~\ref{fig:vars}). 

\begin{figure*}
	\centering  
\includegraphics[width=0.9\linewidth]{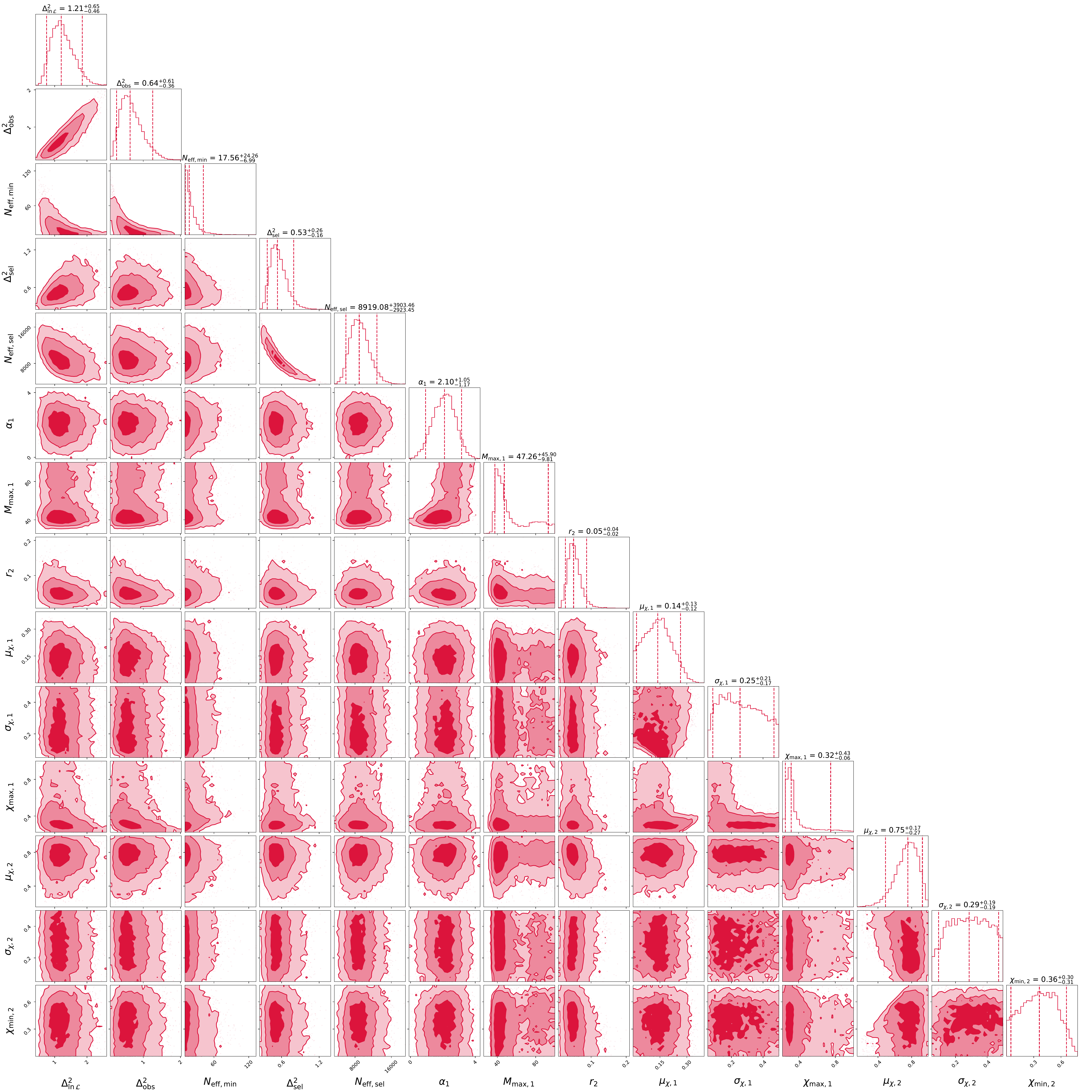}
\caption{ Distributions of the variance in the estimate of log-likelihood, and the posterior distributions of hyper-parameters describing the population model. The dashed lines in the marginal distribution represent the $90\%$ credible intervals. }
\label{fig:vars}
\end{figure*}

\section{Comparing the spin-orientation distribution to other works.}
With the same $\cos\theta_{1,2}$ distribution model, we obtain a different $\cos\theta_{1,2}$ distribution from that of \citet{2023PhRvX..13a1048A}, as reported in the main text. 
Additionally, \citet{2022A&A...668L...2V} find that the $\cos\theta$ distribution may not peak at $\sim 1$. For comparison, we also perform an inference with a variable $\mu_{\rm t}$ for the Two-component model, and obtain $0.79^{+0.19}_{-0.29}$ (see Figure~\ref{fig:mut_corner}. The two sub-populations are unchanged, as shown in Figure~\ref{fig:compare_pair_orientation_dist}.

\begin{figure}
	\centering  
\includegraphics[width=0.6\linewidth]{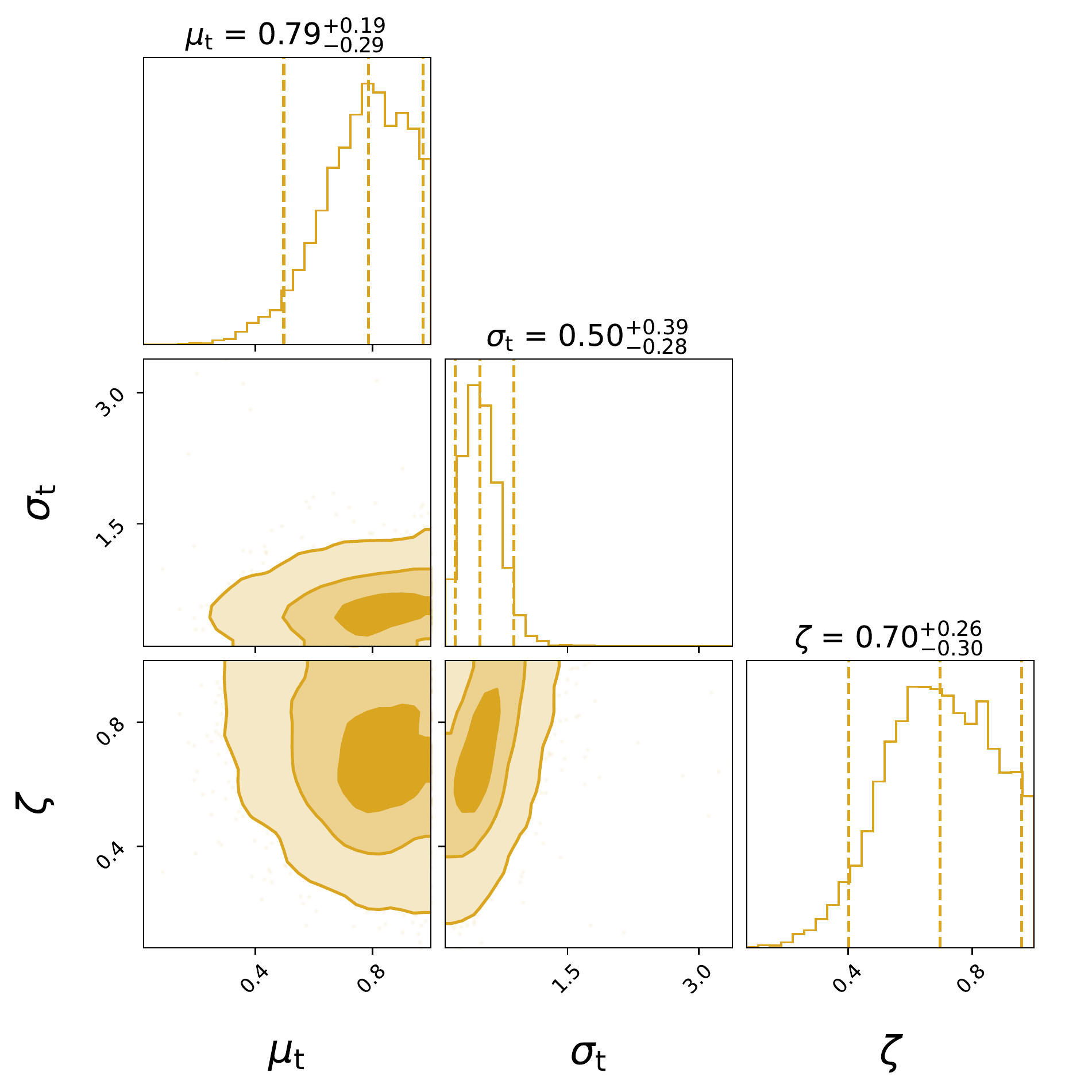}
\caption{Posterior distributions of $\sigma_{\rm t}$, $\mu_{\rm t}$ and $\zeta$. The dashed and solid contours mark the central 50\% and 90\% posterior credible regions, respectively.}
\label{fig:mut_corner}
\end{figure}

\begin{figure*}
	\centering  
\includegraphics[width=0.8\linewidth]{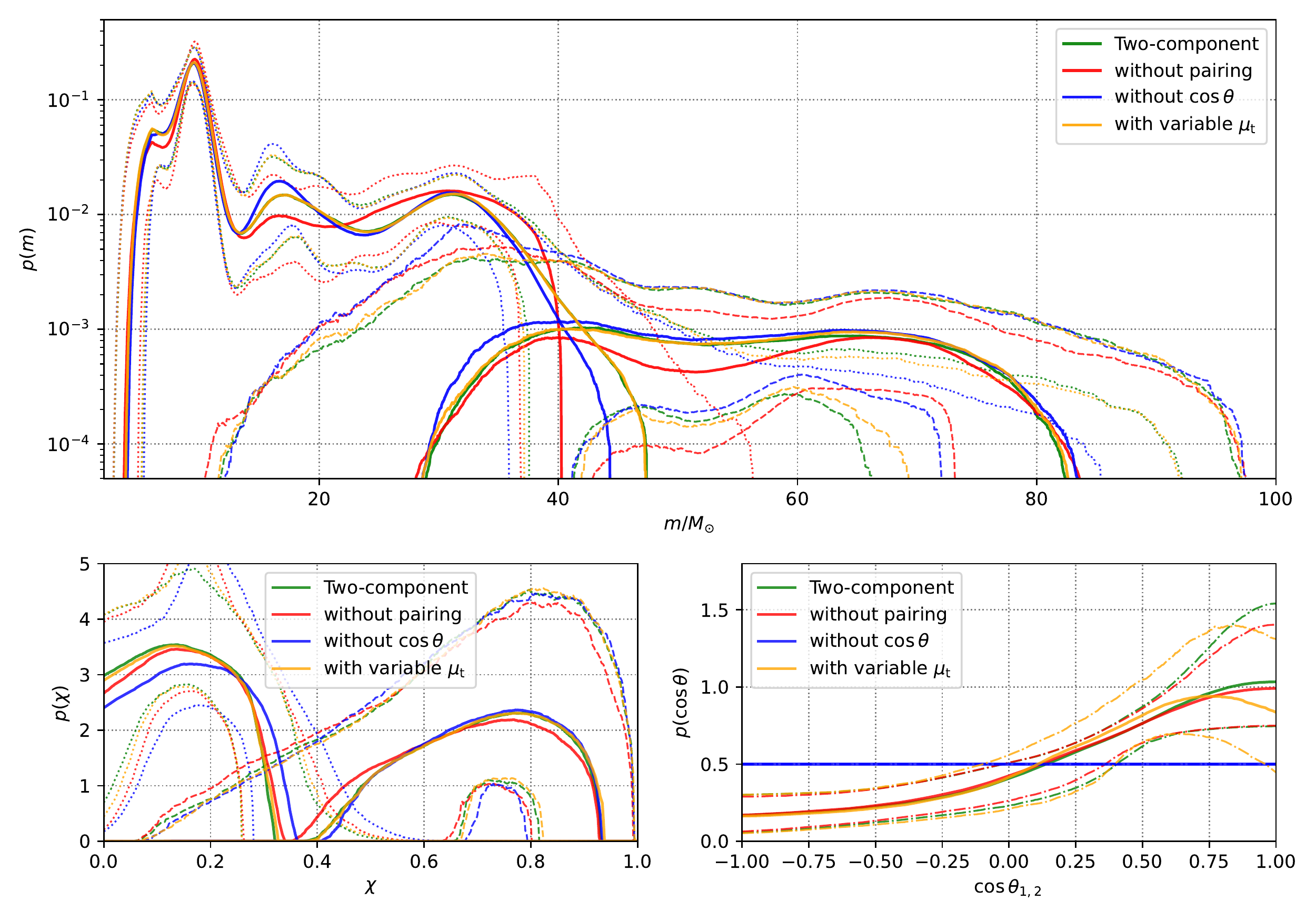}
\caption{Results inferred with the cases: variable $\mu_{\rm t}$, without $\cos\theta$, and without pairing function, compared to result of the fiducial case. It shows the identification of the two sub-population is insensitive to the $\cos\theta$ distribution and pairing function.
}
\label{fig:compare_pair_orientation_dist}
\end{figure*}

We investigate the possible causes of the different results.
{Firstly, it may be attributed to the different configurations for analysis, as illustrated in the last paragraph. See the $\cos\theta_{1,2}$ distributions in Figure~\ref{fig:compare_size_dist} for comparison. Though we are not fully aware of the per-event sample size used by \citet{2023PhRvX..13a1048A}, their convergence condition ($N_{{\rm eff},i} >69$) is much stricter than ours. See comparisons between the result of PP\&Default and that of \citet{2023PhRvX..13a1048A} in Figure~\ref{fig:compare_tilt}.}
We have also reproduce the results of \citet{2023PhRvX..13a1048A} using the same model with our samples, to identify the differences in the per-event samples (see Figure~\ref{fig:compare_LVK}). It turns out that the results are largely consistent, though the result with ($N_{{\rm eff},i} >69$) shows a flatter spin-magnitude distribution. Note that spin-induced selection effect is included for the reproduce.

\begin{figure}
	\centering  
\includegraphics[width=0.9\linewidth]{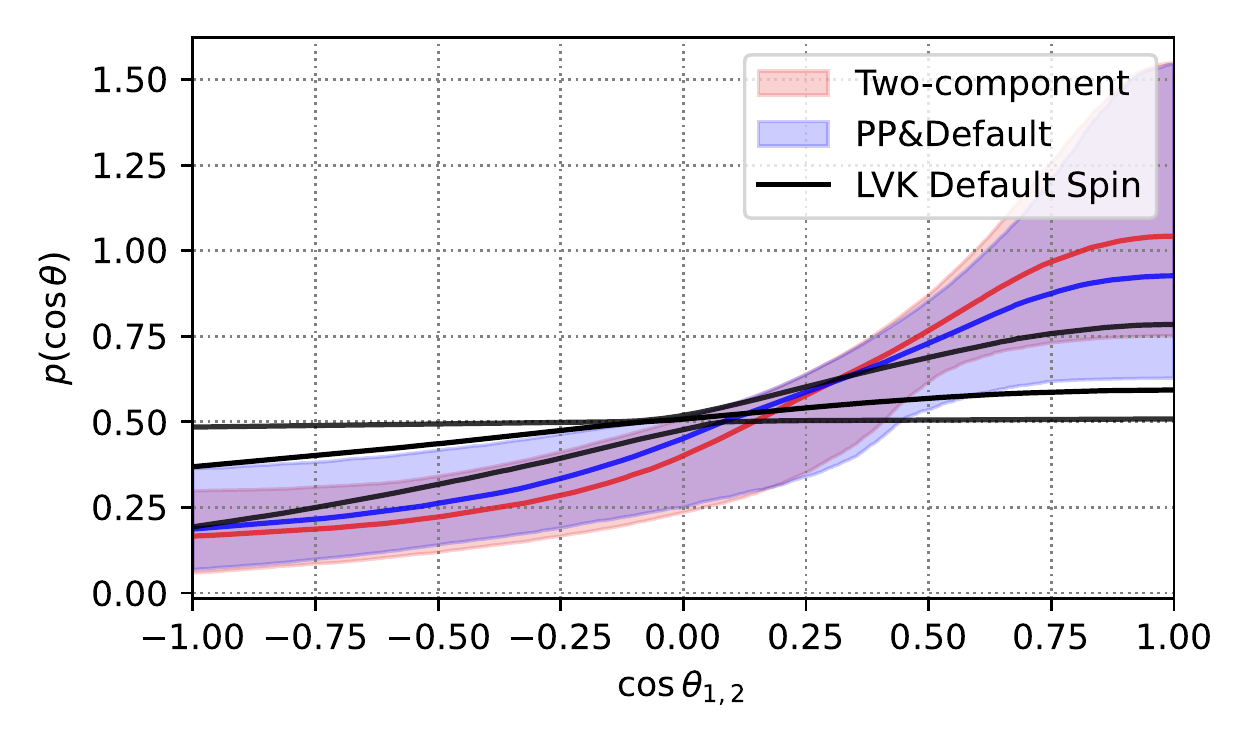}
\caption{Spin-orientation distributions inferred with the Two-component model and the PS\&Default model in this work comparing to that of Default Spin model in \citet{2023PhRvX..13a1048A}.
}
\label{fig:compare_tilt}
\end{figure}

\begin{figure*}
	\centering  
\includegraphics[width=0.8\linewidth]{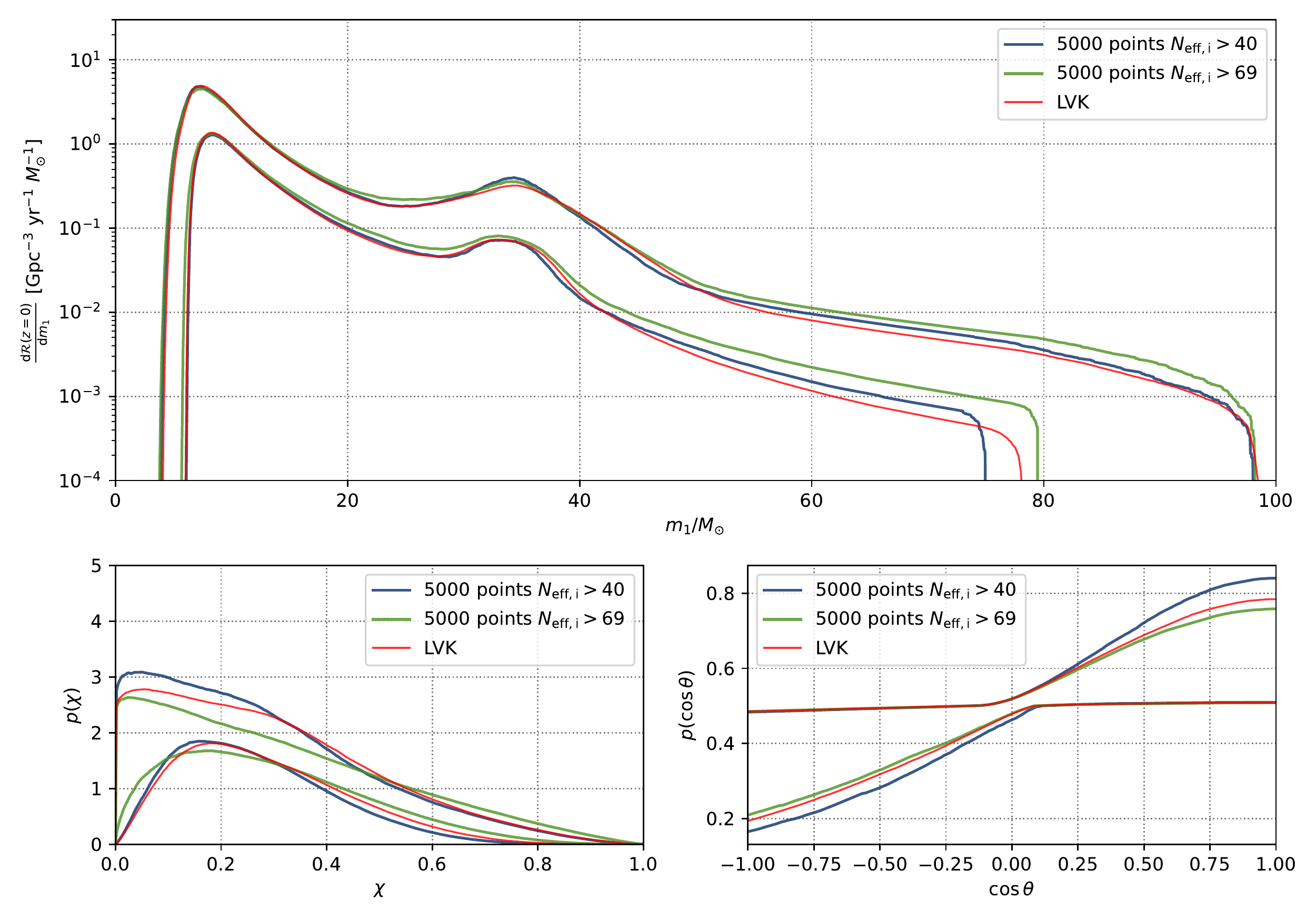}
\caption{Reproduced results using the model in \citet{2023PhRvX..13a1048A} with our samples, compared to the results of \citet{2023PhRvX..13a1048A}.
}
\label{fig:compare_LVK}
\end{figure*}

Secondary, the different modeling of spin-magnitude versus mass distribution can also modulate the spin-orientation distribution, see comparisons between the results of Two-component model, PP\&Default in Figure~\ref{fig:compare_tilt}. This is because, for the Two-component model, the LSG and HSG have different/independent spin-magnitude distributions, and the spin-magnitude distribution of LSG (i.e., the majority of the BHs) concentrates in the low value region, which prevents the $\cos\theta$ distribution from being flatter. Note that the aligned spin $s_z\equiv \chi \cos\theta$ is better measured, while the $\chi$ and $\cos\theta$ are degenerated  \cite{2024arXiv240105613M}. 

\citet{2022PhRvD.105b4076M} suggested that ignoring spins in estimating selection effects would lead to an overprediction of spin alignment in the underlying astrophysical distribution of merging black holes. 
To examine the impact of this factor on the $\cos\theta$ distribution, we plot the $\cos\theta$ distributions of the injection samples and the observable samples in Figure~\ref{fig:ct_inj}. There is a detection preference for aligned events. It is, however, not strong enough to account for the difference between our result and that of \citet{2023PhRvX..13a1048A}. More intuitively, we present the $\cos\theta$ distribution modified (re-weighed) by the detecting preference in Figure~\ref{fig:ct_inj}, which is still significantly different from that of \citet{2023PhRvX..13a1048A}, see Figure~\ref{fig:reweighed_tilt}. 

\begin{figure}
	\centering  
\includegraphics[width=0.9\linewidth]{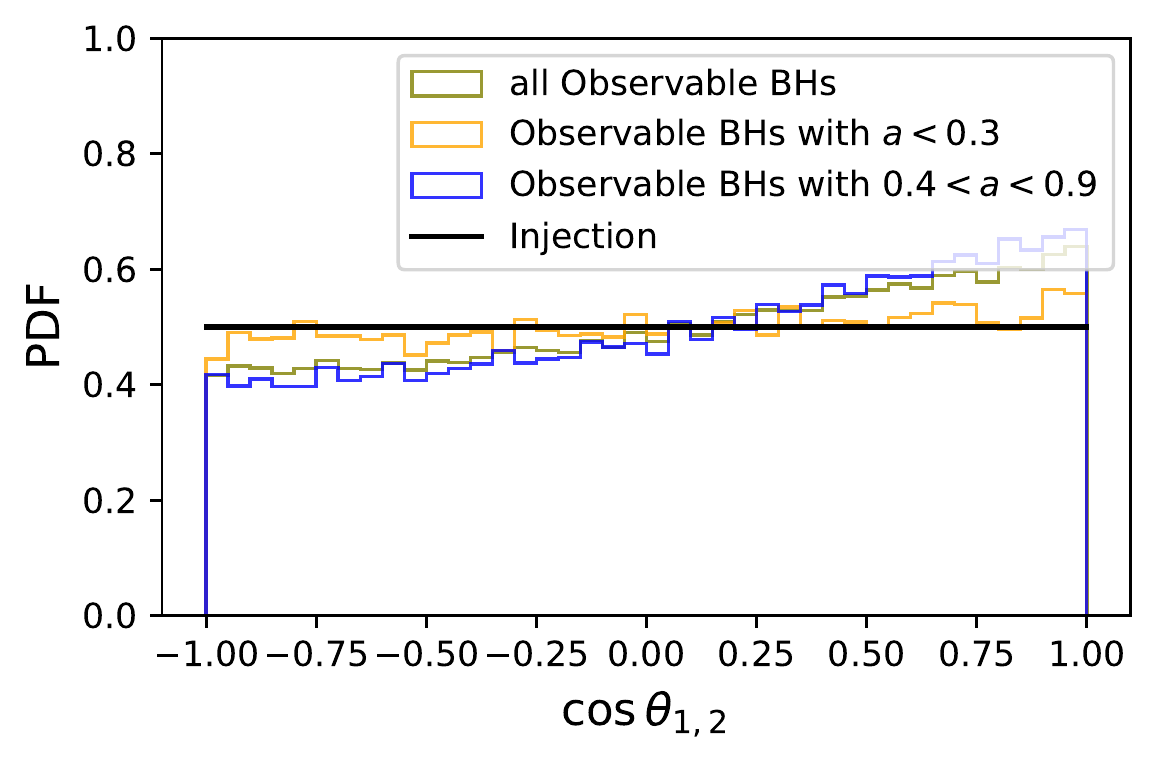}
\caption{The spin-orientation distributions of BHs for the injection samples and the observable samples. {It shows that the samples with $a<0.3$ (which are the majority of the events detected by LIGO/Virgo/KAGRA) have negligible detecting preference for aligned events.
}}
\label{fig:ct_inj}
\end{figure}

\begin{figure}
	\centering  
\includegraphics[width=1\linewidth]{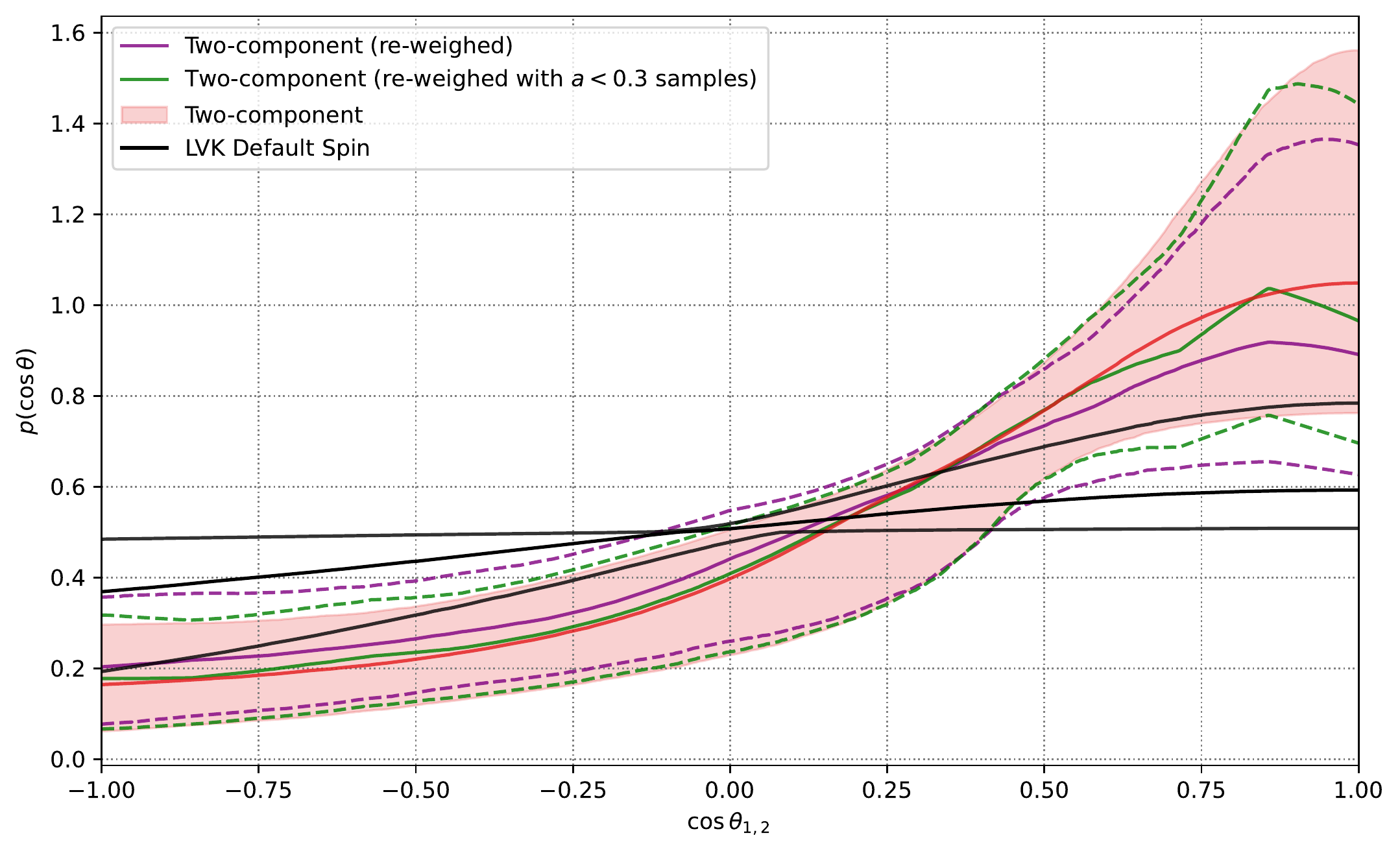}
\caption{Spin-orientation distribution inferred with the Two-component model re-weighed by the detecting preference (for aligned events) comparing to that of Default Spin model in LVK\cite{2023PhRvX..13a1048A}. It shows that the distribution re-weighed by the low-spin (i.e., $a<0.3$, the majority of observed BHs) samples is rather close to the initial distribution, indicating minor impact of the selection effects on our spin-orientation distribution.}
\label{fig:reweighed_tilt}
\end{figure}

In view of above facts, we conclude that the difference of spin-orientation distributions between our result and that of \citet{2023PhRvX..13a1048A} is mainly attributed to the different configurations (the sample sizes and the thresholds of effective numbers) for analysis and the different modeling on spin-magnitude versus mass distribution.

\section{The identification of the two BH sub-populations is insensitive to the pairing function and the tilt-angle distribution}\label{sec:app_independent}
{The pairing function and the tilt-angle distribution used in our model may be not exact. For instances, the pairing functions in different formation channels may be different \cite{2022PhR...955....1M}, and the mass-ratio distribution is not constant in the whole mass range \cite{2022ApJ...933L..14L}. For the $\cos\theta$ distribution, both the isolated field BBHs and the dynamical BBHs formed in the AGN disks are expected to have nearly aligned spin orientations, but their distributions may be different.}

To test whether the probable model mis-specification in the pairing function or the $\cos\theta$ distribution dominates the identification of the two sub-populations of BHs in this work, we have carried out analysis as following. We firstly performed an inference without the information about the spin tilt angles. In practice, the $\cos\theta$ distribution is set just the same as the prior distribution, i.e., $p(\cos\theta_{1},\cos\theta_2) =  1/4$. Secondary, we performed an inference without the pairing function, i.e, the Two-component model without $(m_2/m_1)^{\beta}$.
In both inferences, two sub-populations of BHs are successfully identified, as shown in Figure~\ref{fig:compare_pair_orientation_dist}. The mass and spin-magnitude distributions are only slightly changed, except for the mass distribution in the case of without pairing function.

\section{Results inferred with the Two-component model}
\subsection{The prior and the posterior distributions.}\label{sec:prior_post}
The posteriors distribution of the hyper-parameters describing the mass distributions and the spin distributions introduced in the main text are displayed in Figure~\ref{fig:mass} and Figure~\ref{fig:spin}, respectively.

\begin{figure}
	\centering  
\includegraphics[width=1.1\linewidth]{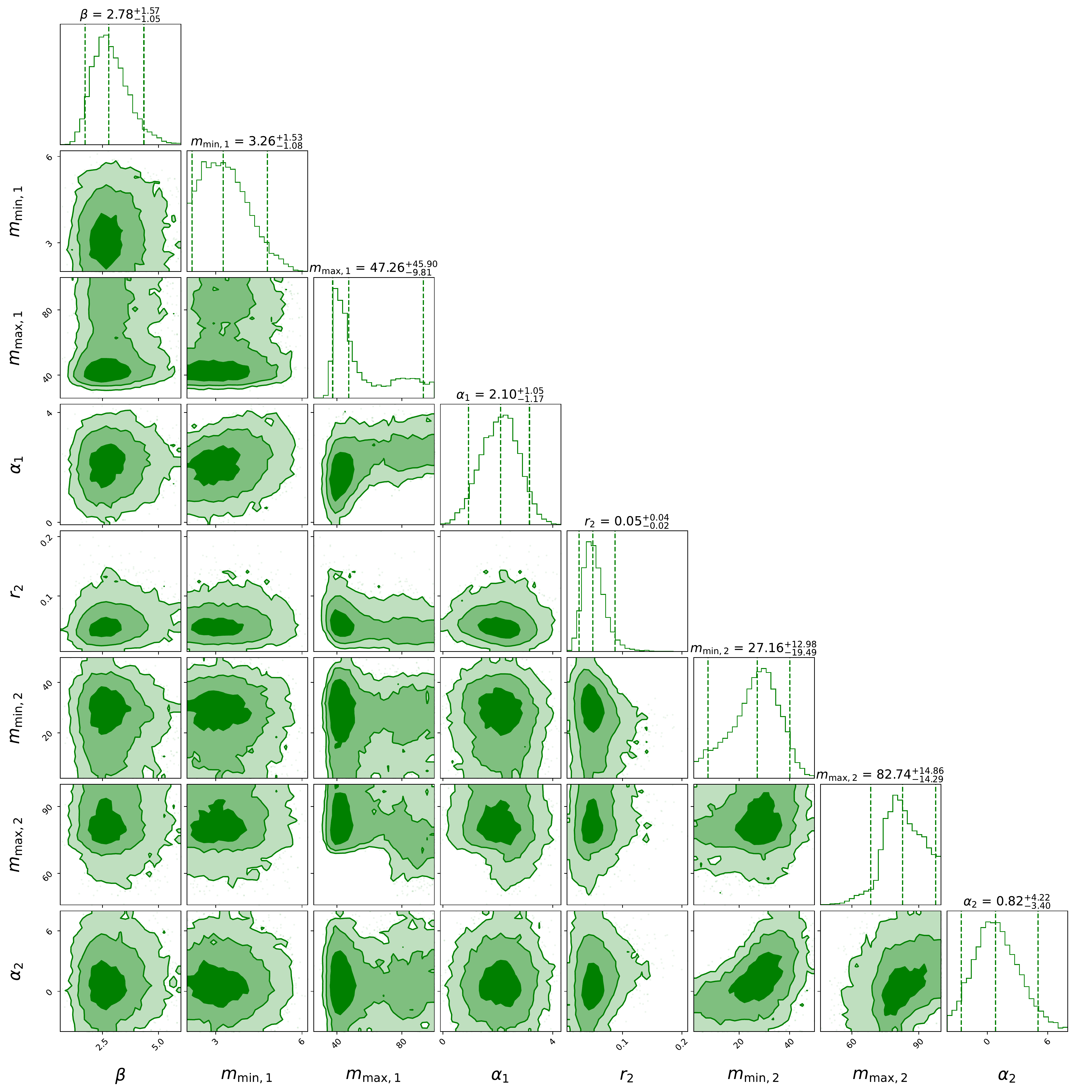}
\caption{Posterior distributions of all the parameters describing the mass distribution. The dashed lines in the marginal distributions represent the $90\%$ credible intervals. }
\label{fig:mass}
\end{figure}

\begin{figure*}
	\centering  
\includegraphics[width=0.6\linewidth]{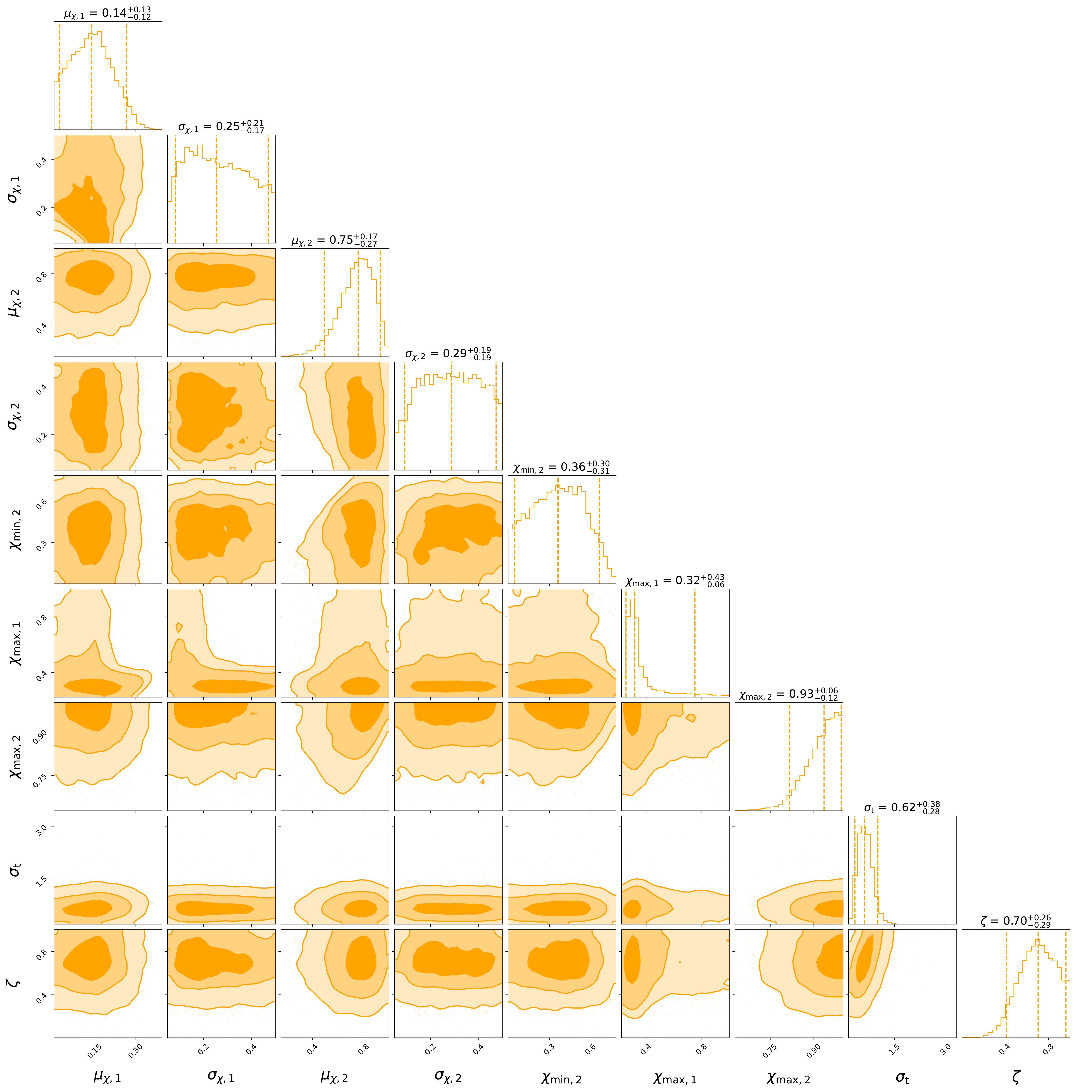}
\caption{Posterior distributions of the parameters describing the spin distribution. The dashed lines in the marginal distribution represent the $90\%$ credible intervals. }
\label{fig:spin}
\end{figure*}

\begin{figure*}
	\centering  
\includegraphics[width=0.7\linewidth]{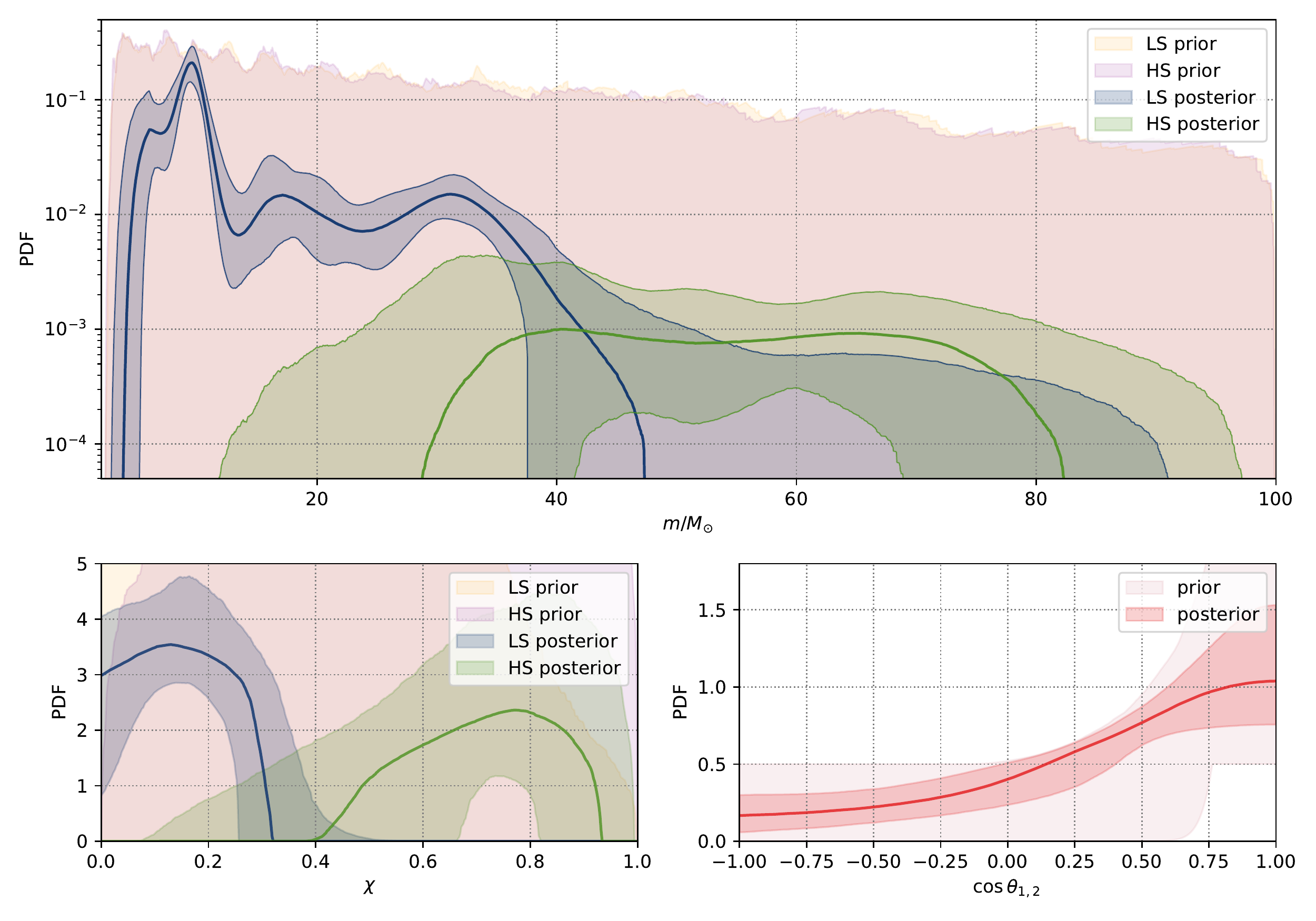}
\caption{Prior distributions of the mass and spin comparing to the posterior distributions of the Two-component model. }
\label{fig:prior}
\end{figure*}

The prior and posterior distributions of mass and spin for the Two-component model are shown in Figure~\ref{fig:prior}. The prior distributions are broad enough, so we can conclude that the GW data are indeed informing the model.

\subsection{The maximum mass of the first-generation BHs}
The posterior distribution of $m_{\rm max,1}$ has a tail extending to the higher mass range, see Fig.~3 in the main text. This may be because that, in our population model, a large $m_{\rm max,1}$ (e.g. $80M_{\odot}$) with a steep power-law slope $\alpha_1$ and low values of the spline for the high-mass range can also construct a mass function with low probability density in the high-mass range. Therefore the posterior of $m_{\rm max,1}$ is degenerated with the perturbation function in the high-mass range, as shown in Figure~\ref{fig:mmax1}. However the masses of the 99th and 99.5th percentiles are better measured as shown in Figure~\ref{fig:BBH_quantiles}.

\begin{figure}
	\centering  
\includegraphics[width=1\linewidth]{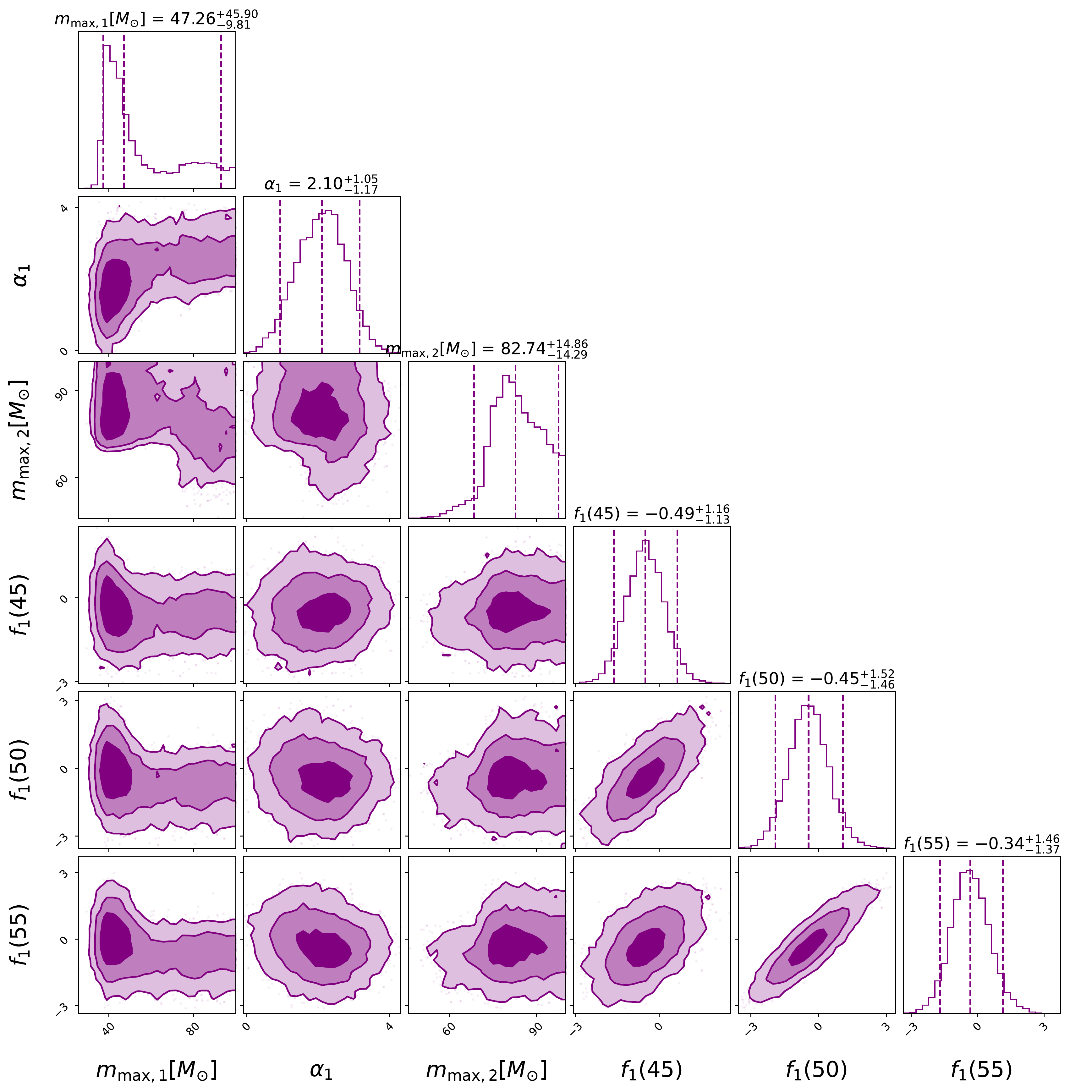}
\caption{{ Posterior distributions of the maximum-mass cutoff of LSG and the perturbation function. The high $m_{\rm max,1}$ values are associated with the large $\alpha_1$ values and small $f(m)$ in the high-mass range, note that the prior distribution of $f_i^j$ peaks at 0.} 
 }
\label{fig:mmax1}
\end{figure}

\begin{figure}[!h]
	\centering  
\includegraphics[width=0.8\linewidth]{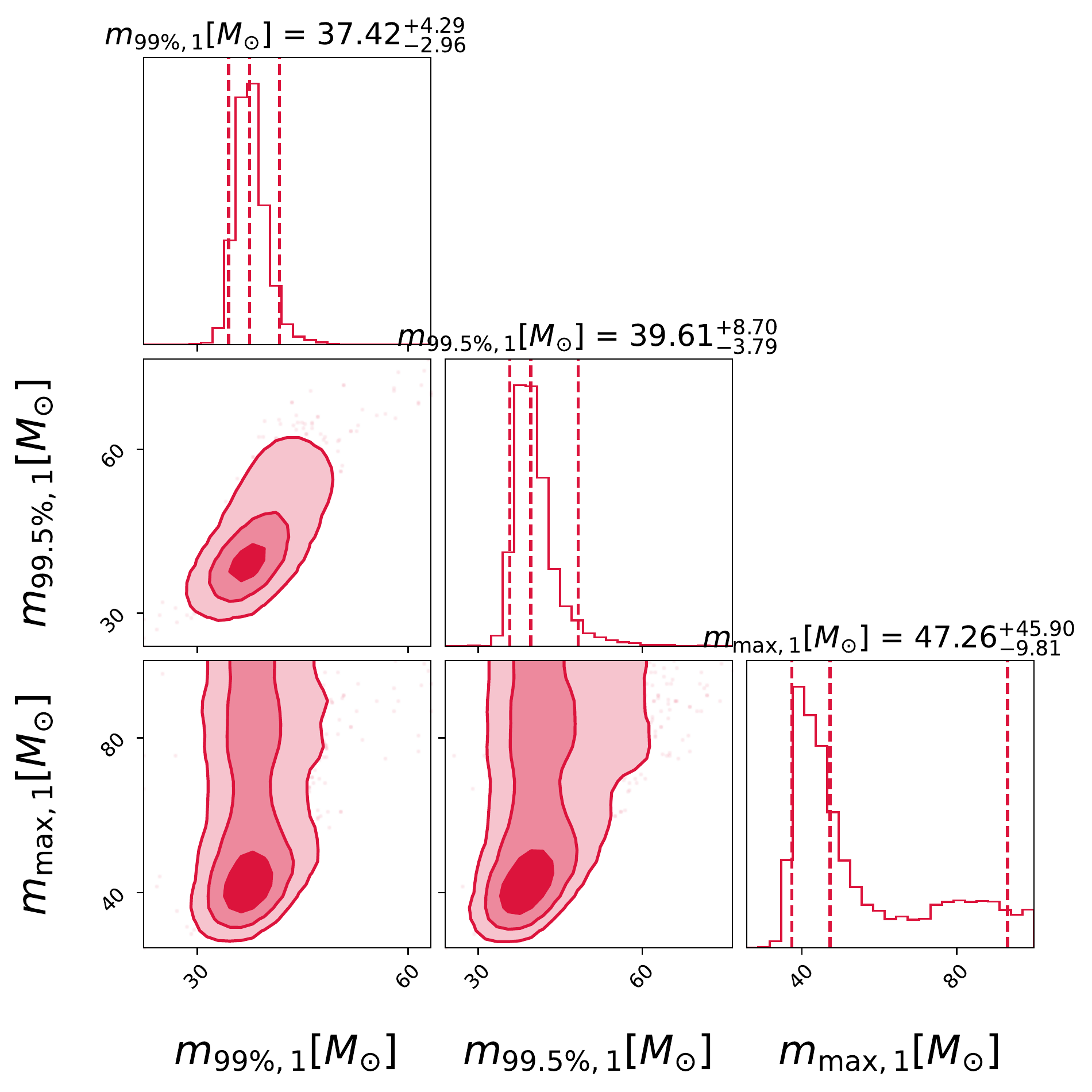}
\caption{{ Posterior distributions of the maximum-mass cutoff of the first-generation BHs, as well as the masses of the 99th and 99.5th percentiles 
of the BBH mass.} 
 }
\label{fig:BBH_quantiles}
\end{figure}

\subsection{The primary and secondary mass distribution}

Figure~\ref{fig:m1m2_dist} shows the primary-mass and secondary-mass function of BBHs, comparing to those of \citet{2023PhRvX..13a1048A}. It shows that the primary-mass function inferred with the Two-component model is consistent with that of \citet{2023PhRvX..13a1048A} inferred with the `PS' model.
Note that the component-mass function displayed in the main text indicates the underlying (or unpaired) mass function, whereas the primary- and secondary- mass functions displayed here are the products undergo pairing function.

\begin{figure*}
	\centering  
\includegraphics[width=0.8\linewidth]{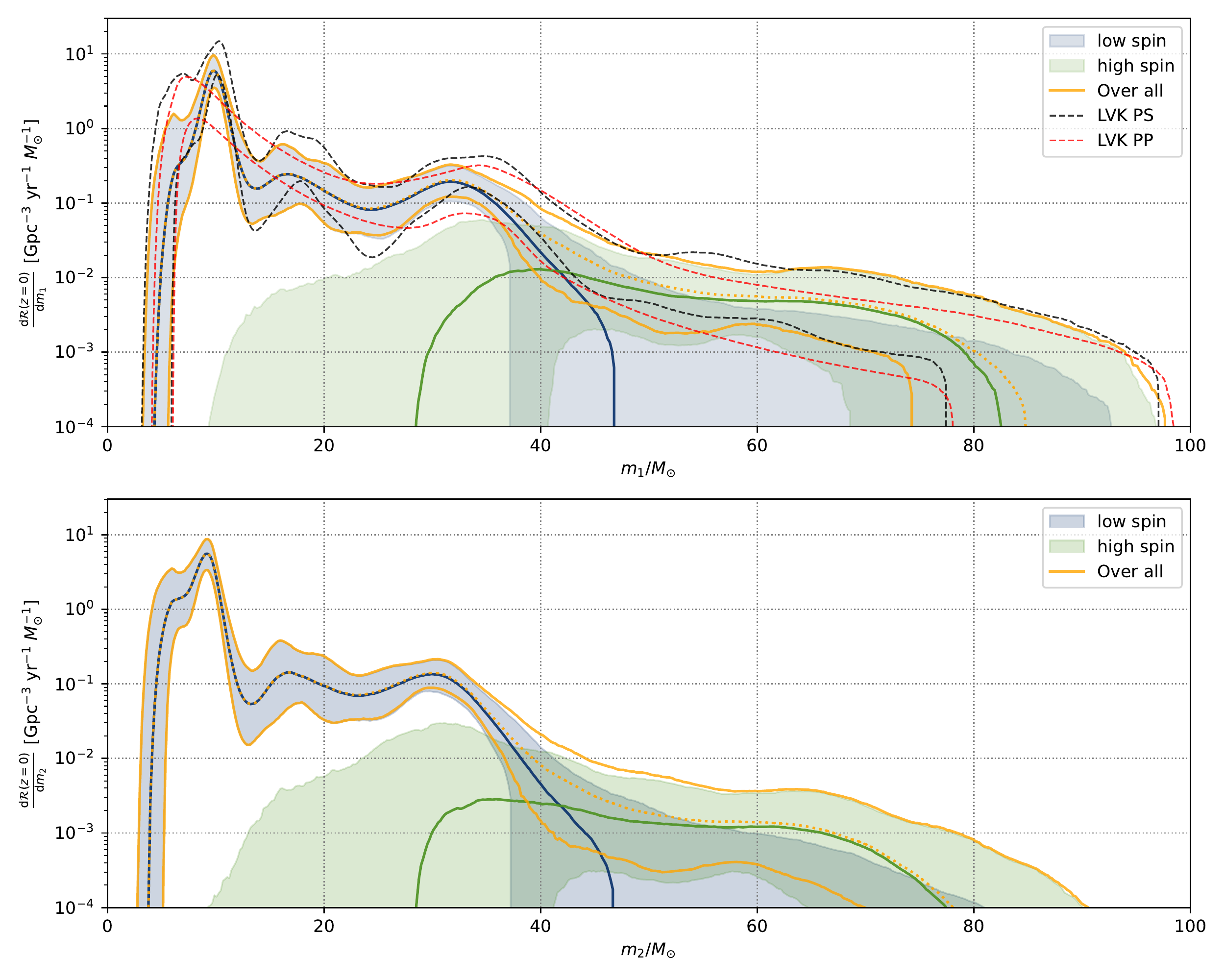}
\caption{Reconstructed primary-mass (top) and secondary-mass (bottom) distributions of BBHs, compared to the results in \citet{2023PhRvX..13a1048A}; the regions between the solid / dashed curves are the 90\% credible intervals, respectively.
}
\label{fig:m1m2_dist}
\end{figure*}

\subsection{The branch ratios of the two categories.}\label{sec:channels}
Using the results from the fiducial model in the main text, we calculate the fractions of the BBH mergers with component BHs belonging to the two categories (see Figure~\ref{fig:ratios}). We find {$97.4^{+1.5}_{-3.7}\%$} of the sources have both BHs in the LSG, while {$1.9^{+2.3}_{-1.0}\%$} have the primary components in the HSG, and {$0.4^{+0.7}_{-0.3}\%$} have both components in the HSG. We assume that the HSG is made of HG BHs, then {$2.6^{+3.7}_{-1.5}\%$} of the sources are hierarchical mergers, which is consistent with the results in \citet{2022ApJ...941L..39W} where an astrophysical-motivated model was used.
We further calculate the local merger rate density of the sources including at least one high-spin component to be {$0.46^{+0.61}_{-0.24}~{\rm Gpc^{-3}yr^{-1}}$}.  

\begin{figure}
	\centering  
\includegraphics[width=0.8\linewidth]{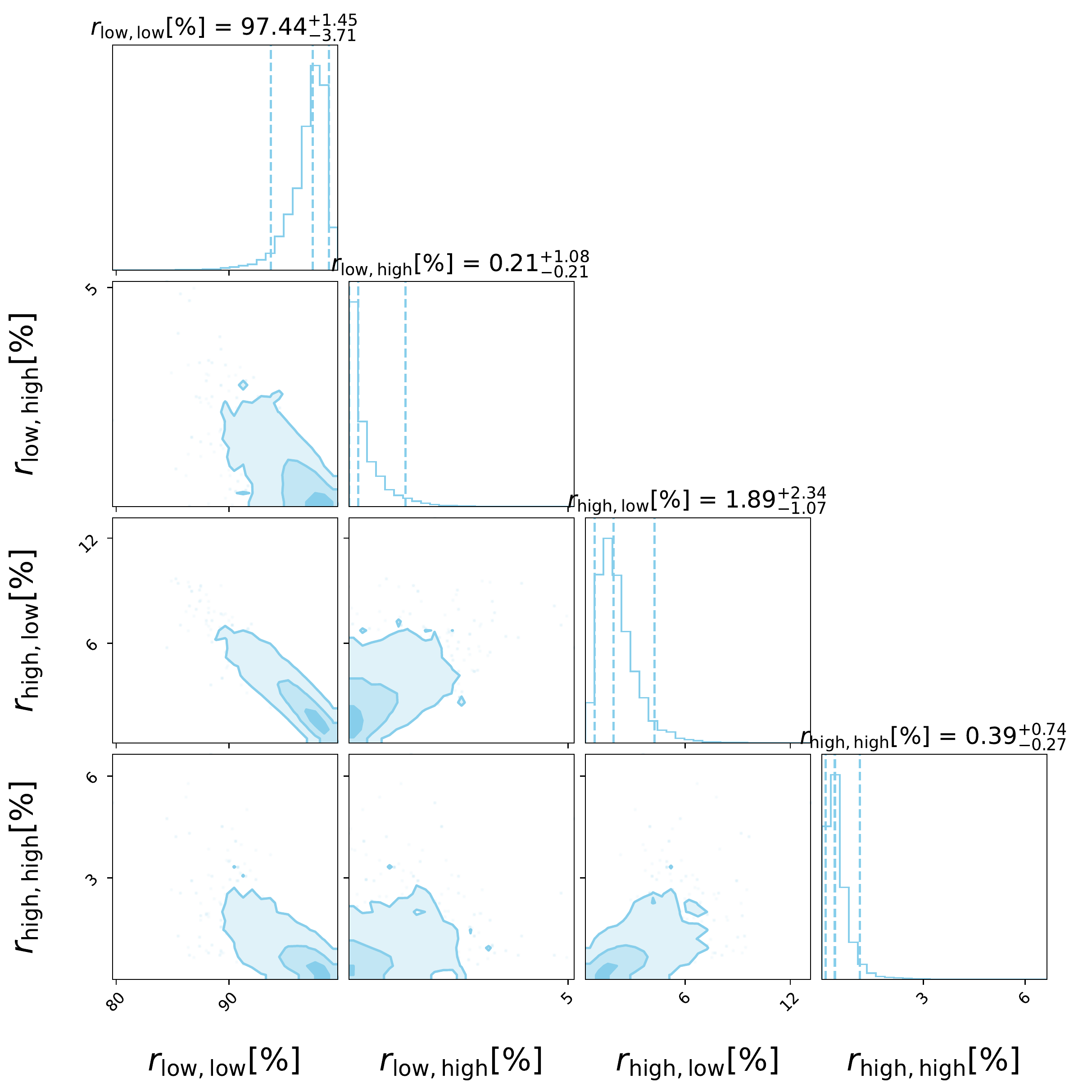}
\caption{Distributions of the branch ratios for the BBHs with the primary and secondary objects belonging to the LSG and HSG. The dashed lines represent the 90\% credible intervals}
\label{fig:ratios}
\end{figure}

\subsection{Events from hierarchical mergers} 
We also calculate the probabilities for each BBH event that belongs to a specific combination of the categories, and select the events with probabilities of $>0.5$ to contain at least one BH belonging to the high-spin category. The results are summarized in Table~\ref{tab:psub}.
Though the results in \citet{2021ApJ...915L..35K} depend on the cluster escape velocity, we find their classification of the hierarchical mergers is broadly consistent with ours. E.g., GW190521 is favored to contain two second-generation BHs, while GW190519\_153544, GW190602\_175927, GW190620\_030421, and GW190706\_222641 are mixed-generation binaries. However, in our classification, GW190517\_055101 is strongly favored as a hierarchical merger, which only has 0.004 chance to be first-generation BBH, though the two component masses may be below the PIMG. The difference may be attributed to the flexibility of our model, where the underlying mass distributions are fitted with a non-parametric model.

\begin{table}[htpb]\label{tab:psub}
\caption{Probabilities of some events belonging to a specific category.}
\begin{tabular}{ccccc}
\hline
\hline
Events   &  low+low & low+high & high+low & high+high \\ 
\cline{1-5}


GW170729&0.13&0.079&0.645&0.147 \\
GW190517\_055101&0.004&0.025&0.71&0.262 \\
GW190519\_153544&0.032&0.058&0.562&0.348 \\
GW190521&0.105&0.062&0.062&0.771 \\
GW190602\_175927&0.105&0.062&0.375&0.457 \\
GW190620\_030421&0.034&0.094&0.64&0.232 \\
GW190701\_203306&0.248&0.053&0.509&0.19 \\
GW190706\_222641&0.044&0.042&0.562&0.353 \\
GW190929\_012149&0.427&0.06&0.447&0.066 \\
GW190805\_211137&0.38&0.101&0.364&0.155 \\
GW191109\_010717&0.012&0.02&0.39&0.577 \\

\hline
\hline
\end{tabular}
\begin{tabular}{l}
Note: Here `low+high' means the primary object belong to the LSG\\
while the secondary object belong to the HSG. We only display events\\
with probabilities of low+low less than 0.5.
\end{tabular}
\end{table}

\subsection{Results with different numbers of knots in the mass function}
Figure~\ref{fig:Nnodes} shows that the number of knots in the mass function only affects the smoothness of the mass distribution of the LSG, but does not affect the identification of the two sub-populations and their spin distributions. 

\begin{figure*}
	\centering  
\includegraphics[width=0.8\linewidth]{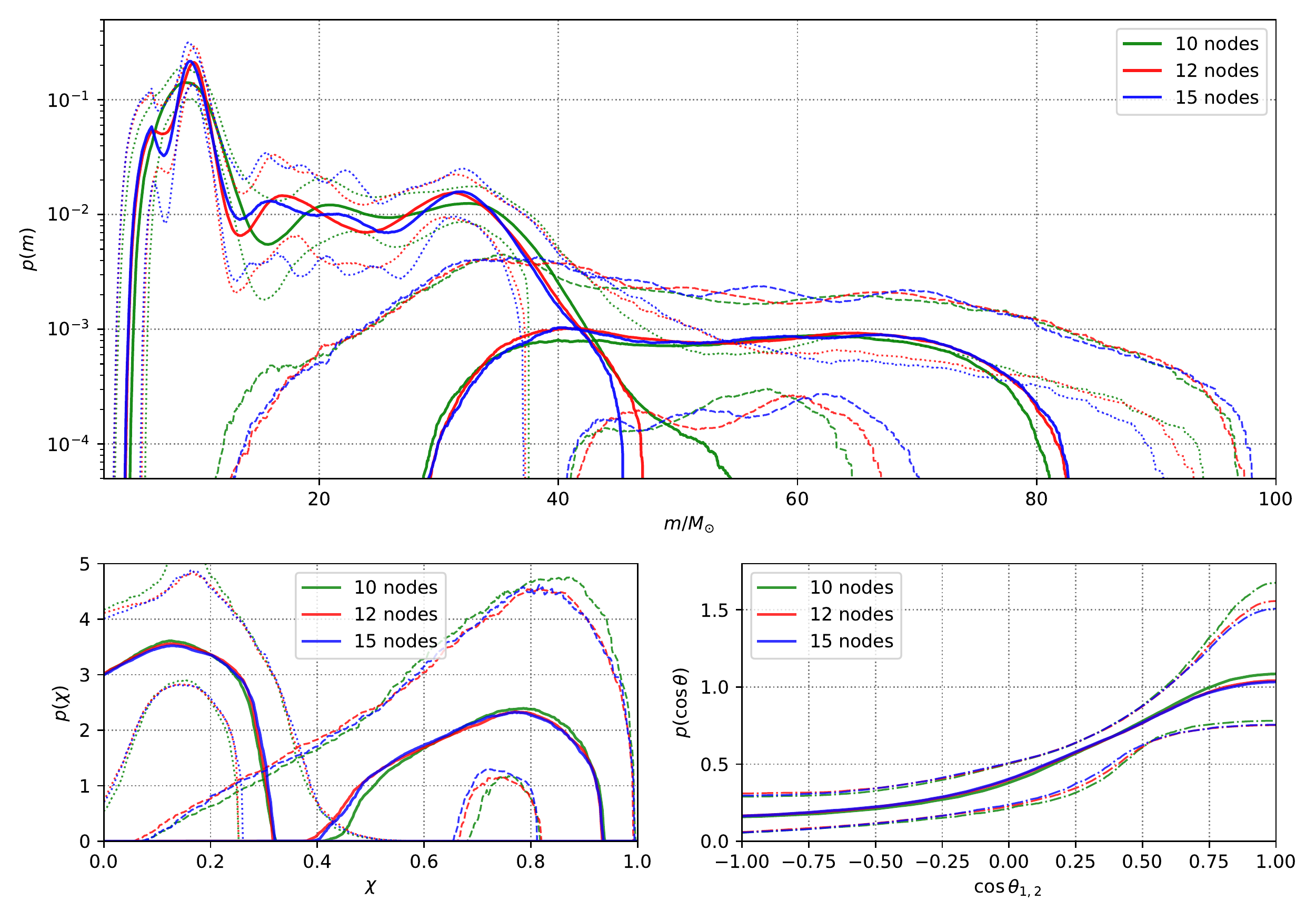}
\caption{Comparison of the results inferred with different different numbers of knots in the mass function.}
\label{fig:Nnodes}
\end{figure*}

\section{Further investigation of the spin-orientation distributions}\label{sec:further}

{Here we modify the mixture model for Two-component scenario, and the spin-orientation distribution of BHs for each component reads
\begin{equation}
\begin{aligned}
\mathcal{GU}(\cos\theta | \zeta_i, \sigma_{{\rm t,}i})&=\mathcal{G}(\cos\theta|-1, 1,1 , \sigma_{{\rm t}, i})\times \zeta_i\\
&+(1-\zeta_i)\times \mathcal{U}(\cos\theta|-1,1).
\end{aligned}
\end{equation} 
Then the mass, spin-magnitude, and spin-orientation distribution for the $i$th component is 
\begin{equation}
\begin{aligned}
&\pi_i(m, \chi, \cos\theta | {\bf \Lambda}_i)=\\
&\mathcal{PS}(m|\alpha_i,m_{{\rm min},i},m_{{\rm max},i},\delta_i,f_i(m;\{f_i^j\}_{j=0}^{N_{\rm knot}}))\\
&\times  \mathcal{G}(\chi |\chi_{{\rm min},i}, \chi_{{\rm max},i}, \mu_{\chi,i}, \sigma_{\chi,i})\times \mathcal{GU}(\cos\theta | \zeta_i, \sigma_{{\rm t,}i}).
\end{aligned}
\end{equation}
Therefore, the distribution of the final BBH population is
\begin{equation}\label{eq:iid_pop}
\begin{aligned}
&\pi({\bf \lambda} | {\bf \Lambda}; \beta) =\\
& C({\bf \Lambda};\beta) \times \pi(m_1, \chi_1,\cos\theta_1| {\bf \Lambda})\times\pi(m_2, \chi_2, \cos\theta_2| {\bf \Lambda})\\
&\times (m_2/m_1)^\beta \times \Theta(m_1-m_2) \times p(z|\gamma=2.7),
\end{aligned}
\end{equation} 
Note that, different from the Two-component model in the main text, here the spin tilt angles of the two objects in one system $\cos\theta_1$ and $\cos\theta_2$ are independently and identically distributed. While in the Two-component model, the two spin orientations of BBHs are from either the isotropic or the aligned component, which is the same as the popular \textsc{Default} spin model \cite{2023PhRvX..13a1048A}.
}
Figure~\ref{fig:iidct} shows the mass and spin distribution of the BHs in the two sub-populations, the mass and spin-orientation distributions are similar to those inferred in the Two-component model. Beside the first-generation sub-population, the HG sub-population may also host a fraction of nearly-aligned assembly (see also Figure~\ref{fig:zeta}), suggesting that some of the hierarchical mergers may be from the gas-rich environments, like the AGN disks \cite{2019PhRvL.123r1101Y,2021MNRAS.507.3362T}. 
It should be noted that with the current data, the purely isotropic distribution ($\zeta_2=0$ or $\sigma_{\rm t,2} \gg 1$) for the HG sub-population can not be ruled out yet.

\begin{figure*}
	\centering  
\includegraphics[width=0.8\linewidth]{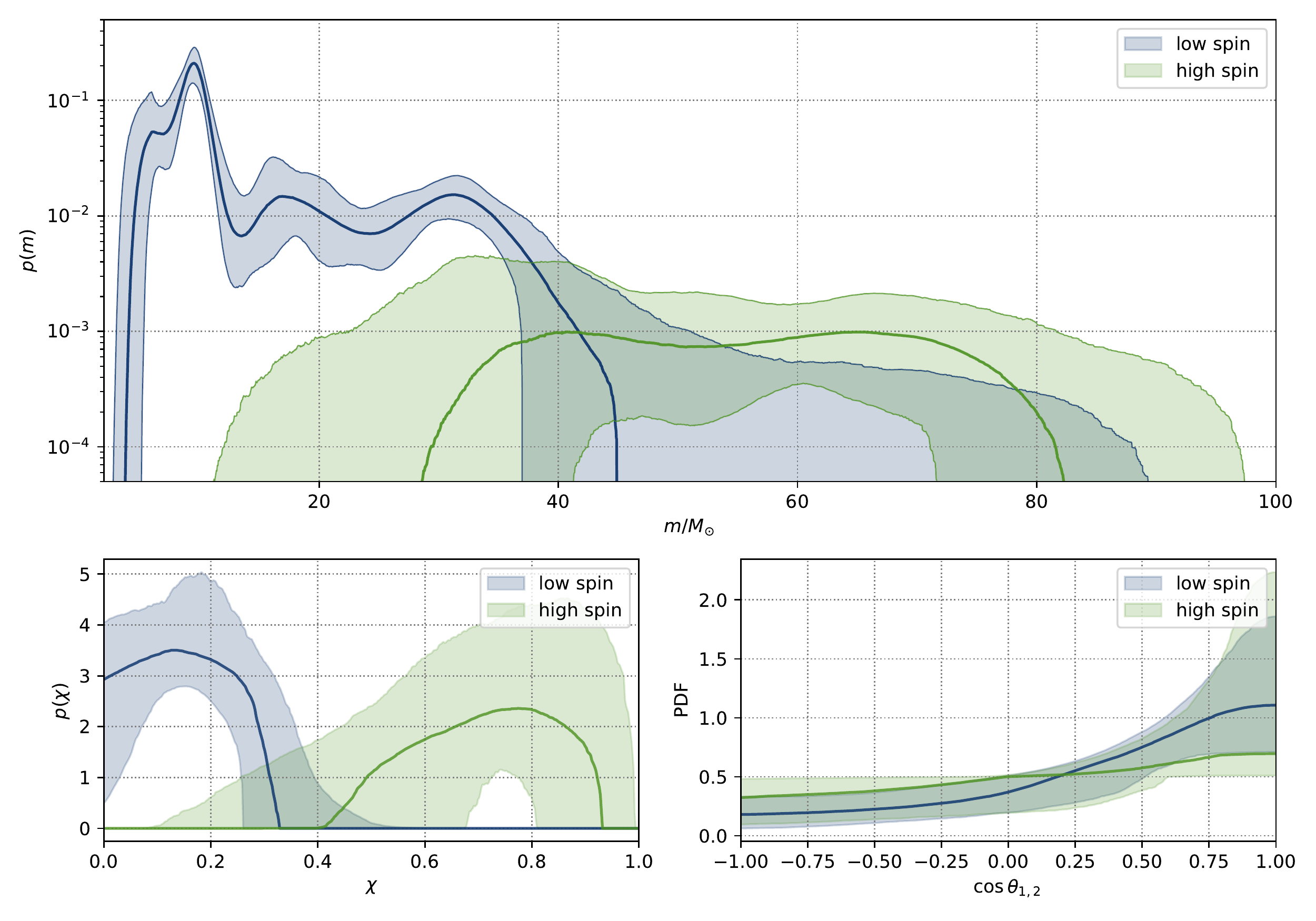}
\caption{Reconstructed distributions of BHs for each group with Two-component\&iid$\cos\theta$. The solid curves are the medians and the colored bands are the 90\% credible intervals.
}
\label{fig:iidct}
\end{figure*}

\begin{figure}
	\centering  
\includegraphics[width=0.45\linewidth]{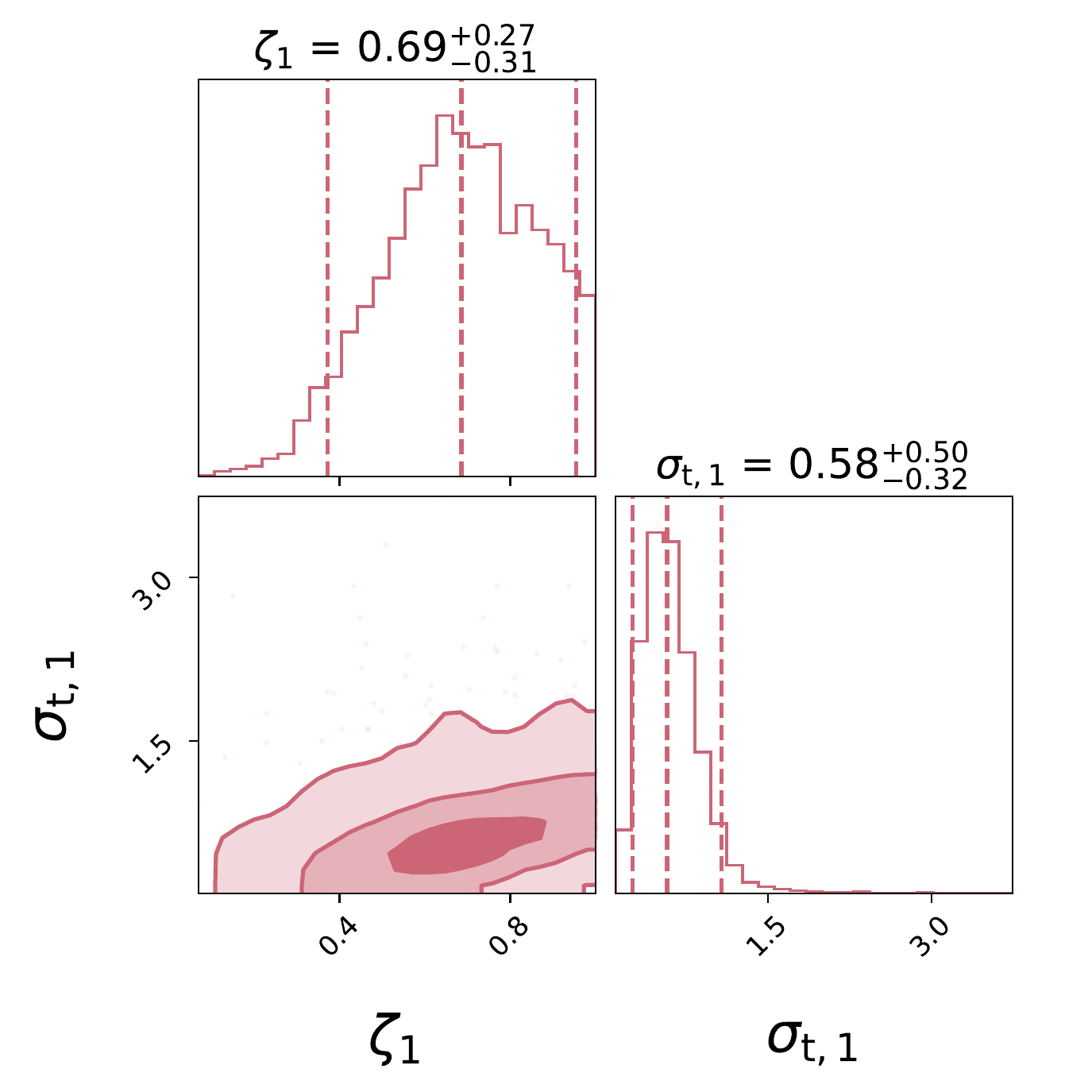}
\includegraphics[width=0.45\linewidth]{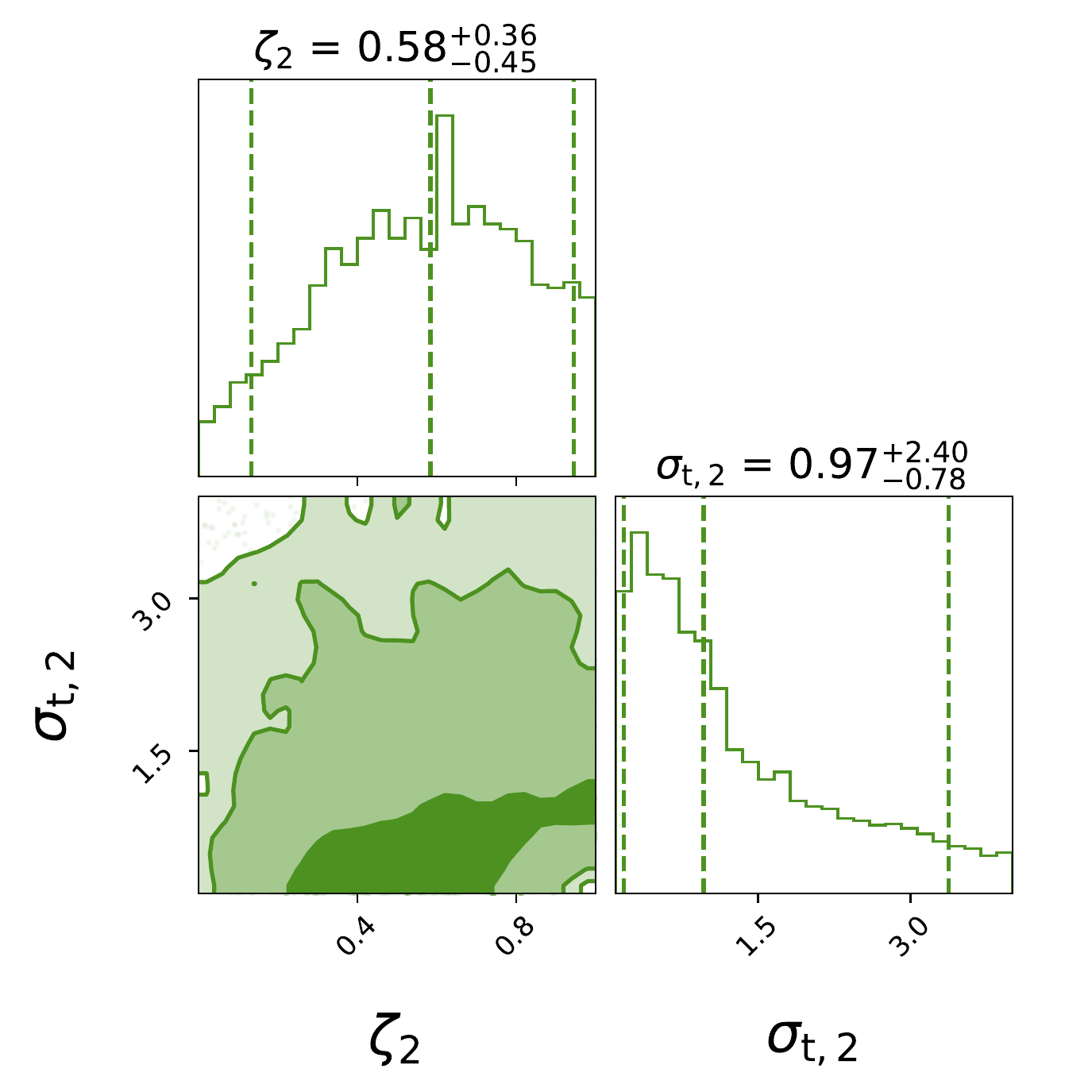}
\caption{Posterior distribution for the mixing fraction $\zeta_i$ and the width $\sigma_{{\rm t},i}$ of the nearly aligned assemblies in the first (left) and the second (right) components. The dashed lines represent the 90\% credible intervals.}
\label{fig:zeta}
\end{figure}

\section{{The contribution of hierarchical mergers to the anti-correlation between $q$-$\chi_{\rm eff}$ of BBHs.}}\label{sec:chieff}
\citet{2021ApJ...922L...5C} firstly found that the unequal-mass BBHs have larger effective spins with GWTC-2 \cite{2019PhRvX...9c1040A,2021PhRvX..11b1053A}, which was further confirmed by \citet{2023PhRvX..13a1048A} using GWTC-3 \cite{2019PhRvX...9c1040A,2021PhRvX..11b1053A,2021arXiv210801045T,2021arXiv211103606T}. Here we plot the posterior $q$-$\chi_{\rm eff}$ distributions of BBHs informed by the Two-component model, as shown in Figure~\ref{fig:qchieff}, the events with probabilities $>50\%$ to be hierarchical mergers are mark by green regions and orange point. The hierarchical mergers (containing one or two higher-generation BHs) may contribute to the anti-correlation between $q$-$\chi_{\rm eff}$ of BBHs. Because the hierarchical mergers have preference for positive $\chi_{\rm eff}$ (though the $\chi_{\rm eff}$ distribution is broad), and have less preference for symmetric systems, as shown in Figure~\ref{fig:qdist}, since most hierarchical mergers are `low + high' systems, i.e., containing only one higher-generation BH in each.

\begin{figure}
	\centering  
\includegraphics[width=0.9\linewidth]{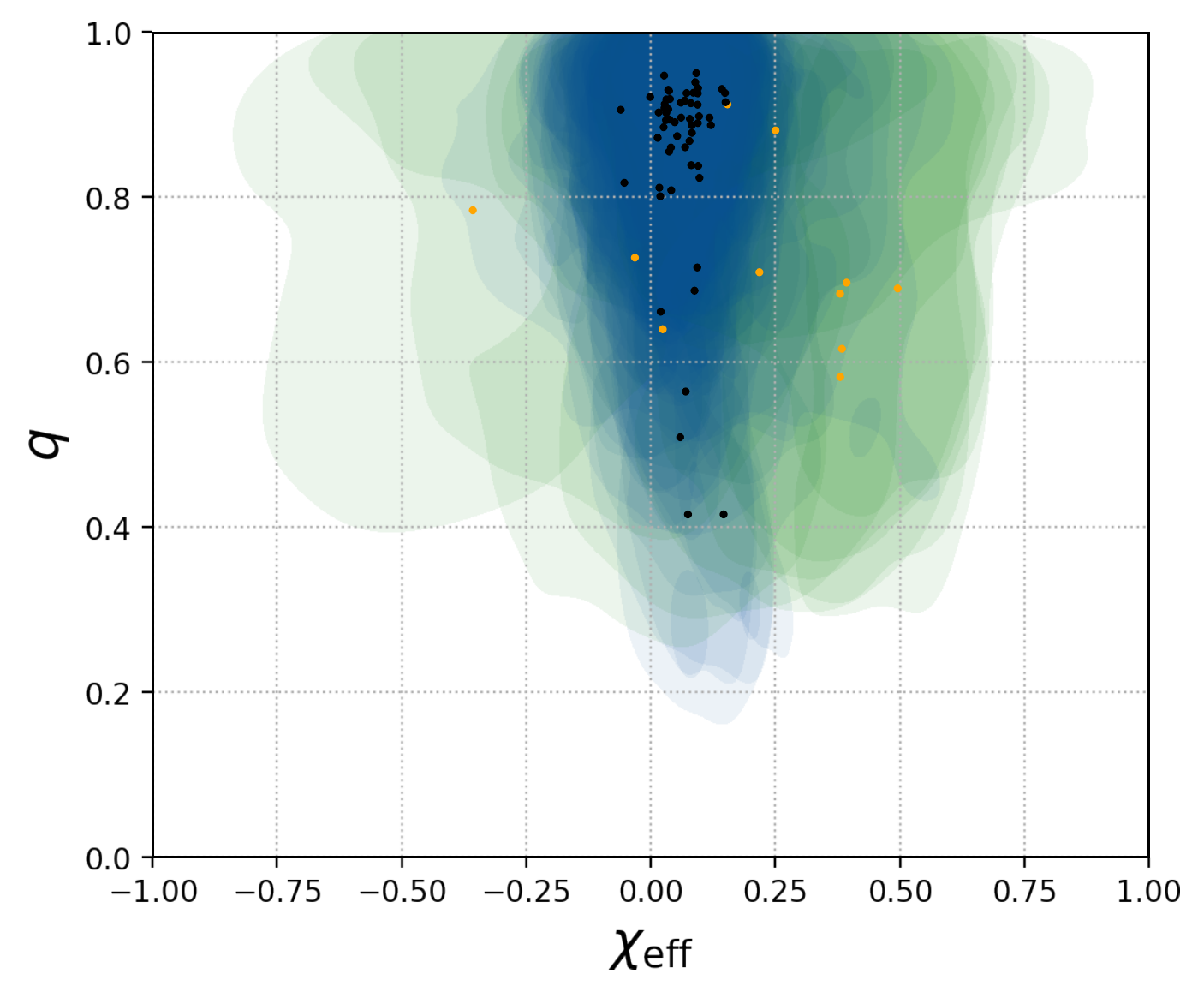}
\caption{Posteriors of effective spin and mass ratio of individual BBHs in GWTC-3, re-weighted to a population-informed prior inferred by the fiducial model. The blue (green) shaded areas mark the $90\%$ credible regions and the black (orange) points stand for the mean values for the first-generation (hierarchical-merger) BBHs. {It's likely that the correlation between the effective spin and mass ratio of BBHs (found by Callister {\it et al}.\cite{2021ApJ...922L...5C}) is mainly attributed to the hierarchical mergers.}
}
\label{fig:qchieff}
\end{figure}

\begin{figure}
	\centering  
\includegraphics[width=0.9\linewidth]{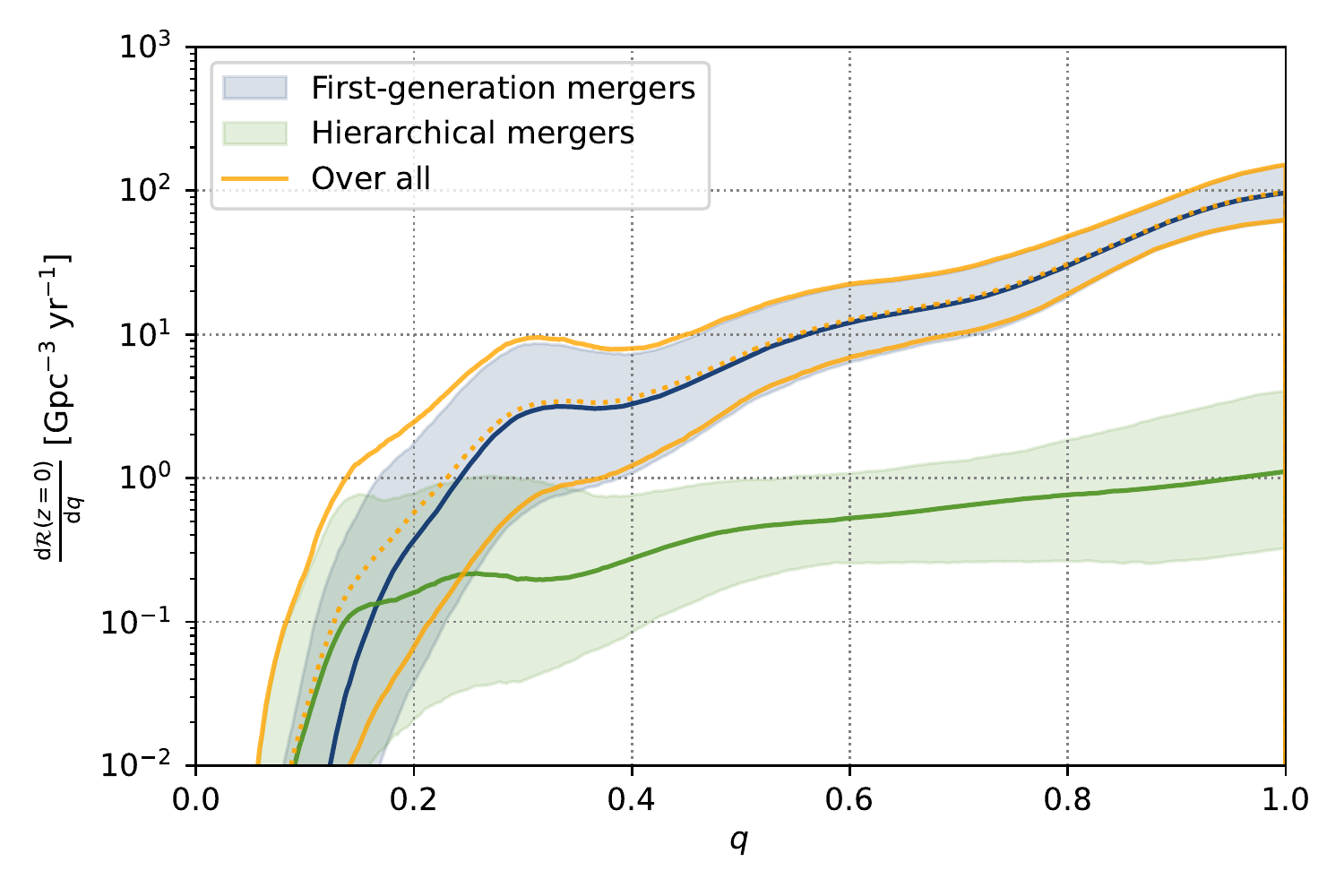}
\caption{Reconstructed mass-ratio distribution of BBHs; the solid curves are the medians and the colored bands are the 90\% credible intervals.
}
\label{fig:qdist}
\end{figure}

\clearpage
\bibliographystyle{apsrev4-1}
\bibliography{export-bibtex}